%% file: Shimajiri2015_astro-ph.tex
\documentclass[12pt,preprint]{aastex}

\usepackage{color}
\slugcomment{Draft ver. \today}

\shorttitle{Catalog of Dense Cores in the Orion A Giant Molecular Cloud}
\shortauthors{Shimajiri et al.}

\begin{document}
\title{Catalog of dense cores in the Orion A giant molecular cloud\thanks{The data sets and annotation files for $MIRIAD$ and $KARMA$ of Tables \ref{table_cores} and \ref{table_c18o_cores} are available at the NRO star-formation project web site via http://th.nao.ac.jp/MEMBER/nakamrfm/sflegacy/data.html .}}


\author{YOSHITO SHIMAJIRI\altaffilmark{1,2,3},
Y. KITAMURA\altaffilmark{4}, 
F. NAKAMURA\altaffilmark{2}, 
M. MOMOSE\altaffilmark{5},
M. SAITO\altaffilmark{2,3},
T. TSUKAGOSHI\altaffilmark{5}, 
M. HIRAMATSU\altaffilmark{2},
T. SHIMOIKURA\altaffilmark{6},
K. DOBASHI\altaffilmark{6},
C. HARA\altaffilmark{7},
and
R. KAWABE\altaffilmark{2}}

%\affil{Space Telescope Science Institute, Baltimore, MD 21218}

\email{Yoshito.Shimajiri@cea.fr}
\altaffiltext{1}{Laboratoire AIM, CEA/DSM-CNRS-Universit${\rm \acute{e}}$ Paris Diderot, IRFU/Service d'Astrophysique, CEA Saclay, F-91191 Gif-sur-Yvette, France}
\altaffiltext{2}{National Astronomical Observatory of Japan, 2-21-1 Osawa, Mitaka, Tokyo 181-8588, Japan}
\altaffiltext{3}{Nobeyama Radio Observatory, 462-2 Nobeyama, Minamimaki, Minamisaku, Nagano 384-1305, Japan}
\altaffiltext{4}{Institute of Space and Astronautical Science, Japan Aerospace Exploration Agency, 3-1-1 Yoshinodai, Chuo-ku, Sagamihara, Kanagawa 252-5210, Japan}
\altaffiltext{5}{College of Science, Ibaraki University, 2-1-1 Bunkyo, Mito, Ibaraki 310-8512, Japan}
\altaffiltext{6}{Department of Astronomy and Earth Sciences, Tokyo Gakugei University, Koganei, Tokyo 184-8501, Japan}
\altaffiltext{7}{The University of Tokyo, 7-3-1 Hongo Bunkyo, Tokyo 113-0033, Japan}


\begin{abstract}
We present Orion A giant molecular cloud core catalogs, which are based on 1.1 mm map with an angular resolution of 36$\arcsec$ ($\sim$ 0.07 pc) and C$^{18}$O ($J$=1--0) data with an angular resolution of 26.4$\arcsec$ ($\sim$ 0.05 pc). 
We have cataloged 619 dust cores in the 1.1 mm map using the Clumpfind method. The ranges of the radius, mass, and density of these cores are estimated to be 0.01 -- 0.20 pc, 0.6 -- 1.2 $\times$ 10$^2$ $M_{\odot}$, 
and 0.3 $\times$ 10$^4$ –- 9.2 $\times$ 10$^6$ cm$^{-3}$, respectively. We have identified 235 cores from the C$^{18}$O data. The ranges of the radius, velocity width, LTE mass, and density are 0.13 -- 0.34 pc, 0.31 -- 1.31 km s$^{-1}$, 1.0 -- 61.8 $M_{\odot}$, and (0.8 -- 17.5) $\times$ 10$^3$ cm$^{-3}$, respectively. 
From the comparison of the spatial distributions between the dust and C$^{18}$O cores, four types of spatial relations were revealed: (1) the peak positions of the dust and C$^{18}$O cores agree with each other (32.4\% of the C$^{18}$O cores), (2) two or more C$^{18}$O cores are distributed around the peak position of one dust core (10.8\% of the C$^{18}$O cores), (3) 56.8\% of the C$^{18}$O cores are not associated with any dust cores, and (4) 69.3\% of the dust cores are not associated with any C$^{18}$O cores. The data sets and analysis are public.
\end{abstract}





\section{Introduction}
Stars are born in dark clouds and giant molecular clouds (GMCs) which consist primarily of molecular hydrogen \citep[e.g.,][]{Cohen79,Tatematsu93, Myers95,Mardones97,Ohashi97, Dobashi05, Buckle12,Liu12}. The densest portions of the clouds are known as dense cores or molecular cloud cores where stars form through the gravitational collapse. To investigate such dense cores, the millimeter dust continuum emission is a good tracer \citep{Motte98, Johnstone06,Kauffmann08,Belloche11}. The observations in molecular lines are also required to investigate the dynamical states of the dense cores and to divide the cores overlapped in the same line of sight. Thus, a dense core survey both in the dust continuum and molecular line emission toward the overall cloud is essential to understand the formation and evolution of the dense cores.

In the last twenty years, many authors have investigated the dense cores in the dust continuum or molecular line emission toward the $\rho$ Ophiuchi, Taurus, L1333,  Chameleon I, Lupus III, CrA, Pipe nebula, and Southern Coalsack regions \citep{Motte98, Tachihara02, Onishi02, Dzib13, Alvaro13}. 
However, dense core surveys covering the entire extent of one continuous cloud have been limited.

There are mapping data both in the dust continuum and molecular line emission toward the Orion-A giant molecular cloud  (Orion-A GMC) which is the nearest GMC \citep[$D$= 400 pc][]{Menten07, Sandstrom07, Hirota08} and the best-studied one. Thus, the Orion-A GMC is one of the best regions to investigate the distributions and physical properties of the dense cores. 
In the Orion-A GMC, a large filamentary structure with a length of several $\times$ 10 pc is seen along the north-south direction \citep{Bally87,Nagahama98}; the filamentary structure is known as the integral-shaped filament. Recently, \citet{Polychroni13} identified filaments and dense cores in L 1641 N using the Herschel PACS 70/160 $\mu$m and SPIRE 250/350/500 $\mu$m data and found that most (71\%) of the prestellar cores are located along the filaments. Many authors have carried out observations in the dust continuum emission at 850 $\mu$m, 1.2 mm, and 1.3 mm to investigate the physical properties of dense cores in the Orion-A GMC \citep{Chini97, Johnstone99, Nutter07, Davis09}. Observations in dense-gas tracers such as the H$^{13}$CO$^{+}$, N$_{2}$H$^{+}$, C$^{18}$O, and CS emission lines have also been carried out \citep{Tatematsu93,Tatematsu98,Tatematsu08,Ikeda07}. These observations have, however, focused only on the integral-shaped filament. Hence, the distribution of the dense cores on the outside of the integral-shaped filament has not been revealed.

Previous authors \citep{Shimajiri11, Nakamura12, Shimajiri14} have presented wide, sensitive 1.1-mm dust-continuum and C$^{18}$O ($J$ = 1--0) line maps of the northern part of the Orion-A GMC with the AzTEC camera mounted on the Atacama Submillimeter Telescope Experiment (ASTE) 10-m telescope and with BEARS mounted on the Nobeyama Radio Observatory (NRO) 45 m telescope, respectively. In the 1.1 mm dust continuum map, we found the following new substructures in addition to the well-known integral-shaped filament. In the OMC-2/3 region, a filamentary structure is found to the east of the integral-shape filament. A shell-like structure around the HII region, M 43, and a filamentary structure associated with Dark Lane South Filament (DLSF), which is know as the photon-dominated region (PDR), \citep{Rodriguez01},  are detected. In the southmost region where an active cluster-forming region of L 1641 N is located, four filamentary structures with a length of $\sim$ 0.5 -- 2.0 pc are seen almost aligned with each other. In the C$^{18}$O ($J$=1--0) map, we found that the overall distribution of C$^{18}$O is similar to that of the 1.1 mm dust continuum emission. 

In this paper, we present a catalog of the identified cores in the 1.1-mm dust continuum and C$^{18}$O ($J$=1--0) emission line using the Clumpfind method and their properties of the peak flux density, radius, mass, density, and aspect ratio. The 1.1 mm dust continuum and C$^{18}$O emission data are from \citet{Shimajiri11} and \citet{Shimajiri14}, respectively. We compare the physical properties of the identified C$^{18}$O cores in the OMC-1, OMC-2/3, OMC-4, Dark Lane South Filament (DLSF), and bending structure regions to investigate region-to-region variations.  In addition, we compare the spatial distributions of the AzTEC/ASTE 1.1 mm dust, BEARS/NRO 45m C$^{18}$O, SCUBA/JCMT\footnote[2]{James Clerk Maxwell Telescope} 850 $\mu$m \citep{Nutter07}, BEARS/NRO 45m H$^{13}$CO$^+$ \citep{Ikeda07}, and BEARS/NRO 45m N$_2$H$^+$ \citep{Tatematsu08} cores to unveil their physical relationships.

\section{Catalogs}
\subsection{1.1-mm dust continuum emission}\label{aztec}
\subsubsection{Core identification in the AzTEC 1.1 mm dust continuum map}\label{1.1mm_ID}

The overall distribution of the AzTEC 1.1-mm dust continuum emission was revealed by \citet{Shimajiri11}. 
Here, we identify cores using the two-dimensional Clumpfind method \citep{Williams94}. This algorithm has been widely used for the identification of cores and clumps \citep[e.g.,][]{Kirk06, Rathborne09, Ikeda11,Tanaka13}. 
The algorithm works well with reasonable parameters to identify cores or clumps, although several authors have pointed out some shortcomings of the clumpfind algorithm \citep{Pineda09}. \citet{Pineda09} examined the behavior of the algorithm by changing the threshold level from 3$\sigma$ to 20$\sigma$, a wider range than \citet{Williams94} did, and found that the power-law index of the core mass function (CMF) sensitively depends on the threshold for the higher threshold range $>$ 5$\sigma$. \citet{Ikeda09}, however, demonstrated the weak dependence of core properties and CMF in Orion A on the threshold in a reasonable range from 2$\sigma$ to 5$\sigma$ levels, which was also shown by \citet{Pineda09}.  
As described in Sections \ref{sect:properties_aztec} and \ref{sect:properties_c18o}, the physical properties (radius, mass, velocity width, etc.) of the condensations identified by the Clumpfind in this paper are similar to those of the star forming dense cores traced by the H$^{13}$CO$^+$ line \citep[ and references therein. See also Appendix \ref{sect:dependence} of this paper]{Ikeda07}. In this paper, 
we will therefore define cores identified by the Clumpfind with a similar threshold value as they used.

The noise distribution of the AzTEC 1.1 mm continuum emission is not uniform over the image and is higher in the outer region. Before applying the Clumpfind \citep{Williams94}, we cut off the outer edge of the AzTEC image, where the coverage is less than 30\% and the noise level is higher by a factor of 1.3 than that in the central part. This is because the observations were made by the raster scan that boresights in azimuth and elevation and because the AzTEC 1.1 mm dust continuum image was obtained by mosaicing observations of two fields. The noise level is $\sim$ 9 mJy beam$^{-1}$ in the central region and $\sim$ 12 mJy beam$^{-1}$ in the outer region of the trimmed image. We applied the Clumpfind method to the AzTEC 1.1 mm dust continuum image with the criteria that the threshold should be the 2$\sigma$ level and the depth of the valley between adjacent peaks should be larger than the 2$\sigma$  interval. We adopted 9 mJy beam$^{-1}$ which is the noise level (= 1$\sigma$) in the central region as the noise level in the Clumpfind. Next, we removed cores whose FWHM sizes are less than the effective angular resolution ($\sim$ 36$\arcsec$). Furthermore, we only took cores having peak intensities above the 4$\sigma$ noise level. As described by \citet{Shimajiri11}, the emission around the central Orion-KL region could not be reconstructed as an accurate structure with the AzTEC data-reduction technique, because the continuum emission around Orion-KL is too bright. Thus, we removed cores in the central Orion-KL region.  As a result, we identified 619 dust cores(see also Appendix \ref{sect:c18oID}), as shown in Figure \ref{mass}. Here, we note that the 257 cores are located in the C$^{18}$O observed region, excluding the central part of the Orion-KL region (see Figure \ref{fig:c18ocores}).

To investigate the performance of the FRUIT data reduction, \citet{Shimajiri11} performed a simulated source extraction in which Gaussian sources with various FWHM sizes and total flux densities were artificially embedded in the Orion data, and obtained the following result: 
The larger FWHM size of the model source, the lower the recovered fraction of the input total flux density of the source.  In the case that the FWHM size of the input source is under 150$\arcsec$ ($\sim$0.3pc), the output total flux density is underestimated by less than 20\%. The total flux density of the source with a peak flux density under 20 Jy is underestimated by less than 10\%. Moreover, the restored image of the 1.1 mm dust continuum emission is consistent with that of the SCUBA 850 $\mu$m dust continuum emission. Consequently, the total flux densities of the sources in our map should be recovered by more than 80\%, because the peak flux density FWHM size of the identified 1.1 mm dust cores are smaller than 20 Jy and 0.2 pc, respectively, as described in Section \ref{sect:properties_aztec} (see also Appendix \ref{sect:negative}).

\subsubsection{Physical properties of the 1.1 mm dust cores}\label{sect:properties_aztec}

The mass of the 1.1 mm dust core ($\equiv$ $M_{\rm H_2}$) was derived from the total flux density at 1.1 mm, $F_{\nu}$, on the assumption that all the 1.1 mm continuum emission arises from dust and that the emission is optically thin, using the formula, 

\begin{equation} 
M_{\rm H_2}=\frac{F_{\nu}D^2}{\kappa _{\nu}B_{\nu}(T_{\rm d})}. \label{dustmass}
\end{equation}

\noindent We adopted the dust mass opacity of 
$\kappa _{\nu }=0.1 \left(\frac{\nu}{10^{12} {\rm Hz}} \right) ^\beta$ cm$^{2}$ g$^{-1}$ with $\beta $=2 \citep{Hildebrand83,Chini97} and $D$ = 400 pc. For the dust temperature, we adopted $T_{\rm d}$=20 K \citep{Cesaroni94}. 
We determined the apparent core radius $R_{\rm obs}$ as 

\begin{equation} 
R_{\rm obs}=\left(\frac{A}{\pi}\right)^{\frac{1}{2}}, \label{Robs}
\end{equation}

\noindent assuming that the core is a sphere. Here, $A$ is the projected area of the core, derived by the Clumpfind. We further estimated the core radius $R_{\rm core}$ corrected for the beam size on the assumption that the core has a Gaussian intensity profile as, 

\begin{equation} \label{eq:Rcore}
R_{\rm core}= \left\{\vbox to 24pt{} R_{\rm obs}^2 - \left[\vbox to 21pt{} \frac{\Delta \theta /2}{\sqrt{2\ln2}}(2\ln \frac{T_{\rm peak}}{\Delta I})^{1/2}\right]^2 \right\} ^{1/2}, \label{Rcore}
\end{equation}

\noindent where $\Delta \theta$ (=36$\arcsec$) is the effective beam size of the AzTEC 1.1 mm dust continuum map, $T_{\rm peak}$ is the peak intensity of the core, and $\Delta I$ is the threshold level in the core identification \citep[see][]{Williams94}. We note that the grid size of the map was set to the effective angular resolution in the Clumpfind analysis. The mean gas density ($\equiv$ $n$) of the core was derived as,

\begin{equation}\label{eq:n}
n=\frac{3M_{\rm H_2}}{4\pi \mu m_{\rm H} R_{\rm core}^3}, \label{density}
\end{equation}

\noindent where $\mu$ is the mean molecular weight per free particle taken to be 2.33 and $m_{\rm H}$ is the mass of a hydrogen atom. The range of the radius $R_{\rm core}$, mass $M_{\rm H_2}$, and density $n$ of the 1.1 mm dust cores are estimated to be 0.01 -- 0.2 pc,  0.6 -- 1.2 $\times$10$^2$ $M_{\odot}$, and 0.3 $\times$ 10$^4$ -- 9.2 $\times$ 10$^6$ cm$^{-3}$, respectively. 
Figure \ref{dust_core_hist} shows histograms of the radius, mass, density of the 1.1 mm dust cores. In the distributions of $R_{\rm core}$, $M_{\rm H_2}$, and $n$ of the 1.1 mm dust cores, peaks are seen at 0.09 pc, 1.3 $M_{\odot}$, and 1.0 $\times$10$^4$ cm$^{-3}$, respectively. 
The uncertainty of $R_{\rm core}$ is 0.07 pc, which is derived from the uncertainty in the estimation of the core projected area. 
As estimated in Section \ref{1.1mm_ID}, the uncertainty of the total flux density of the identified 1.1mm core should be less than 20\%, since the size of the core is less than 0.2 pc. Since the uncertainty of the total flux density of the identified 1.1mm core should be less than 20\% (see Section \ref{1.1mm_ID}), the uncertainty of $M_{\rm H_2}$ is 20\%.
We summarized the mean, minimum, and maximum values of each physical parameter in Table \ref{Physical_properties_1mm}.
Table \ref{table_cores} shows the physical properties of all the identified cores.

\subsection{The C$^{18}$O ($J$=1--0) emission line} \label{c18o}
\subsubsection{Core identification in the C$^{18}$O map} \label{clumpfind}
The overall distribution of the C$^{18}$O emission and the velocity structure are described by \citet{Shimajiri14}. To identify cores from the C$^{18}$O ($J$=1--0) data, we applied the Clumpfind algorithm \citep{Williams94} to the C$^{18}$O ($J$=1--0) cube data with an angular resolution of 26$\arcsec$.4 and a velocity channel width of 0.104 km s$^{-1}$. Here, note that we applied the Gaussian gridding convolution function (GCF) with 22$\arcsec$.5 FWHM size to the original C$^{18}$O data, resulting in the effective angular resolution of 26$\arcsec$.4, and did no smoothing in the velocity space. 
We adopted the criteria that the threshold should be the 2$\sigma$ level (1$\sigma$= 0.19 K in $T_{\rm MB}$) and the depth of the valley between adjacent peaks should be larger than the 2$\sigma$ interval as suggested by \citet{Williams94}.
We  also followed the additional criteria introduced by \citet{Ikeda07} and rejected ambiguous or fake core candidates whose sizes and velocity widths are smaller than the spatial and velocity resolutions, respectively: a core must contain two or more continuous velocity channels, 
and they must have at least 3 pixels whose intensities are above the 3$\sigma$ level, and 
in addition the pixels must be connected to one another in both the space and velocity domains. 
As a result, we identified 235 cores in total (see Figure \ref{fig:c18ocores}).

\subsubsection{Physical properties of the C$^{18}$O cores} \label{sect:properties_c18o}
We estimated the radius $R_{\rm core}$, velocity width in FWHM $dV_{\rm core}$, LTE (Local Thermodynamic Equilibrium) mass $M_{\rm LTE}$, virial mass $M_{\rm VIR}$, and mean density $n$ of the C$^{18}$O cores. The definitions of these parameters are the same as those used by \citet{Ikeda07} and \citet{Ikeda09}. {We estimated the core radius $R_{\rm core}$ using equation (\ref{eq:Rcore}).
To obtain the observed velocity width $dV_{\rm obs}$, we calculated a velocity dispersion within the C$^{18}$O cores, and then we multiplied the factor $\sqrt{8 {\rm ln} 2}$ to convert the core velocity dispersion to the FWHM line width on the assumption of a Gaussian profile. Thus, the observed velocity width $dV_{\rm obs}$ is given by

\begin{equation}
dV_{\rm obs} = \sqrt{8 {\rm ln} 2} \Biggr[\frac{\Sigma_i v_i^2 I_i}{\Sigma_i I_i} - \Biggl(\frac{\Sigma_i v_i I_i}{\Sigma_i I_i}\Biggr)^2\Biggr]^{1/2}, 
\end{equation}

\noindent where $v_i$ and $I_i$ are the radial velocity and intensity of the $i$-th pixel in each core, respectively.
The velocity width $dV_{\rm core}$ should be corrected for the velocity resolution, $dV_{\rm spec}$=0.104 km s$^{-1}$ as

\begin{equation}
dV_{\rm core} = \sqrt{dV_{\rm obs}^2 - dV_{\rm spec}^2}.
\end{equation}

In the mass estimation, we adopted $T_{\rm ex}$ = 20 K \citep{Cesaroni94}. For the fractional abundance of C$^{18}$O relative to H$_2$, $X_{\rm C^{18}O}$, we adopted 1.7 $\times$ 10$^{-7}$ \citep{Frerking82}. Assuming that the C$^{18}$O($J$ =1--0) emission is optically thin, we have

\begin{equation}\label{eq:LTE}
M_{\rm LTE} = 3.47 \times 10^{-2} 
\Biggl(\frac{X_{\rm C^{18}O}}{1.7 \times 10^{-7}} \Biggr)^{-1}
T_{\rm ex}e^{5.27/T_{\rm ex}}\\
\Biggl(\frac{D}{400\ {\rm pc}}\Biggr)^2
\Biggl(\frac{\Delta \theta}{26\arcsec.4}\Biggr)^2 
\Biggl(\frac{\eta_{\rm MB}}{0.4}\Biggr)^{-1} 
\Biggl(\frac{\Sigma_{i}T_{\rm A}^{\ast}\Delta V_{i}}{\rm K\ km s^{-1}}\Biggr) M_{\odot},
\end{equation}

\noindent where $\Sigma$$_i$$T_{\rm A}$$^{\ast}$$\Delta$$V_{i}$ is the total integrated intensity of the core. We adopted a main beam efficiency $\eta_{\rm MB}$ of 38\% for the 2010 season data and 36\% for the 2013 season data \citep[see][]{Shimajiri14}.  Note that we adopted the grid spacing of the data cube, $\Delta \theta$, set to the effective angular resolution of 26$\arcsec$.4. The mean gas density of the core was derived by

\begin{equation}\label{eq:n}
n=\frac{3M_{\rm LTE}}{4\pi \mu m_{\rm H} R_{\rm core}^3}. \label{density}
\end{equation}

The virial mass assuming that the cores have spherical shapes and the virial ratio are estimated as

\begin{equation}
M_{\rm VIR} = 209 \Biggl(\frac{R_{\rm core}}{\rm pc}\Biggr) \Biggl(\frac{dV_{\rm core}}{\rm km\ s^{-1}}\Biggr)^2  M_{\odot}
\end{equation}

\noindent and

\begin{equation}
\mathcal{R}_{\rm vir} = \frac{M_{\rm VIR}}{M_{\rm LTE}},
\end{equation}

\noindent respectively. The range of $R_{\rm core}$, $dV_{\rm core}$, $M_{\rm LTE}$, and $n$ are 0.13 -- 0.34 pc, 0.31 -- 1.31 km s$^{-1}$, 1.0 -- 61.8 $M_{\odot}$, and (0.8 -- 17.5) $\times$ 10$^{3}$ cm$^{-3}$, respectively (see Table \ref{Physical_properties_c18o}). Figure \ref{dust_core_hist} shows histograms of $R_{\rm core}$, $M_{\rm LTE}$, and $n$ of the C$^{18}$O cores: peaks are seen at 0.22 pc, 15.1 $M_{\odot}$, and 2.9 $\times$10$^3$ cm$^{-3}$, respectively. 
The uncertainty of $R_{\rm core}$ is 0.05 pc derived from the uncertainty in the estimation of the core projected area. 
The uncertainty of $dV_{\rm core}$ is 0.104 km s$^{-1}$, corresponding to the velocity resolution. The uncertainty of $M_{\rm LTE}$ is a factor of 6, which is derived from the uncertainty in $X_{\rm C^{18}O}$ \citep{Shimajiri14}. 
The physical properties of the individual C$^{18}$O cores are listed in Table \ref{table_c18o_cores}. 

Figures \ref{fig5} (a) and (b) show virial ratio - LTE mass and virial ratio - density relations of the identified C$^{18}$O cores, respectively. We consider that the cores with 
a $\mathcal{R}_{\rm vir}$ value
 less than three are under virial equilibrium according to \citet{Ikeda09}
 With the increasing LTE mass and density, the virial ratio decreases; a similar trend between the LTE mass and the virial ratio was also found by \citet{Dobashi96}, \citet{Yonekura97}, and \citet{Ikeda07}. Most of the bound C$^{18}$O cores are distributed in the filamentary structure. On the other hand, most of the unbound C$^{18}$O cores are distributed outside the filamentary structure (see Section \ref{compare}). Figure \ref{fig5} (c) shows a LTE mass - $R_{\rm core}$ relation of the C$^{18}$O cores. The best-fit power-law functions for the unbound and bound cores are log$_{10}$($M_{\rm LTE}$/$M_{\odot}$) = (3.8 $\pm$ 13.6)log$_{10}$($R_{\rm core}$/pc)+(3.2 $\pm$ 9.3) and  log$_{10}$($M_{\rm LTE}$/$M_{\odot}$) = (6.6 $\pm$ 16.5)log$_{10}$($R_{\rm core}$/pc)+(5.3 $\pm$ 10.5), respectively.\footnote[3]{The $\chi$-square fitting method with uncertainties in both the x and y coordinates was applied by using the IDL MPFITEXY tool \citep{Williams10}.} No significant difference can be seen between the data distributions for the unbound and bound core due to the larger uncertainty in the estimation of the LTE mass, although the trend that the LTE mass of the bound core is larger than that of the unbound core can be recognized in Figure \ref{fig5} (c). 
Figure \ref{fig5} (d) shows a $dV_{\rm core}$ - $R_{\rm core}$ relation of the identified C$^{18}$O cores. 
The best-fit power-law functions for the unbound and bound cores are 
log$_{10}$($dV_{\rm core}$/km s$^{-1}$) = (2.7 $\pm$ 14.6)log$_{10}$($R_{\rm core}$/pc)+(1.6 $\pm$ 10.0) and  log$_{10}$($dV_{\rm core}$/km s$^{-1}$) = (2.4 $\pm$ 8.3)log$_{10}$($R_{\rm core}$/pc)+(1.3 $\pm$ 5.3), respectively. No significant difference can be seen between the data distributions for the unbound and bound cores. }

Figure \ref{fig:dec_vsys} shows the change of the systemic velocity of the cores (i.e., the peak C$^{18}$O velocity) along the declination, which is best fitted by the relation ($V_{\rm sys}$/km s$^{-1}$)= (5.1 $\pm$ 0.3) (Dec/deg) + (36.5 $\pm$ 1.8). This result suggests a presence of a large-scale velocity gradient along the south-north direction ($\sim$ 0.7 km s$^{-1}$ pc$^{-1}$) (cf., the velocity gradient along the integral-shaped filament is estimated to be 1.0 km s$^{-1}$ pc$^{-1}$ in the $^{12}$CO, $^{13}$CO, H$^{13}$CO$^+$, CS lines \citep{Bally87, Tatematsu93, Ikeda07, Shimajiri11, Buckle12, Shimajiri14}).

\section{Discussion}
\subsection{Mass distribution of the 1.1 mm dust cores}
\begin{figure*}
\centering
\includegraphics[width=75mm,angle=0]{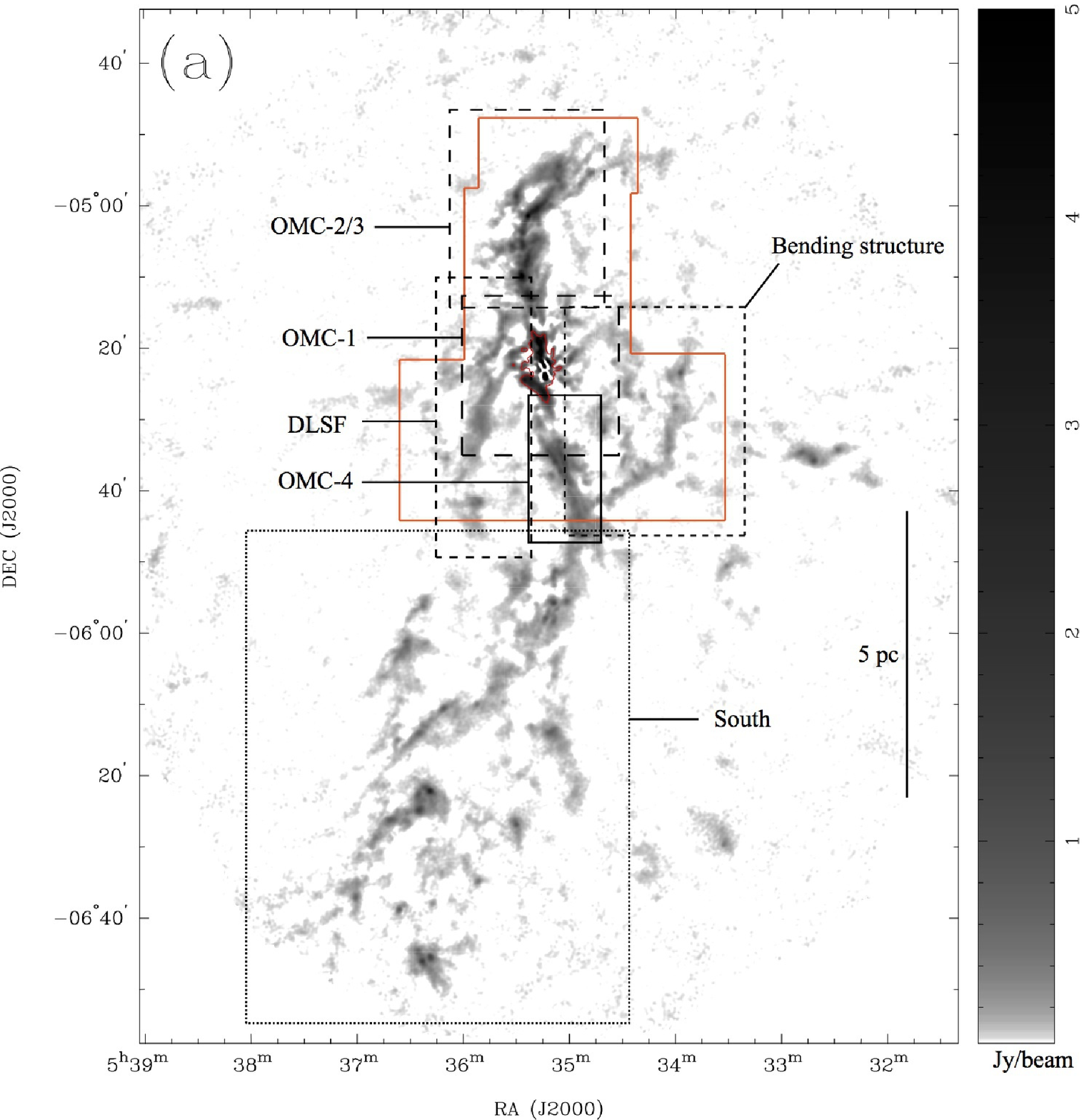}
\includegraphics[width=67mm,angle=0]{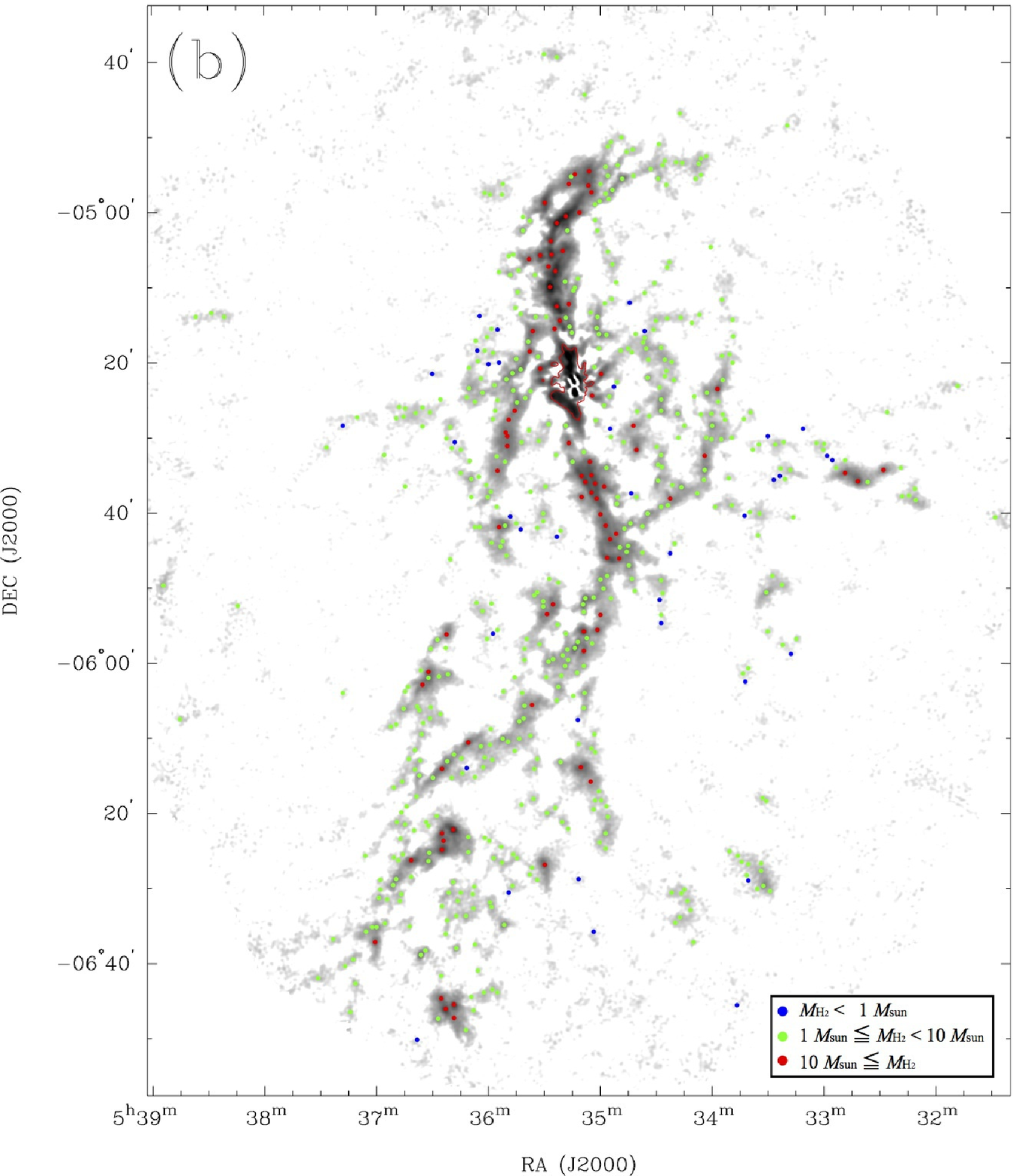}
\caption{
(a) AzTEC 1.1 mm dust continuum emission map of the Orion A GMC in gray scale. The five distinct regions of the OMC-1, OMC-2/3, OMC-4, DLSF, bending structure, and south regions are indicated by the boxes with different black broken lines, corresponding to the regions shown in Figs \ref{omc23_id} -- \ref{orion_s_id}. The 1.1 mm emission around the central Orion-KL region could not be reconstructed as an accurate structure with the AzTEC data-reduction technique, because the continuum emission around Orion-KL is too bright. The red polygon shows the region mapped in C$^{18}$O ($J$=1--0). The linear scale of 1 pc is shown on the right side. (b) The identified dust cores in the AzTEC 1.1 mm map. The red, green, and blue filled circles show the positions of the dust cores with masses of $\ge$ 10 $M_{\odot}$, 1 -- 10 $M_{\odot}$, and $<$ 1 $M_{\odot}$, respectively (see also Fig \ref{dust_core_hist}.).}
\label{mass}
\end{figure*}

\begin{figure*}
\centering
\includegraphics[width=80mm,angle=0]{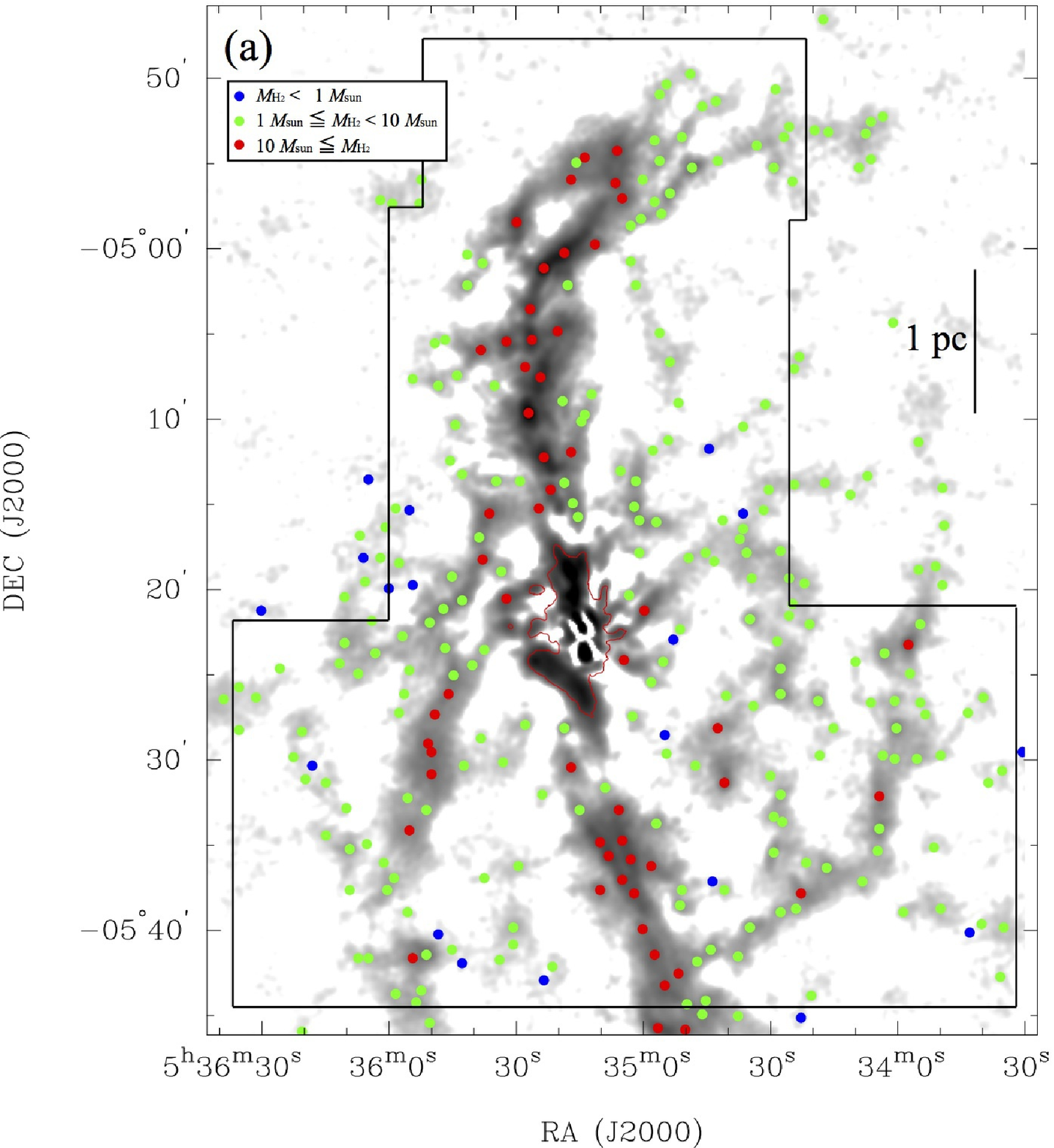}
\includegraphics[width=80mm,angle=0]{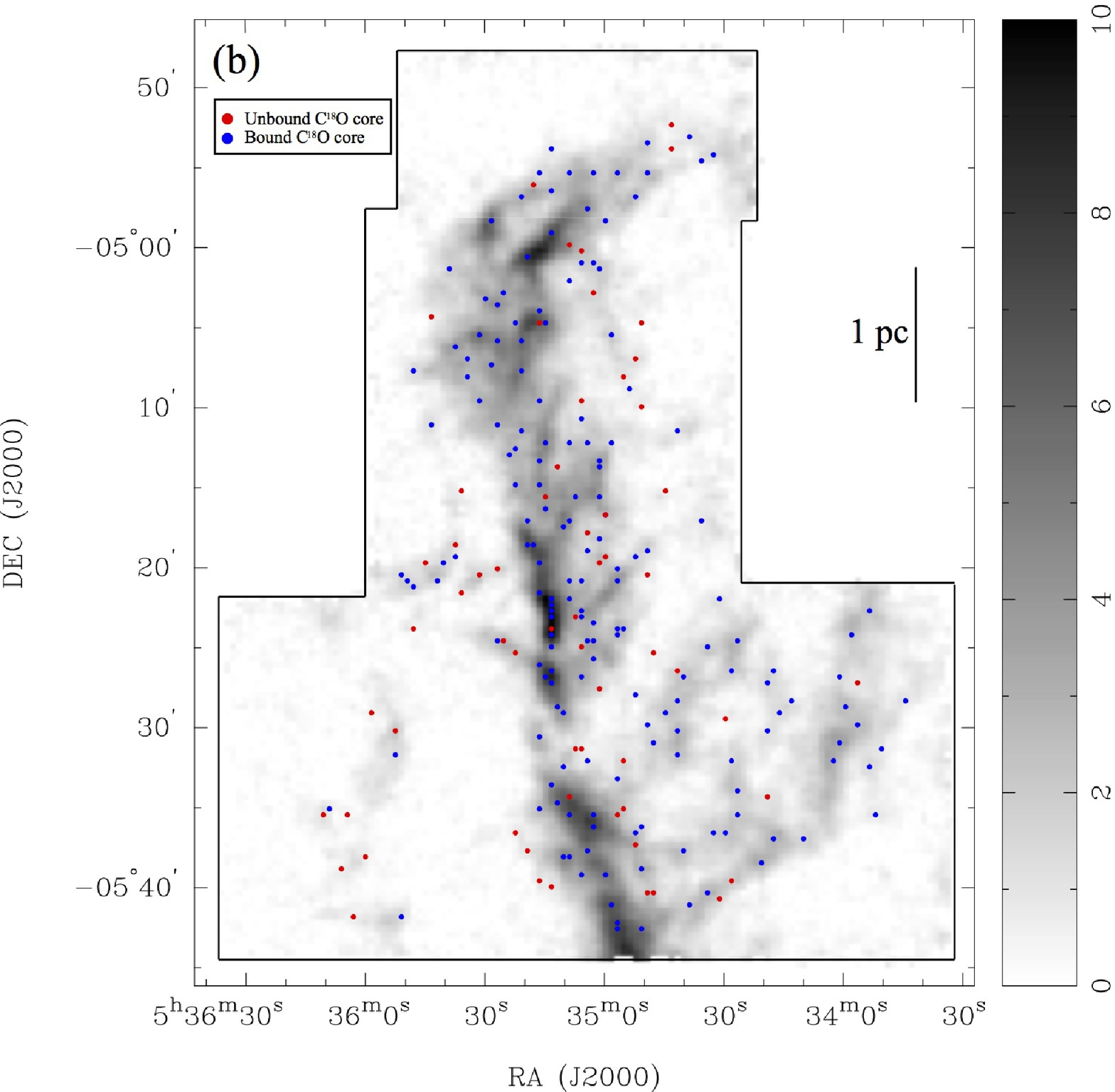}
\caption{(a) Close-up view of the upper half of Fig. \ref{mass} (b). The black polygon shows the the area mapped in C$^{18}$O. (b) Positions of the identified C$^{18}$O cores shown on the integrated intensity map (gray scale). The blue and red filled circles correspond to gravitationally bound and unbound cores, respectively. The gray scale bar on the right side of the panel is for the C$^{18}$O intensity in K km s$^{-1}$.}
\label{fig:c18ocores}
\end{figure*}

Region-to-region variation of the dust core mass distribution can be recognized in Figure \ref{mass}. The high-mass cores ($M_{\rm H_2}$ $\ge$ 10.0 $M_{\odot}$) are mainly located in the integral-shaped filament. On the other hand, the intermediate-mass (1.0 $M_{\odot}$ $\le$ $M_{\rm H_2}$ $<$ 10.0 $M_{\odot}$) and low-mass ($M_{\rm H_2}$ $<$ 1.0 $M_{\odot}$) cores tend to appear on the outside of the integral-shaped filament. One of the reasons why the high-mass cores are concentrated in the integral-shaped filament is that the 1.1 mm dust cores in the integral-shaped filament could not be resolved in the AzTEC observations with an angular resolution of $\sim$ 36$\arcsec$ (corresponding to $\sim$ 0.07 pc at 400 pc). In fact, previous SCUBA 850 $\mu$m observations with an angular resolution of $\sim$ 14$\arcsec$ \citep{Nutter07}  resolved each 13 1.1 mm dust core into two or three smaller cores (also see Table \ref{table_cores}). 
Furthermore, interferometer observations with a high angular resolution of $\sim$ 1 -- 3$\arcsec$ revealed that  cores in OMC-2/FIR 4, OMC-2/FIR 6, and L1641 N consist of several smaller cores \citep{Shimajiri08, Shimajiri09, Stanke07}. Observations with interferometers such as the Atacama Large Millimeter/Submilimeter Array (ALMA) are crucial to unveil the internal structures of the cores and confirm whether the trend of the region-to-region variation of the dust core mass distribution is significant.

\subsection{Cross identification between the 1.1 mm dust cores and the cataloged YSOs}\label{spitzer}

Recently, \citet{Megeath12} identified young stellar objects (YSOs) in the Orion A and B molecular clouds from the infrared array camera (IRAC)/Spitzer and multi-band imaging photometer for Spitzer (MIPS)/Spitzer data. They classified 2991 YSOs as pre-main sequence stars with disks and 488 YSOs as protostars using the criterion that the spectral index $\alpha$ (=$\lambda F_{\lambda}$/$d\lambda$) $\ge$ -0.3 for protostars and $\alpha$ $<$ -0.3 for pre-main sequence (PMS) stars with disks.  The active galactic nucleus (AGN), galaxies with polycyclic aromatic hydrocarbons (PAH) emission, outflow shock knots and stars contaminated by PAH emission are excluded from these YSO catalog \citep[see][]{Megeath12}. The AzTEC map includes 1801 of  2991 pre-main sequence stars with disks and 202 of 488 protostars. The central part of the Orion-KL region, which could not be well reconstructed by the  AzTEC, includes 122 pre-main sequence stars with disks and 22 protostars. We removed these pre-main sequence stars with disks and protostars from the following comparison with our cores.

Figure \ref{fig:cores_YSOs} shows a histogram of 
the separations between the YSOs cataloged by \citet{Megeath12} and the peak positions of the 1.1 mm dust cores nearest to them. In the figure, only the protostars and PMSs with separations less than 36$\arcsec$, which is the angular resolution of the AzTEC 1.1 mm dust continuum data, are counted. The distribution of the protostars has a peak at a separation of 7.5$\arcsec$ and decreases to 15$\arcsec$. Thus, we adopted the 15$\arcsec$ separation as a criterion to identify the YSOs associated with the 1.1 mm dust cores.
As a result, 50/1679 (3.0\%) of the pre-main sequence stars with disks are associated with the 1.1 mm dust cores and 49/180 (36.1\%) of the protostars associated with the cores. Figures \ref{postageA}--\ref{postageC} show close-up images of the 1.1 mm dust cores associated with the YSOs. We note that the spatial resolution of the AzTEC 1.1 mm dust continuum map is $\sim$ 36$\arcsec$ (corresponding to $\sim$ 0.07 pc at 400 pc), which cannot resolve each dense core in a cluster-forming region where many dense cores are concentrated in small areas. This effect should lower the detection rate. For example,  previous dust continuum observations in the 1.3-mm dust continuum emission with an angular resolution of 11$\arcsec$ found eleven dust cores in the OMC-2 region \citep{Chini97}, but only three cores are identified in our AzTEC 1.1 mm dust continuum map. In addition, we cannot exclude the possibility that the dust cores and the YSOs overlap by chance on the same line of sight. However, this possibility is thought to be small, since the YSOs are likely to be in the Orion-A GMC.

To investigate the region-to-region variation of the environments and evolutionary phases of star formation, we compared the number density of the 1.1 mm dust cores, protostars, and pre-main sequence stars in OMC-1, OMC-2/3, OMC-4, DLSF, the bending structure, and the southern part in the 1.1 mm dust continuum map. The OMC-1 region is known to be a high-mass star-forming region \citep{Furuya09,Bally11,Lee13}. The OMC-2/3 region is known to be an intermediate-mass star-forming region \citep{Takahashi06,Takahashi08,Takahashi09,Takahashi12,Takahashi13, Shimajiri08, Shimajiri09}. The DLSF is influenced by the far ultraviolet (FUV) radiation from the trapezium cluster \citep{Rodriguez01, Shimajiri11,Shimajiri13, Shimajiri14}. In the southern area of the 1.1 mm dust continuum map, cloud-cloud collision is suggested to be occurring by \citet{Nakamura12}. We summarize numbers and number densities of the 1.1 mm dust cores, protostars, and pre-main sequence stars in each region in Table \ref{Spitzer_dist}. Although the six areas/regions are not uniquely and rigorously defined and this is likely to introduce some uncertainties in statistical analyses and discussions, the number density of the 1.1 mm dust cores in the southern part of the 1.1 mm dust continuum map (4.2 cores pc$^{-2}$) is found to be the lowest. 
In the integral-shaped filament, the number densities of the protostars in OMC-1, OMC-2/3, and OMC-4 (6.3 -- 8.0 protostars pc$^{-2}$) are similar. On the other hand, the number densities of protostars in the other regions of DLSF, the bending structure, and the southern part are 3 - 27 times lower than those in the integral-shaped filament. This is consistent with the fact that the OMC-1, OMC-2/3, and OMC-4 regions are more active star-forming regions than the DLSF, bending structure, and southern regions. The number density ratios of the 1.1 mm dust cores without YSOs to YSOs including protostars and pre-main sequence stars with disk in OMC-2/3 (6.8 cores pc$^{-2}$/41.8 YSOs pc$^{-2}$ = 0.16) and OMC-4 (7.4 cores pc$^{-2}$/81.3 YSOs pc$^{-2}$ = 0.09) are 2.6--6.0 times lower than those in the other regions of DLSF (11.1 cores pc$^{-2}$/20.3 YSOs pc$^{-2}$= 0.54), the bending structure (8.4 cores pc$^{-2}$/18.1 YSOs pc$^{-2}$= 0.46), and the southern part (4.2 cores pc$^{-2}$/8.6 YSOs pc$^{-2}$ = 0.49). Thus, we speculate that the DLSF, bending structure, and southern regions are in younger evolutionary stages than that of the integral-shaped filament.

This interpretation is supported by the spatial variation of the mean density of the 1.1mm dust core.
The mean densities of the 1.1mm dust cores in OMC-2/3 and OMC-4 are 6.2 $\times$ 10$^4$ and 18.9 $\times$ 10$^4$ cm$^{-3}$, respectively, while those in the bending, DLSF, and south regions are 1.7 $\times$ 10$^4$, 2.1 $\times$ 10$^4$, and 3.0 $\times$ 10$^4$ cm$^{-3}$, respectively. Consequently, the mean densities of the 1.1 mm dust cores in the integral-shaped filament is 2.1 -- 11.1 times higher than those in the DLSF, bending structure, and southern regions. 
Since the gas density is generally thought to increase as star formation progresses, the difference in mean density suggests that the integral-shaped filament is most evolved. 
We must consider another possibility that the OMC-2/3 and OMC-4 regions are forming higher mass stars. In fact, the OMC-2/3 region is known as an intermediate star forming region \citep{Takahashi06}. However,  as shown in Fig. \ref{fig6} (c), the difference in the $M_{\rm LTE}$ distribution between OMC-2/3 and Bending as well as between DLSF and OMC-4 is not significant, suggesting that the stars with similar masses will form on the assumption of the same star formation rate among the regions.

\subsection{Comparison of the core properties in the OMC-1, OMC-2/3, OMC-4, DLSF, and bending structure regions} \label{}
In the Orion-A GMC, the environments and evolutional phases of star formation have considerable region-to-region variations as discussed in Section \ref{spitzer}. To investigate the influence of the different environment and evolutional phase on physical properties of the dense cores, we compare the physical properties of the C$^{18}$O cores in the OMC-1, OMC-2/3, OMC-4, DLSF, and bending structure regions. 
Figure \ref{fig6} shows histograms of the radius, velocity width, LTE mass, and virial ratio of the C$^{18}$O cores, respectively, in the five regions. 
To quantitatively examine the similarities of the physical properties among the five regions, we applied the Kolmogorov-Smirnov (KS) test, which considers the maximum deviation between the distributions of two samples \citep[e.g.,][]{Wall12}, to the five histograms in each panel (see Tables \ref{KStest_Rcore_dV} and \ref{KStest_Ratio_MLTE}). 
The results of the KS test show that there is no significant difference among the five regions for most of the physical properties. 
Although the $R_{\rm core}$ values in DLSF are relatively small as shown in Fig. \ref{fig6} (a), the KS test shows no significant difference for the $R_{\rm core}$ values due to the small sample for DLSF.
The $dV_{\rm core}$ distribution in DLSF is significantly different from that in the OMC-2/3 region with a significant level of five percent ($p$-value = 3.1\%). The $dV_{\rm core}$ value in DLSF is relatively small as shown in Fig. \ref{fig6} (b).
The LTE-mass distribution in DLSF is significantly different from those in OMC-1 and OMC-2/3 with a significant level of five percent ($p$-value = 0.7\%). 
As described in the above, the LTE-mass distributions in DLSF and OMC-4 have two distinct peaks at $M_{\rm LTE}$ = 2.9 $M_{\odot}$ and 24.2 $M_{\odot}$.

\subsection{Effects of external pressure and internal magnetic field on the dynamical states of the C$^{18}$O cores}
The external pressure and internal magnetic ($B$) field in the cores are important factors to determine their dynamical states. Recently, \citet{Li13} performed high angular (5$\arcsec$) resolution observations with a velocity resolution of 0.6 km s$^{-1}$ in NH$_3$ using the Very Large Array (VLA) and Green Bank Telescope (GBT) toward the OMC-2/3 region (c.f., the mean velocity width of the C$^{18}$O cores is 0.61 km s$^{-1}$). They found that most of the massive cores are supercritical from the comparison between mass and critical mass, suggesting that cores will collapse or fragment. The critical mass $M_{\rm critical}$ is defined as $M_{\rm critical}$=$M_{\rm J}$ + $M_{\Phi}$, where $M_{\rm J}$ and $M_{\Phi}$ are the Jeans mass and the maximum mass that can be supported by a steady $B$ field \citep[e.g.,][]{McKee92}. Here, we investigate the dynamical states of the C$^{18}$O cores with the external pressure and the internal magnetic field according to \citet{Li13}.

The Jeans mass is estimated as 

\begin{equation}
M_{\rm J} = 1.182 \frac{\sigma^4}{G^{3/2}P_{\rm ic}^{1/2}},
\end{equation}

\noindent 
where $G$ is the gravitational constant, $\sigma$ is the one-dimensional velocity dispersion within the core, and $P_{\rm ic}$ is the external pressure. The pressure $P_{\rm ic}$ can be expressed as $P_{\rm ic} =n_{\rm ic} \mu m_{\rm H} \sigma_{\rm ic}^2$.
We considered the tenuous gas traced by the $^{13}$CO (1--0) emission line as the external gas of the C$^{18}$O cores. Thus, we adopted the density $n_{\rm ic}$ of  2.0 $\times$ 10$^3$ cm$^{-3}$ \citep{Nagahama98} and the velocity dispersion $\sigma_{\rm ic}$ of 0.67 km s$^{-1}$ \citep{Shimajiri14}. The velocity dispersion in the core, $\sigma$, can be estimated from the core velocity width $dV_{\rm core}$ as $\sigma$ = $dV_{\rm core}$ / $\sqrt{8 {\rm ln}2}$, assuming a Gaussian velocity profile. As a result, the Jeans mass $M_{\rm J}$ of the C$^{18}$O cores is estimated to be  0.2 -- 55.6 $M_{\odot}$ (see Table \ref{Physical_properties_c18o}). On the other hand, the maximum mass can be estimated using the formula,

\begin{equation}
M_{\Phi} = c_{\Phi}\frac{\pi B R_{\rm core}^2}{G^{1/2}}, 
\end{equation}

\noindent where $c_{\Phi}$ is a non-dimensional  scaling factor and is 0.12 for an axisymmetric isothermal cloud \citep{Tomisaka88} and $B$ is the magnetic field strength. \citet{Crutcher99} derived the field strength from the observations of the Zeeman effect in CN toward OMC-1 and found $B$ $\sim$ 0.19 -- 0.36 mG. 
Here, we adopted 0.1 mG as the field strength according to \citet{Li13}.
 Thus, the estimated maximum mass could be the lower limit. As a result, the maximum mass $M_{\Phi}$ of the C$^{18}$O cores is estimated to be  0.1 -- 0.8 $M_{\odot}$ (see Table \ref{Physical_properties_c18o}).

\noindent Here we define the critical mass ratio $\mathcal{R}_{\rm c}$ as,

\begin{equation}
\mathcal{R}_{\rm c} = \frac{M_{\rm LTE}}{M_{\rm J} + M_{\Phi}}.
\end{equation}

We list $M_{\rm J}$, $M_{\Phi}$, and $\mathcal{R}_{\rm c}$ of each C$^{18}$O core in Table \ref{table_c18o_cores}.
Figure \ref{fig:virial_critical} shows a relation between the virial and critical mass ratios of the C$^{18}$O cores. 
The $\mathcal{R}_{\rm c}$ value decreases with the increasing the $\mathcal{R}_{\rm vir}$ value. The difference between $\mathcal{R}_{\rm vir}$ -- $\mathcal{R}_{\rm c}$ relations for the bound and unbound C$^{18}$O cores are recognized in Fig. \ref{fig:virial_critical}. The slopes for the bound C$^{18}$O cores are larger than those for the unbound C$^{18}$O cores.
The best-fit power-law functions are $\mathcal{R}_{\rm c}$ = 6.6 $\pm$ 0.12 $\mathcal{R}_{\rm vir}$$^{-1.16 \pm 0.04 }$ for the unbound C$^{18}$O cores and $\mathcal{R}_{\rm c}$ = 12.3 $\pm$ 2.9 $\mathcal{R}_{\rm vir}$$^{-1.56 \pm 0.18}$ for the bound C$^{18}$O cores. 
We found that all the bound C$^{18}$O cores are supercritical ($\mathcal{R}_{\rm c}$ $\ge$ 1) and 25 of the 61 unbound C$^{18}$O cores are subcritical ($\mathcal{R}_{\rm c}$ $<$ 1) as seen in Fig. \ref{fig:virial_critical}. 
Here, we note that there are large uncertainties in the estimates of the Jeans mass owing to the relation of $M_{\rm J}$ $\propto$ $\sigma^4$ and in the estimation of the maximum mass owing to the large uncertainty in $B$. For further investigation of the dynamical states of the dense cores, observations with higher spectral resolution and measurements of the magnetic field strength would be crucial.

\subsection{Comparison between the 1.1 mm dust and C$^{18}$O cores} \label{compare}
In Sections \ref{aztec} and \ref{c18o},  we have identified the 257 dust and 213 C$^{18}$O cores using the Clumpfind method and estimated their physical properties in the C$^{18}$O observing area. The mean $R_{\rm core}$ value of the C$^{18}$O cores (0.22 $\pm$ 0.04 pc) is 2.4 times larger than that of the 1.1 mm dust cores (0.09 $\pm$ 0.03 pc). The 1.1 mm dust continuum emission probably traces the inner part of the dense cores, while the C$^{18}$O emission line probably traces the outer part of the dense cores. The $n$ value range of the 1.1 mm dust cores ((0.3 -- 915.0) $\times$ 10$^{4}$ cm$^{-3}$) is 13 times larger than that of the C$^{18}$O cores ((0.8 -- 17.5) $\times$ 10$^{3}$ cm$^{-3}$).

We compared the spatial distribution of the 1.1 mm dust cores with that of the C$^{18}$O cores. Figures \ref{omc23_id} -- \ref{orion_w_id} show the positions of the 213 C$^{18}$O cores on the AzTEC 1.1 mm dust continuum map in the OMC-2/3, OMC-4, DLSF, and bending structure regions. 
Figure \ref{orion_omc1_id}  shows the positions of the C$^{18}$O cores in the OMC-1 region where the 1.1 mm image could not be well reconstructed there. Figure \ref{orion_s_id} shows the positions of the 1.1 mm dust cores in the southern part of the 1.1 mm dust continuum map where there is no C$^{18}$O data. We found that the spatial relation between the 1.1 mm dust and C$^{18}$O cores can be categorized into the following four types 
as shown in Figure \ref{category_sample}:

\begin{enumerate}
\item[]{[Category A] The peak positions of the 1.1 mm dust and C$^{18}$O cores agree with each other within the 1.1 mm map resolution of 36$\arcsec$.}
\item[]{[Category B] Several C$^{18}$O cores are distributed around the peak positions of the 1.1 mm dust cores within the 1.1 mm map resolution of 36$\arcsec$.}
\item[]{[Category C] The C$^{18}$O cores not associated with any 1.1 mm dust cores.}
\item[]{[Category D] The 1.1 mm dust cores not associated with any C$^{18}$O cores.}
\end{enumerate}

Table \ref{category_table} summarizes the numbers of the 1.1 mm dust and C$^{18}$O cores in each category.  We found 69 pairs of the dust and C$^{18}$O cores in Category A. In this category, one C$^{18}$O core seems to be associated with one 1.1 mm dust core. In Category B, there are 23 C$^{18}$O cores (10.8\%) and 10 dust cores (3.9\%). There are three possible explanations for the B type cores. 
One is the several cores are overlapped in the same line of sight. The $V_{\rm LSR}$ is different among the C$^{18}$O cores which are associated with the same 1.1 mm dust core (see Tables \ref{table_cores} and \ref{table_c18o_cores}). 
Furthermore, the peak positions of 6/10 dust cores (AzTEC-Ori 110, 232, 234, 242, 272, and 482) coincide with those on the C$^{18}$O integrated intensity map which includes all velocity components as well as the dust continuum map. These results suggest that several cores are distributed on the same line of sight.
Second is due to the poor angular resolution of the AzTEC data. The resolution of the AzTEC data (= 36$\arcsec$) is larger than in the C$^{18}$O data (= 26$\arcsec$). To confirm this possibility, we identified the C$^{18}$O cores using the C$^{18}$O data smoothed to the same angular resolution as the AzTEC 1.1mm map (=36$\arcsec$). As a result, 5 of 10 dust cores (AzTEC-Ori 162, 232, 242,273, and 482) are associated with one C$^{18}$O core identified on the 36$\arcsec$ map, although these dust cores are associated with two C$^{18}$O cores identified on the 26$\arcsec$.4 map. This result suggests that these 1.1 mm dust cores are not resolved due to the poor angular resolution.
Hence, distinct condensations resolved by the C$^{18}$O observations may remain unresolved by the AzTEC observations. 
The other is the depletion of the C$^{18}$O molecules in the central parts of dust cores. Such depletion in the central part of the dust condensation has been reported in the B 68 and L 1498 regions \citep{Bergin02,Tafalla02}. Although the inter-core diffuse gas in the Orion-A GMC is warmer than those in low-mass star forming regions, the dense cores are well-shielded from nearby radiation and become cool enough for CO molecules to freeze onto dust grains. In fact, the CO depletion in the dense cores in the Orion-A GMC has been reported by several authors \citep{Ripple13,Tatematsu14, Ren14}.

\begin{figure*}
\centering
\includegraphics[width=50mm,angle=270]{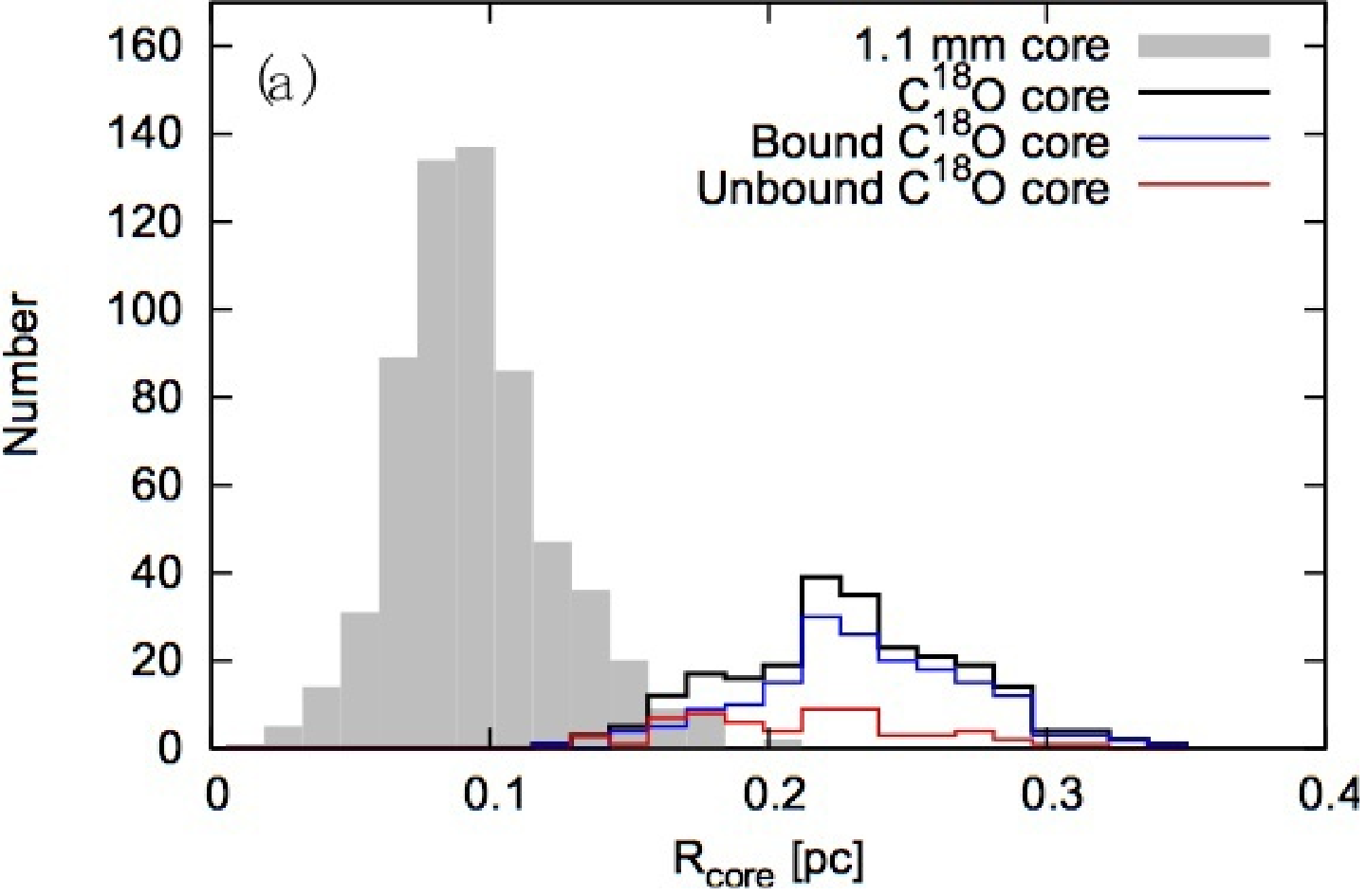}
\includegraphics[width=50mm,angle=270]{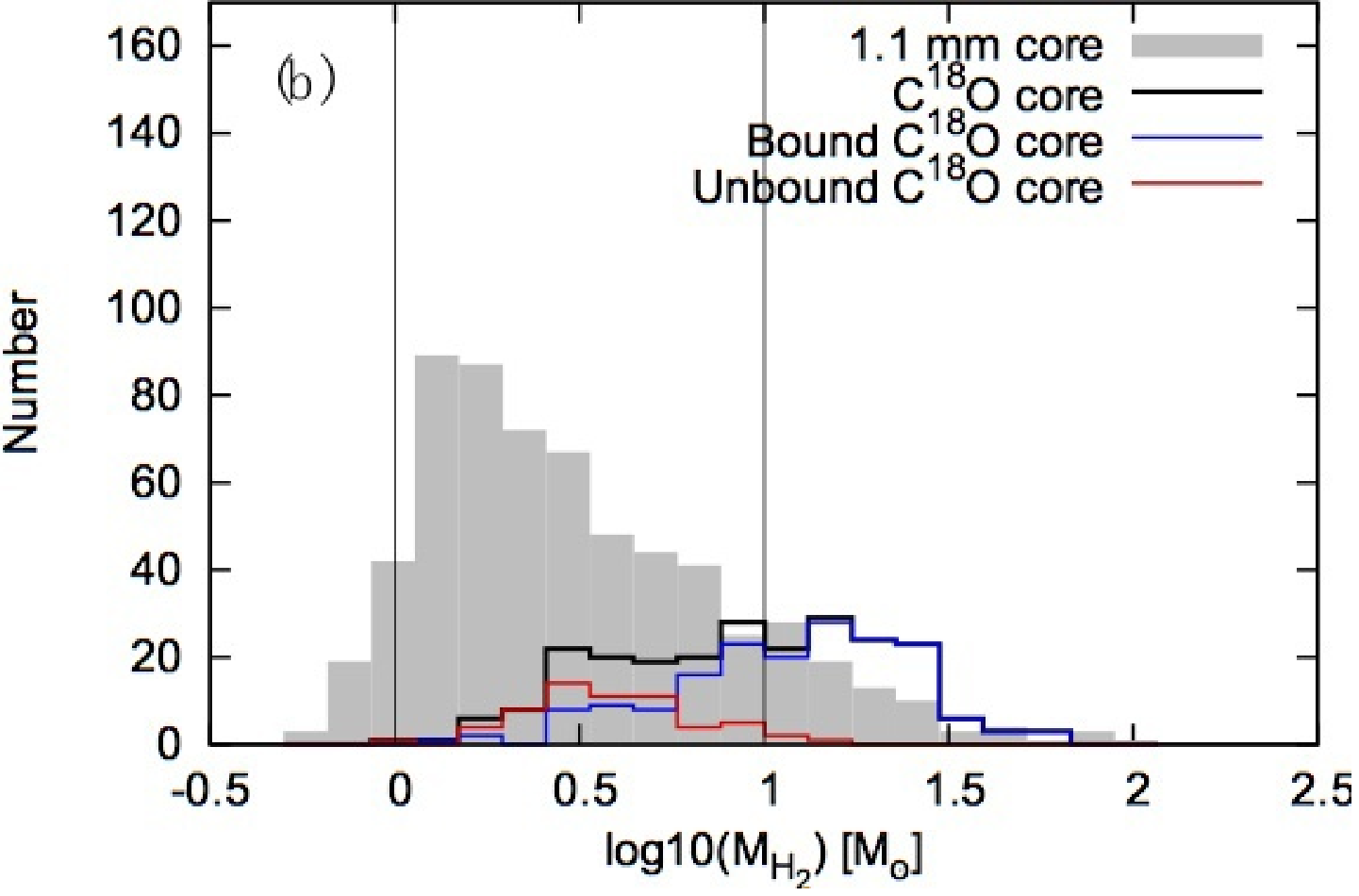}
\includegraphics[width=50mm,angle=270]{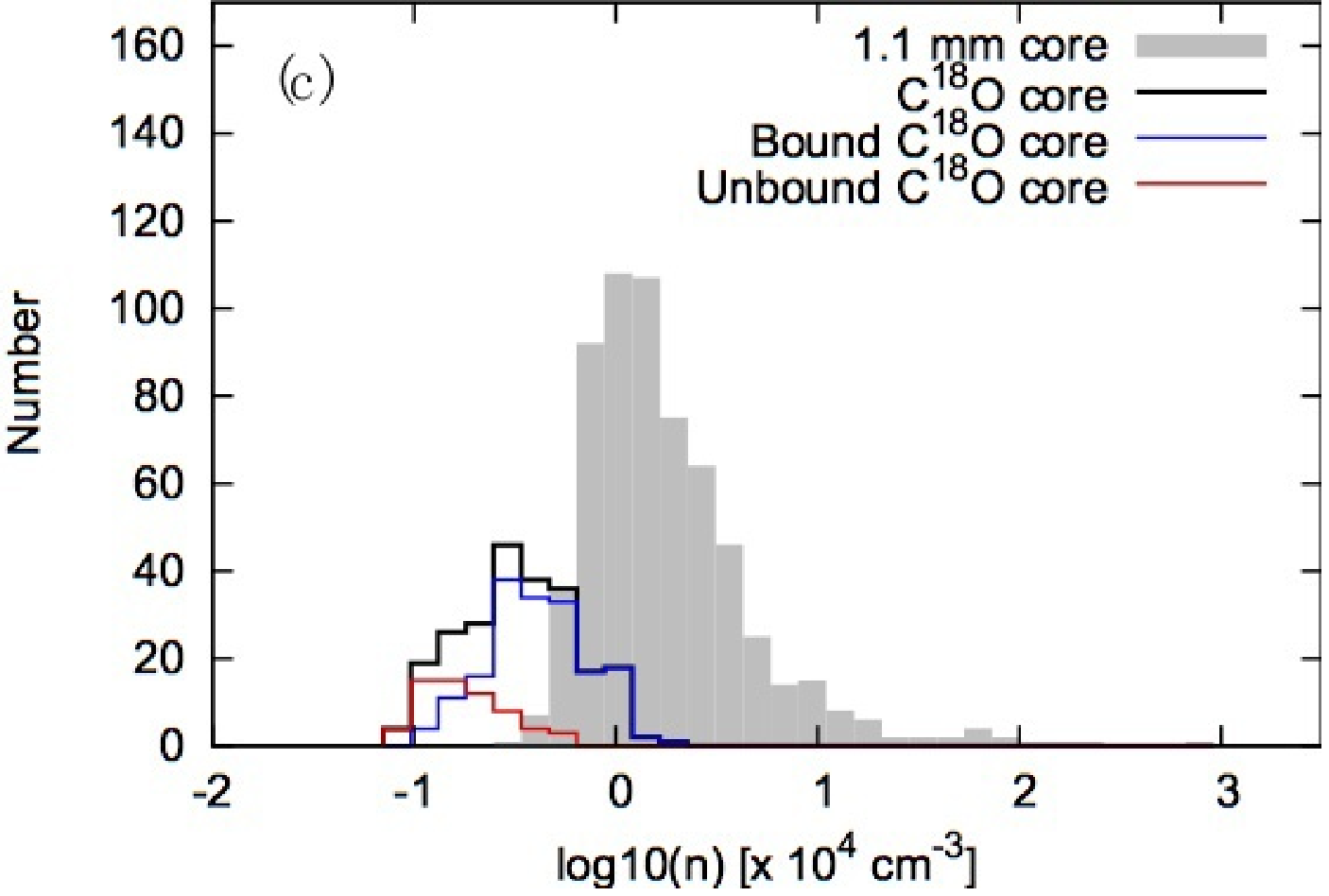}
\caption{Histograms of (a) the radius $R_{\rm core}$ of the 1.1 mm dust and C$^{18}$O cores, (b) the mass $M_{\rm H_2}$, and (c) the density $n$. The gray boxes show the 1.1 mm dust cores, and the black, blue, and red lines correspond to all the C$^{18}$O cores, the gravitationally bound C$^{18}$O ones, and the unbound C$^{18}$O ones, respectively. 
In panel (b), the vertical black broken lines indicate $M_{\rm H_2}$=1 and 10 $M_{\odot}$.}
\label{dust_core_hist}
\end{figure*}

\begin{figure*}
\centering
\includegraphics[width=170mm,angle=0]{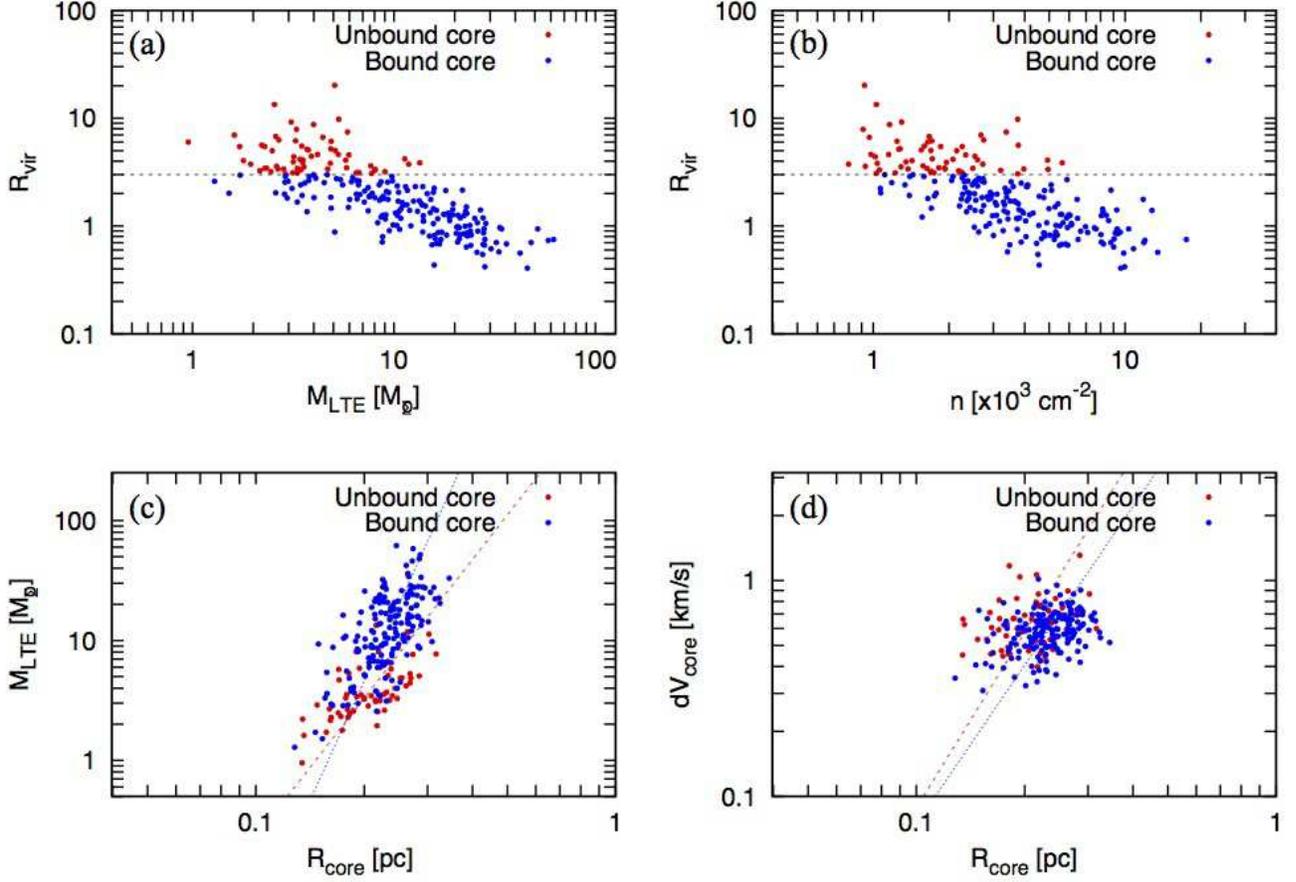}
\caption{The relations of the (a) $\mathcal{R}_{\rm vir}$ vs. $M_{\rm LTE}$, (b) $\mathcal{R}_{\rm vir}$ vs. $n$, (c) $M_{\rm LTE}$ vs. $R_{\rm core}$, and (d) $dV_{\rm core}$ vs. $R_{\rm core}$ of the C$^{18}$O cores.
We assume that the cores with $\mathcal{R}_{\rm vir}$ $<$ 3.0 are under virial equilibrium by considering the uncertainty in $X_{\rm C^{18}O}$ of a factor 3.
The black broken lines in panels (a) and (b) show $\mathcal{R}_{\rm vir}$ = 3.0. 
The red and blue filled circles represent the unbound ($\mathcal{R}_{\rm vir}$ $\ge$ 3.0) and bound ($\mathcal{R}_{\rm vir}$ $<$ 3.0) cores, respectively. 
The red and blue broken lines in the bottom panels show the best-fit power-law functions for the unbound and bound cores, respectively.}
\label{fig5}
\end{figure*}

\begin{figure*}
\centering
\includegraphics[width=80mm,angle=270]{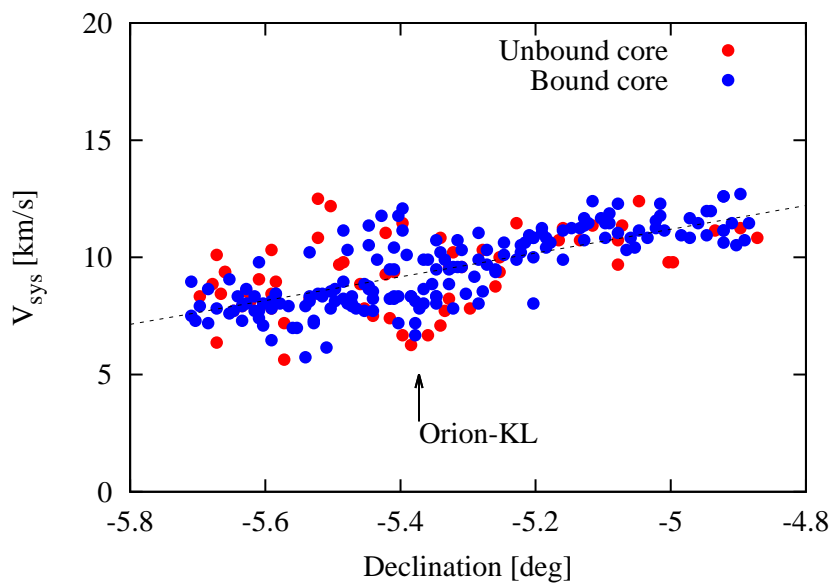}
\caption{The $V_{\rm sys}$ vs. Dec. relation of the C$^{18}$O cores. The red and blue filled circles show the unbound ($\mathcal{R}_{\rm vir}$ $\ge$ 3.0) and bound ($\mathcal{R}_{\rm vir}$ $<$ 3.0) cores, respectively.The black broken line shows the linear function for all the cores. The vertical arrow shows the position of Orion-KL.\label{fig:dec_vsys}}
\end{figure*}

\begin{figure*}
\centering
\includegraphics[width=110mm,angle=0]{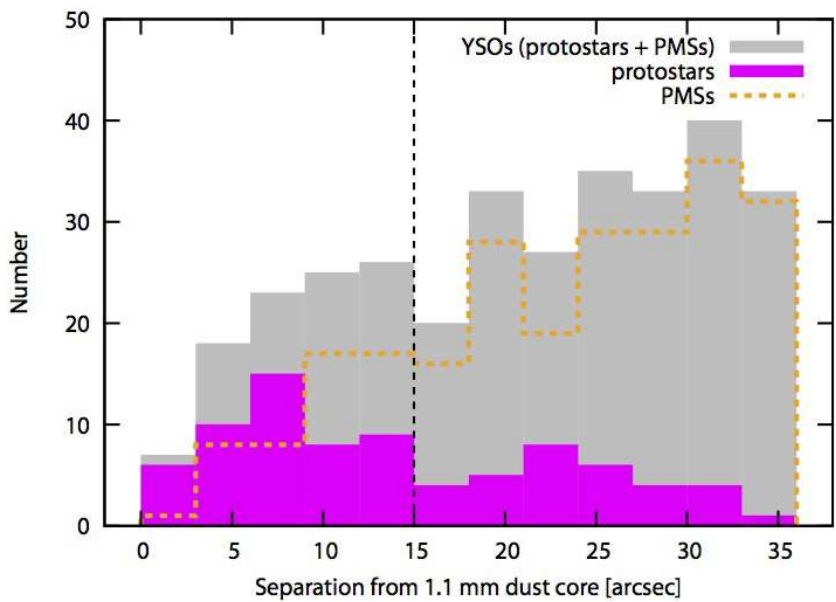}
\caption{Histogram of the separations between the AzTEC 1.1 mm dust cores and YSOs.
\label{fig:cores_YSOs} }
\end{figure*}

In Category C, we identified 121 C$^{18}$O isolated cores. There are two possible origins of these cores. 
First is due to the poor angular resolution of the AzTEC data. 
Some 1.1 mm dust cores are associated with two cores in the 850 $\mu$m data with an angular resolution of 14$\arcsec$, suggesting that the 850 $\mu$m data with the higher angular resolution resolved the 1.1 mm dust cores. 
Some 850 $\mu$m cores associated with the C$^{18}$O cores are not associated with any 1.1 mm dust cores. These cores are located between two 1.1 mm dust cores or on elongated structures in the 1.1 mm dust map as shown in Figs. \ref{omc23_id_all} - \ref{omc4_id_all}. 
The reason why the 1.1 mm dust cores are not associated with any 850 $\mu$m cores is probably that the 850 $\mu$m counterparts in the 1.1 mm map are not identified due to the poor angular resolution. We note that 
the smoothed 850 $\mu$m map with the same angular resolution as in the 1.1 mm map is quite consistent with the 1.1 mm map \citep[see Fig. 20 in][]{Shimajiri11}.
Second is that these C$^{18}$O cores do not have high enough column density to be detected in the dust continuum. Figure \ref{hist_column} shows the histogram of each column densities of the 1.1 mm dust, all C$^{18}$O cores, and C$^{18}$O cores in Category C. The column densities of each cores is estimated from the equation, $N_{\rm H_2}$ = $n$ $\times$ 2 $R_{\rm core}$. 
The minimum column density of the 1.1 mm cores (2.4 $\times$ 10$^{21}$ cm$^{-2}$) is twice larger than that of the C$^{18}$O cores (1.0 $\times$ 10$^{21}$ cm$^{-2}$). The column density sensitivity of the C$^{18}$O data is higher than that of the 1.1 mm data. On the contrary, the mass sensitivity of the C$^{18}$O data is worse than that of the 1.1 mm data as shown in Fig. \ref{dust_core_hist} (b), since the mean radius $R_{\rm core}$ of the C$^{18}$O cores is twice larger than that of the 1.1 mm dust cores.
Although the column density estimation has uncertainties, there is a possibility that the C$^{18}$O cores lacking 1.1 mm cores are due to the lack of the sensitivity of the column density of the 1.1 mm data.
For the 81 bound C$^{18}$O cores (51.9\%), we speculate that 
the central part of the cores have not yet evolved to reach the density $>$ $\sim$ 10$^4$ cm$^{-3}$ and are not detected in the dust continuum, as described in the beginning of this section. For the 40 unbound  C$^{18}$O cores, we speculate that the unbound C$^{18}$O cores are transient structures created by turbulent compression and do not have high enough column density. Most of the unbound C$^{18}$O cores ($\sim$ 70.2\%) are in Category C and are not located on the integral-shaped filament. Here, we define the integral-shaped filament as the area having signal-to-noise ratios above 15 for the 1.1 mm flux density in the OMC 2, 3, and 4 regions for this study.
Recent three-dimensional Magnetohydrodynamic (MHD) simulations have suggested that the turbulent compression creates a local dense part that is gravitationally unbound and cannot produce stars \citep{Nakamura11}.

In Category D, there are 178 dust cores that are not associated with any C$^{18}$O cores
out of the 257 dust cores that were in the C$^{18}$O-observed region (excluding Orion-KL).
We discuss three possible origins of these cores as follows.

The first possible origin is that the C$^{18}$O molecule is selectively dissociated by the FUV radiation from the massive stars in the trapezium cluster and NU Ori. 
The FUV intensity at the wavelengths of the dissociation lines for abundant CO decays rapidly on the surface of molecular clouds owing to very large optical depths of the FUV emission at these wavelengths \citep{Glassgold85,Yurimoto04,Liszt07, Rollig13}. 
\citet{Bethell07} suggested that the photoionization may play a significant role at $A_{\rm v}$  $\sim$ 10 in the case that the cores are sufficiently clumpy using a reverse Monte Carlo radiative transfer code and spectral modeling. They also mentioned that the cosmic-ray ionization is dominant in such high $A_{\rm v}$ regions.
For less abundant C$^{18}$O, which has shifted absorption lines owing to the difference in the vibrational-rotational energy levels, the decay of FUV is much lower. 
The FUV radiation can penetrate the dense region owing to the clumpiness of the cloud. 
\citet{Shimajiri13} have found that the distributions of the [CI] emission coincide with those of the $^{12}$CO emission in the PDRs of Orion bar, M43, and DLSF as well as the entire of the cloud, suggesting that these PDRs and the entire of the Orion A cloud have the clumpy structures \citep{Spaans96,Kramer08}.
For the three PDRs, the ionizing sources are the neighboring OB stars, because the 8 $\mu$m (PAH), 1.1 mm, and $^{12}$CO emission are located sequentially as a function of the distance from the OB stars, i.e., the edge-on view for the OB star/PDR system \citep{Shimajiri11}. This result suggests that these PDRs are located on the plane of the sky against OB stars.
As a result, C$^{18}$O  molecules are expected to be selectively dissociated by FUV photons even in the inner part of the cloud.
It is possible that the structure of the cloud is clumpy and full of holes such that the mean extinction through the cloud from the perspective of the exciting stars is generally $A_{\rm V}$ $<<$ 5, even though the apparent extinction (based on the observed dust emission) is much greater.
Figure \ref{flux_comp} shows the correlation between the 1.1 mm dust and C$^{18}$O intensities at the position of the 1.1 mm dust cores. The 1.1 mm dust cores associated with the C$^{18}$O cores, which are categorized into A or B, have stronger C$^{18}$O intensity. The number of the 1.1 mm dust cores not associated with any C$^{18}$O cores (Category D) increases with decreasing C$^{18}$O peak intensity. 
Most of the 1.1 mm dust cores (56/70 cores) in the PDRs are not associated with any C$^{18}$O cores. 
Here, we defined the area of PDRs, DLSF, M43, and Regions A-D, as listed in Table \ref{area_definition} \citep[also see Regions A, B, C, and D shown by][]{Shimajiri11}.
The C$^{18}$O intensity at the position of the 1.1 mm dust cores in the PDRs is lower than the intensity expected from the best-fit line for the 1.1 mm dust cores associated with the C$^{18}$O cores as shown in Fig. \ref{flux_comp}. 
The C$^{18}$O intensity at the position of several 1.1 mm dust cores not in PDRs is below the 3$\sigma$ level.
These dust cores are located around NGC1977 and DLSF. Thus, these cores seem to be also influenced by the FUV radiation.
These facts suggest that the C$^{18}$O molecule is selectively dissociated by the FUV radiation in the low 1.1 mm flux density range. \citet{Shimajiri14} found that the abundance ratio of $^{13}$CO to C$^{18}$O, $X_{\rm ^{13}CO}$/$X_{\rm C^{18}O}$, in the Orion A GMC decreases with the increasing C$^{18}$O column density,
 implying that the effect of the selective FUV dissociation of C$^{18}$O becomes smaller in the higher 1.1 mm flux density  range where it becomes hard for the FUV radiation to penetrate owing to the dust shielding.
Note that several 1.1 mm dust cores in the large 1.1 mm flux density range are categorized into D. 
This might be because the FUV radiation can penetrate the large 1.1 mm flux density regions owing to the clumpiness of the cloud.
 In fact, the $X_{\rm ^{13}CO}$/$X_{\rm C^{18}O}$ value is larger than the solar system value of 5.5 even in the inner part of the cloud \citep{Shimajiri14}.
Meanwhile, \citet{Shimajiri11} found that the $^{12}$CO peak intensity range in the DLSF region is $\sim$ 20 -- 50 K. Especially, at the outer layers of the cloud surface in DLSF, it increases up to 50 K.  On the assumption that the $^{12}$CO ($J$=1--0) line is optically thick, the result shows that the the temperature is 20 -- 50 K in DLSF, suggesting that the regions with bright dust but faint C$^{18}$O emission have dust temperatures higher than the assumed 20 K.  Thus, there is a possibility that the outer layers of the cloud surface is significantly heated, driving up dust emission and (self-shielded) $^{12}$CO emission, but the FUV radiation could still penetrate far enough into the unshielded C$^{18}$O layer to photodissociate the C$^{18}$O molecule without heating it.
 
The second possibility is the depletion of the C$^{18}$O molecule in the central part of the dust cores. 
In this case, the C$^{18}$O peak positions are distributed around the center positions of the dust cores.
The peak positions of the C$^{18}$O cores disagree with those of the 1.1 mm dust cores \citep{Ripple13,Tatematsu14, Ren14}. 

The third possibility is contaminations from the components by ambient gas and surrounding cores. 
As shown in Fig. \ref{flux_comp}, many 1.1 mm dust cores are not associated with any C$^{18}$O cores, in spite of the fact that these C$^{18}$O intensities are more than 3$\sigma$ and are not located in PDRs. 
In most cases, the C$^{18}$O cores and/or extended emission are distributed around the 1.1 mm dust cores categorized into D \citep[also see Fig. 3 (b) in][]{Shimajiri14}. There is a possibility that the C$^{18}$O cores associated with the 1.1 mm dust cores categorized into D are embedded in the components of other C$^{18}$O cores and/or extended emission and and can not be extracted as cores.

\subsection{Comparison among the 1.1 mm dust, 850 $\mu$m dust, H$^{13}$CO$^+$, and N$_2$H$^+$ cores}
We also compared the spatial distribution of the 1.1 mm dust cores with those of the SCUBA 850 $\mu$m dust \citep{Nutter07}, H$^{13}$CO$^+$ (1--0)\citep{Ikeda07}, and N$_2$H$^+$ (1--0) cores \citep{Tatematsu08}. The H$^{13}$CO$^+$ and N$_2$H$^+$ emission are known as the dense gas tracers \citep[e.g.,][]{Saito01,Takakuwa03,Maruta10, Friesen10,Johnstone10,Tanaka13}. 
The angular resolution of the SCUBA 850 $\mu$m data is 14$\arcsec$ (0.03 pc) and \citet{Nutter07} identified the condensations having a peak flux density more than 5$\sigma$ relative to the local background as cores.
The H$^{13}$CO$^+$ data has an angular resolution of 21$\arcsec$ (0.04 pc) and \citet{Ikeda07} identified the H$^{13}$CO$^+$ dense cores by the Clumpfind.
The N$_2$H$^+$ observations with a telescope beam size of 17$\arcsec$.8 (0.03 pc) were performed with a grid spacing of 20$\arcsec$.55 (0.04 pc) and \citet{Tatematsu08} identified the N$_2$H$^+$ dense cores by eyes from the $F_1,F$ =0, 1--1, 2 component, which is an isolated component of the seven hyperfine components, and the most intense hyperfine component $F_1, F$=2, 3--1, 2.
In Table \ref{table_cores}, we summarize the SCUBA 850 $\mu$m dust, H$^{13}$CO$^+$, and N$_2$H$^+$ cores distributed around the peaks of the 1.1 mm cores within the spatial resolution of 36$\arcsec$.
Figures \ref{omc23_id_all} -- \ref{orion_s_id_all} show the comparison of the spatial distribution among the 1.1 mm dust, C$^{18}$O, 850 $\mu$m dust, H$^{13}$CO$^+$, and N$_2$H$^+$ cores. Table \ref{association_rate} summarizes the comparison.

The AzTEC map excluding the Orion KL region contains 198 850 $\mu$m dust cores and 202 H$^{13}$CO$^+$ cores. The overall spatial distribution of the 1.1 mm dust cores has good agreement with that of the 850 $\mu$m dust cores. We found that 133 of the 215 850 $\mu$m dust cores (61.9\%) are associated with the 1.1 mm dust cores. However, 82 850 $\mu$m dust cores (38.1\%) are not detected in the 1.1 mm continuum emission, probably because of the insufficient sensitivity and beam dilution in the 1.1 mm map. 
The lowest mass of the dense cores detected in the SCUBA 850 $\mu$m observations is 0.13 -- 0.15 $M_{\odot}$, which is four times smaller than those detected in the 1.1 mm dust continuum observations. In addition, the angular resolution of the SCUBA 850 $\mu$m observations is 2.6 times higher than that of the 1.1 mm dust continuum observations. 
These facts indicate that the sensitivity of the SCUBA data is ten times higher than that of the AzTEC 1.1 mm dust continuum data.
We found that 56.7\% (119/210) of the H$^{13}$CO$^+$ cores are associated with the 1.1 mm dust cores. The fraction is 1.3 times higher than that for the C$^{18}$O cores, in spite of the fact that the mass detection limit of the H$^{13}$CO$^+$ observations is twice higher than that of the C$^{18}$O observations. 
Furthermore, the mean density of the 1.1mm dust cores (5.5 $\times$ 10$^4$ cm$^{-3}$) is closer to 
that of the H$^{13}$CO$^+$ cores (1.6 $\times$ 10$^4$ cm$^{-3}$) and smaller than that of the C$^{18}$O cores (0.4 $\times$ 10$^4$ cm$^{-3}$). Thus, the 1.1 mm dust continuum and the H$^{13}$CO$^+$ emission line are thought to trace the similar density area. These facts suggest that the H$^{13}$CO$^+$ emission is a better tracer of the dense cores than the C$^{18}$O emission.

In the comparison of the spatial distributions between the 1.1 mm dust and N$_2$H$^+$ cores, the high fraction of 77.8\% (21/27) of the N$_2$H$^+$ cores are found to be associated with the 1.1 mm dust cores. Such a high fraction implies that both the N$_2$H$^+$ and H$^{13}$CO$^+$ emission can trace well the dense cores compared to the C$^{18}$O emission. The remaining six N$_2$H$^+$ cores are not associated with any 1.1 mm dust cores probably owing to the insufficient spatial resolution of the 1.1 mm map. In fact, the six cores are located in the extended feature in the 1.1 mm map, and are associated with the SCUBA 850 $\mu$m dust cores. We note that the number of the N$_2$H$^+$ cores is the smallest and the lowest mass of the detected N$_2$H$^+$ cores is 7.3 $M_{\odot}$, which is much larger than those of the 1.1 mm, 850 $\mu$m, and H$^{13}$CO$^+$ cores.

\section{Summary}
The main results of this study are summarized as follows:

\begin{enumerate}
\item{We have cataloged 619 dust cores in the AzTEC 1.1 mm dust continuum using the Clumpfind method. The ranges of the radius $R_{\rm core}$, mass $M_{\rm H_2}$, and density $n$ of these cores are estimated to be 0.01--0.2 pc, 0.6 -- 1.2 $\times$ 10$^2$ $M_{\odot}$, and 0.3 $\times$ 10$^4$ -- 9.2 $\times$ 10$^6$ cm$^{-3}$, respectively. The high-mass cores ($M_{\rm H_2}$ $\ge$ 10.0 $M_{\odot}$) are located in the integral-shaped filament. On the other hand, the intermediate-mass (1.0 $M_{\odot}$ $\le$ $M_{\rm H_2}$ $<$ 10.0 $M_{\odot}$) and low-mass ($M_{\rm H_2}$ $<$ 1.0 $M_{\odot}$) cores are mainly located on the outside of the filament.}

\item{The distribution of the C$^{18}$O ($J$=1--0) emission is similar to that of the 1.1 mm dust continuum emission. We have identified 235 C$^{18}$O cores from the C$^{18}$O data by the Clumpfind algorithm. The ranges of $R_{\rm core}$, $dV_{\rm core}$, $M_{\rm LTE}$, and $n$ are 0.13 -- 0.34 pc, 0.31 -- 1.31 km s$^{-1}$, 1.0 -- 61.8 $M_{\odot}$, and (0.8 -- 17.5) $\times$ 10$^3$ cm$^{-3}$, respectively. In the 235 C$^{18}$O cores, 61 cores are gravitationally unbound, while 174 cores are gravitationally bound.}

\item{
We performed the core identification with various step sizes and threshold levels in a reasonable range of 2$\sigma$ to 5$\sigma$ in order to investigate the influence of the Clumpfind parameters on the core properties. The number of the identified C$^{18}$O cores significantly decreases with increasing step size, while the core number weakly depends on the threshold level. 
The $R_{\rm core}$, $dV_{\rm core}$, and $M_{\rm LTE}$ values gradually increase with increasing step size, but do not depend on the threshold level.
}

\item{The LTE mass vs. virial ratio and density vs. virial ratio relations of the identified C$^{18}$O cores show that the virial ratio tends to decrease with the increasing LTE mass and  density. The best-fit power-law functions for the unbound and bound cores are ($M_{\rm LTE}$/$M_{\odot}$) = (58.2 $\pm$ 27.3)($R_{\rm core}$/pc)$^{1.7 \pm 0.3}$ and  ($M_{\rm LTE}$/$M_{\odot}$) = (325.9 $\pm$ 130.4)($R_{\rm core}$/pc)$^{2.1 \pm 0.3}$, respectively. The coefficient for the bound cores is significantly larger than for the unbound cores. The difference between the data distributions for the unbound and bound cores cannot be seen in the $dV$ vs. $R_{\rm core}$ relation. }

\item{We compared the physical properties of the C$^{18}$O cores among the OMC-1, OMC-2/3, OMC-4, DSLF, and bending structure regions. The Kolmogorov-Smirnov (KS) test showed that there is no significant difference among the five regions for each property, although the physical environments in the regions are very different from each other.}

\item{We investigated the dynamical states of the C$^{18}$O cores with the external pressure and internal magnetic field. We found that all the bound C$^{18}$O cores are supercritical ($\mathcal{R}_{\rm c}$ $\ge$ 1) and 25 of the 61 unbound C$^{18}$O cores are subcritical ($\mathcal{R}_{\rm c}$ $<$ 1), although there are large uncertainties in the estimation of the Jeans mass owing to the relation of $M_{\rm J}$ $\propto$ $\sigma^4$.}

\item{We examined the spatial relations between the 1.1 mm dust and C$^{18}$O cores. We found that the relations can be categorized into the following four groups. First, one C$^{18}$O core is associated with one 1.1 mm dust core. Second, two or more C$^{18}$O cores are associated with one dust core. Third, there are isolated C$^{18}$O cores which are not associated with any dust core. Fourth, 1.1 mm dust cores not associated with any C$^{18}$O cores also exist.}

\item{We compared the spatial distributions of the 1.1 mm dust, 850 $\mu$m dust, C$^{18}$O, H$^{13}$CO$^+$, and N$_2$H$^+$ cores. The overall distribution of the 1.1 mm dust cores is found to have good agreement with that of the 850 $\mu$m dust cores. In addition, the N$_2$H$^+$ and H$^{13}$CO$^+$ emission are found to trace well the dense dust cores compared to the C$^{18}$O emission.}

\end{enumerate}

\acknowledgments
We acknowledge the anonymous referee for providing helpful suggestions to improve this paper.
Y. Shimajiri was financially supported by a Research Fellowship from the JSPS for Young Scientists. This work was supported by JSPS KAKENHI Grant Number 90610551. 
Part of this work was supported by the French National Research Agency (Grant no. ANR–11–BS56–0010–STARFICH).
K. S., T. S., and F. N., were supported by JSPS KAKENHI Grant Numbers, 26287030, 26610045, 26350186, and 24244017.
M. M., and T. T., were supported by MEXT KAKENHI No. 23103004.


%% in parentheses, after the keyword, but they will not be verified.

{\it Facilities:} \facility{ASTE (AzTEC)}, \facility{Nobeyama 45m (BEARS)}.

\clearpage

%% the file, uncomment the \plotone, \plottwo, and
%% If you need a layout that cannot be achieved with \plotone or
%% \plottwo, you can invoke the graphicx package directly with the
%% \includegraphics command or use \plotfiddle. For more information,
%% please see the tutorial on "Using Electronic Art with AASTeX" in the
%% documentation section at the AASTeX Web site,
%% http://www.journals.uchicago.edu/AAS/AASTeX.
%%
%% The examples below also include sample markup for submission of
%% supplemental electronic materials. As always, be sure to check
%% the instructions to authors for the journal you are submitting to
%% for specific submissions guidelines as they vary from

%% This example uses \plotone to include an EPS file scaled to
%% 80% of its natural size with \epsscale. Its caption
%% has been written to indicate that additional figure parts will be
%% available in the electronic journal.

\begin{figure*}
\centering
\includegraphics[width=150mm,angle=0]{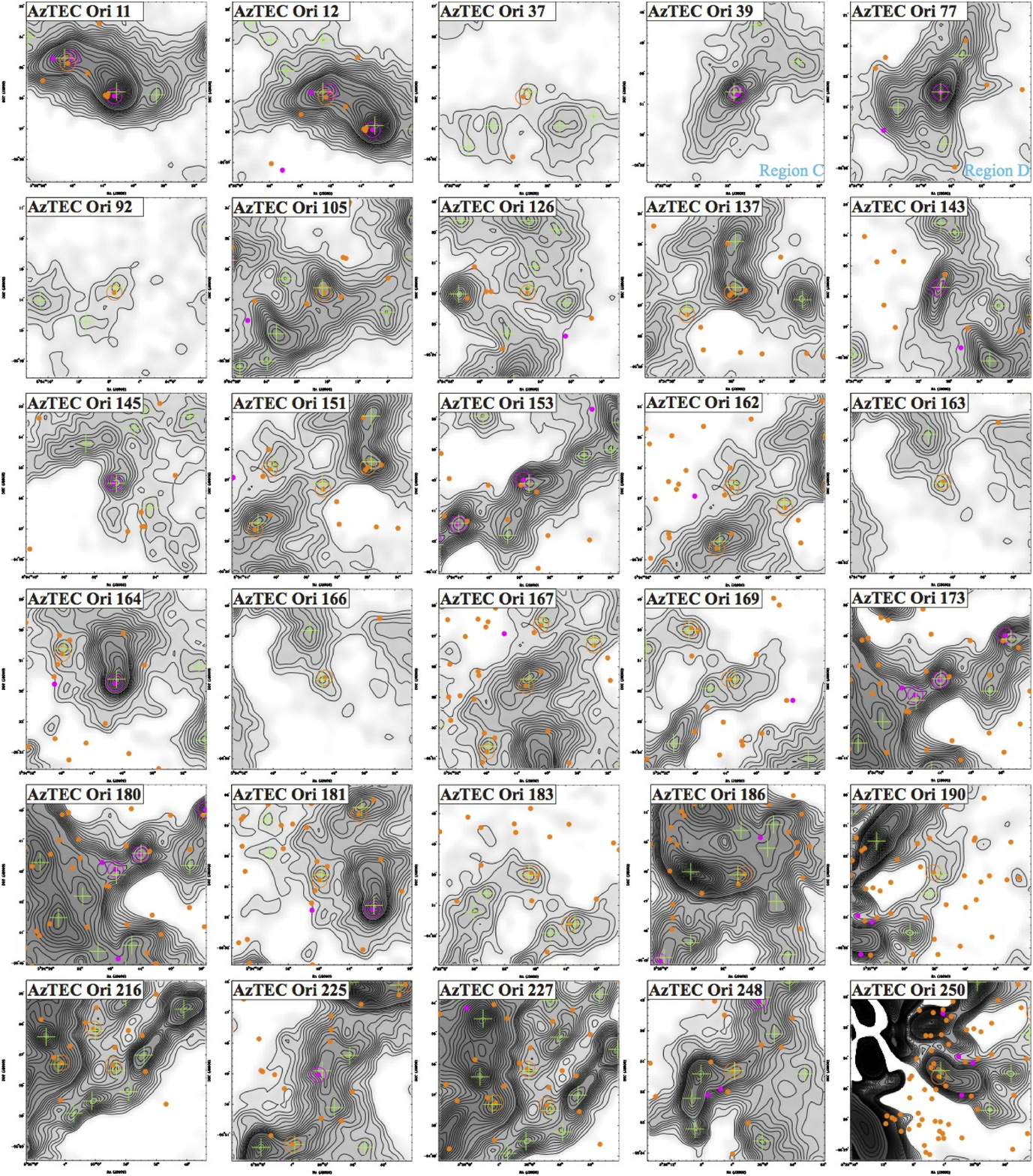}
\caption{Close-up 1.1 mm continuum images of the dust cores associated with protostars and pre-main sequence stars. The green plus signs denote the peak positions of the identified 1.1 mm dust cores, and the ID number of the central core is shown at the top of each panel. 
The labels in light blue show the source names used in other studies \citep{Chini97, Stanke07, Shimajiri11}. 
In each panel, the contours start from the 4$\sigma$ level with an intervals of 2$\sigma$ for the range of 4 -- 30$\sigma$, 5$\sigma$ for the range of 30 -- 305$\sigma$, and 50$\sigma$ for the range $>$ 305$\sigma$. The magenta and orange open circles indicate the positions of the Spitzer protostars and pre-main sequence stars, respectively \citep{Megeath12}, associated with the 1.1 mm dust cores. The magenta and orange small filled circles show the positions of the other protostars and pre-main sequence stars, respectively.}
\label{postageA}
\end{figure*}

\begin{figure*}
\centering
\includegraphics[width=150mm,angle=0]{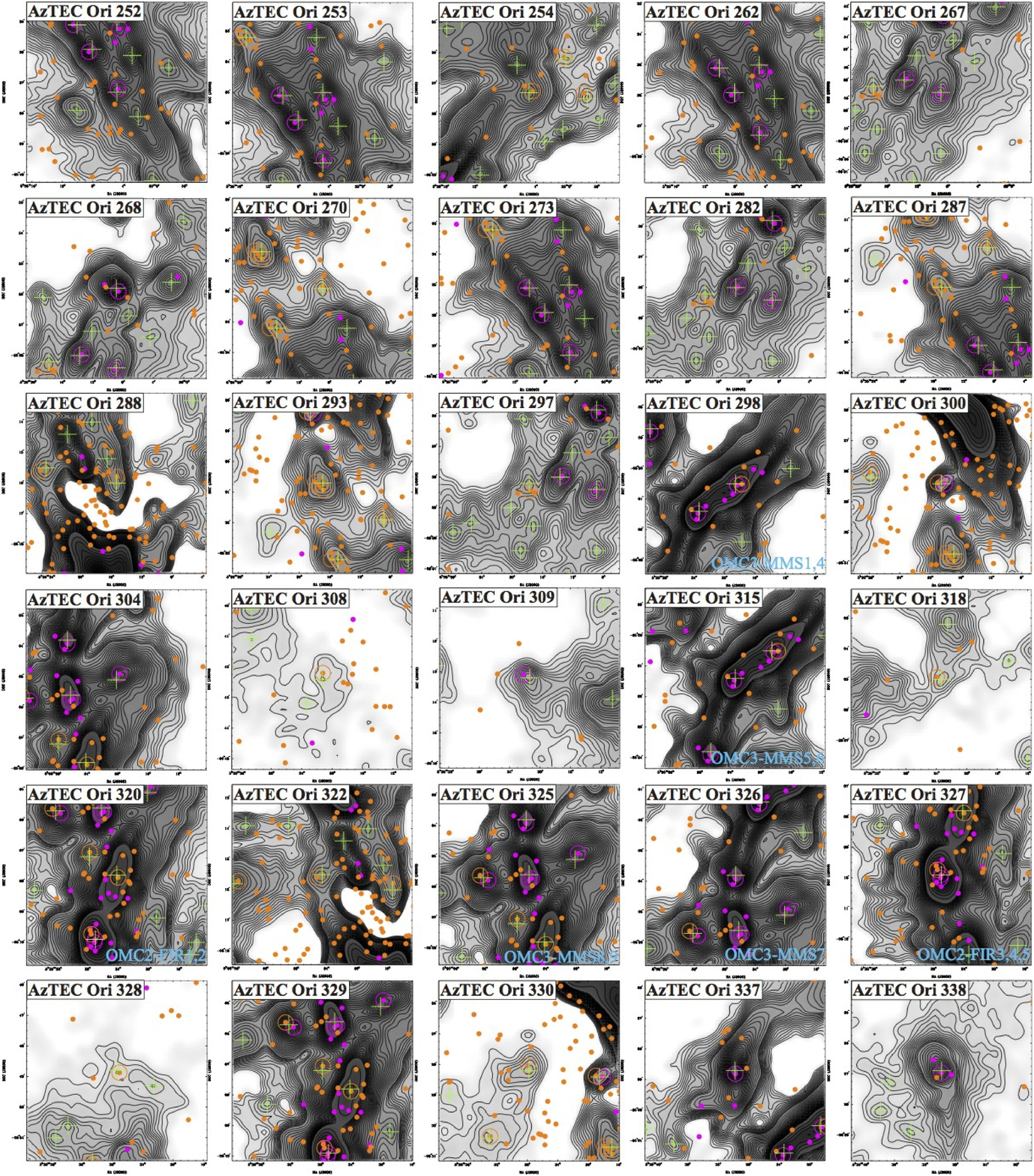}
\caption{Continuation of Fig. \ref{postageA}.}
\label{postageB}
\end{figure*}

\begin{figure*}
\centering
\includegraphics[width=150mm,angle=0]{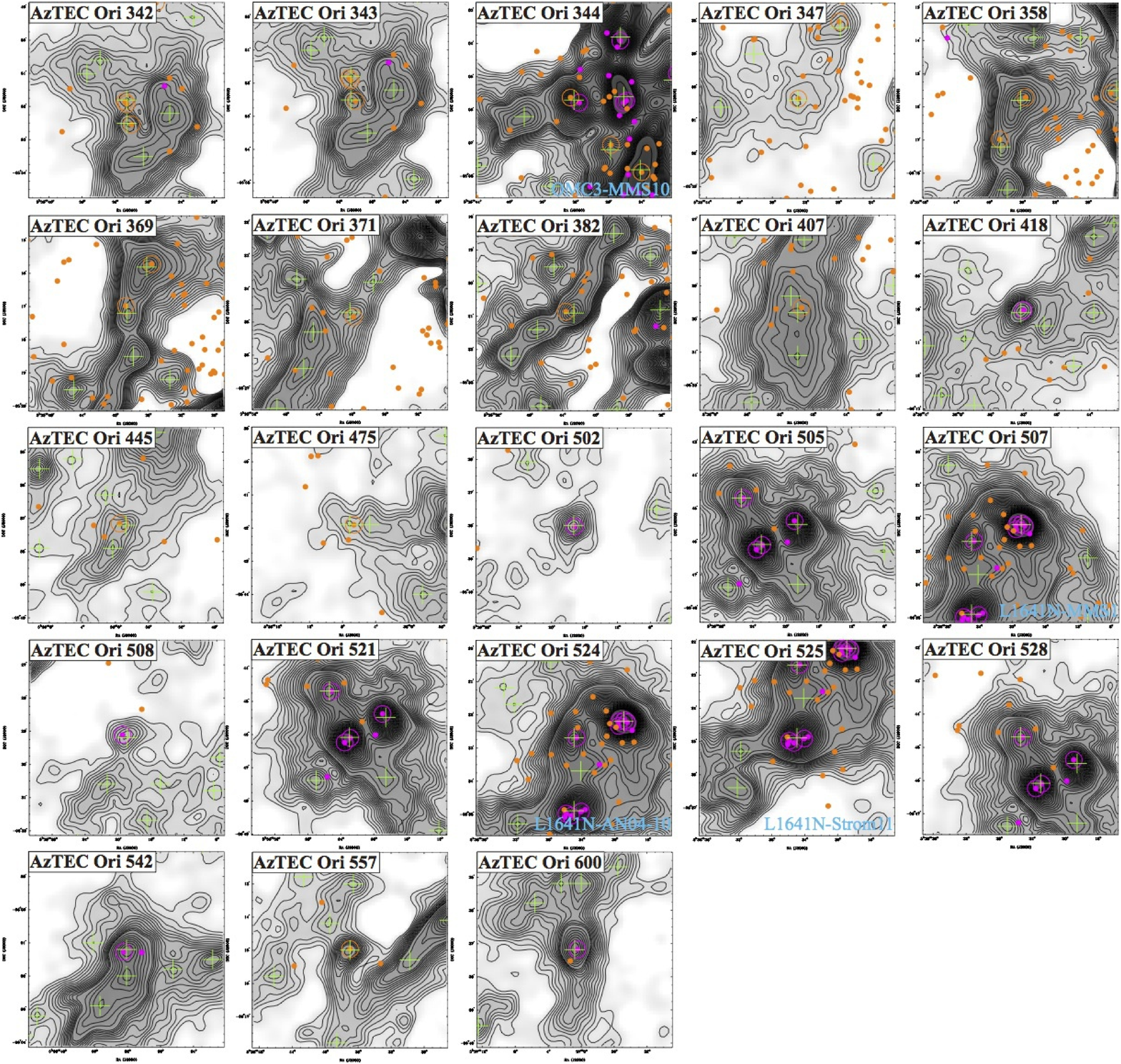}
\caption{Continuation of Fig. \ref{postageA}.}
\label{postageC}
\end{figure*}

\begin{figure*}
\centering
\includegraphics[width=120mm,angle=270]{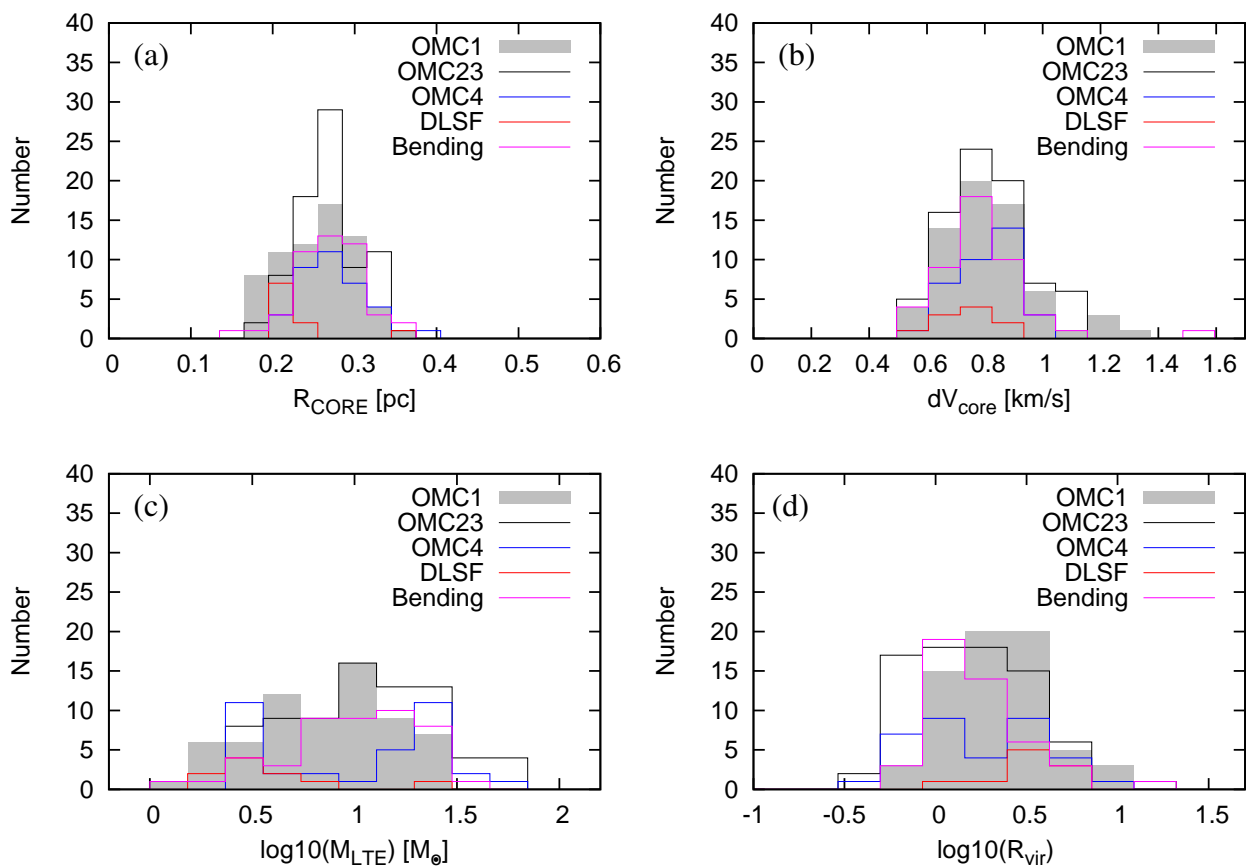}
\caption{Histograms of (a) $R_{\rm core}$, (b) $dV_{\rm core}$, (c) $M_{\rm LTE}$, and (d) $\mathcal{R}_{\rm vir}$ of the C$^{18}$O cores in the OMC-1, OMC-2/3, OMC-4, DLSF, and bending structure regions shown by the gray, black, blue, red, and magenta lines, respectively. \label{fig6}}
\end{figure*}

\begin{figure*}
\centering
\includegraphics[width=100mm,angle=270]{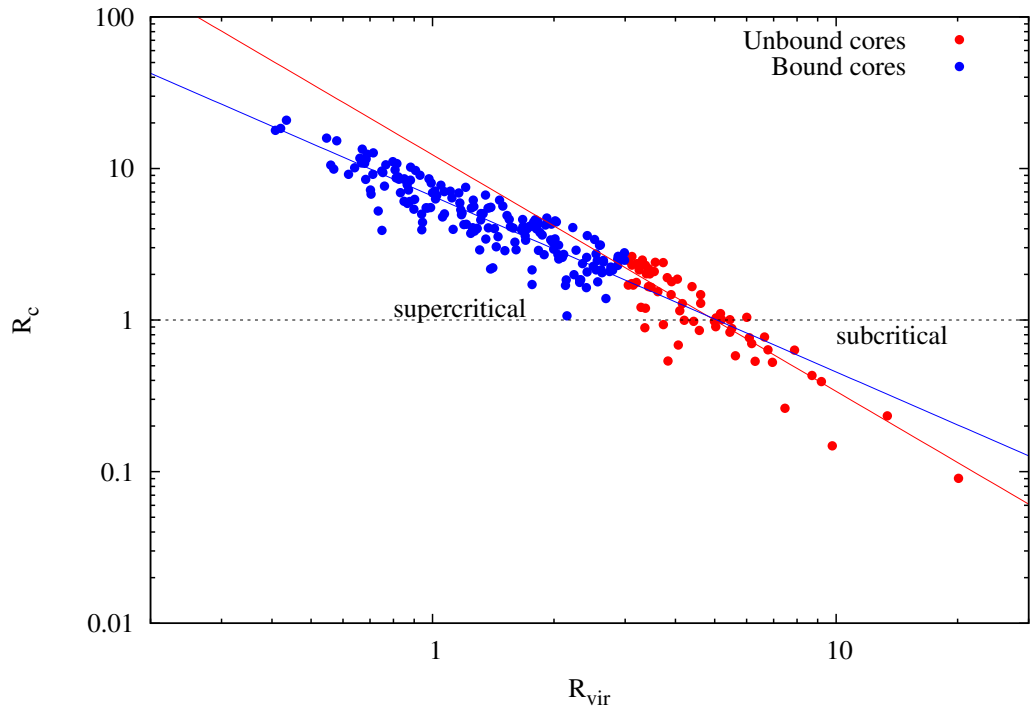}
\caption{
The $\mathcal{R}_{\rm c}$ vs. $\mathcal{R}_{\rm vir}$ relation of the C$^{18}$O cores. The red and blue filled circles show the unbound ($\mathcal{R}_{\rm vir}$ $\ge$ 3.0) and bound ($\mathcal{R}_{\rm vir}$ $<$ 3.0) cores, respectively. The red and blue lines show the best-fit power-law functions for the unbound and bound cores, respectively. The horizontal black broken line denotes $\mathcal{R}_{\rm c}$ = 1. \label{fig:virial_critical}}
\end{figure*}

\begin{figure*}
\centering
\includegraphics[width=170mm,angle=0]{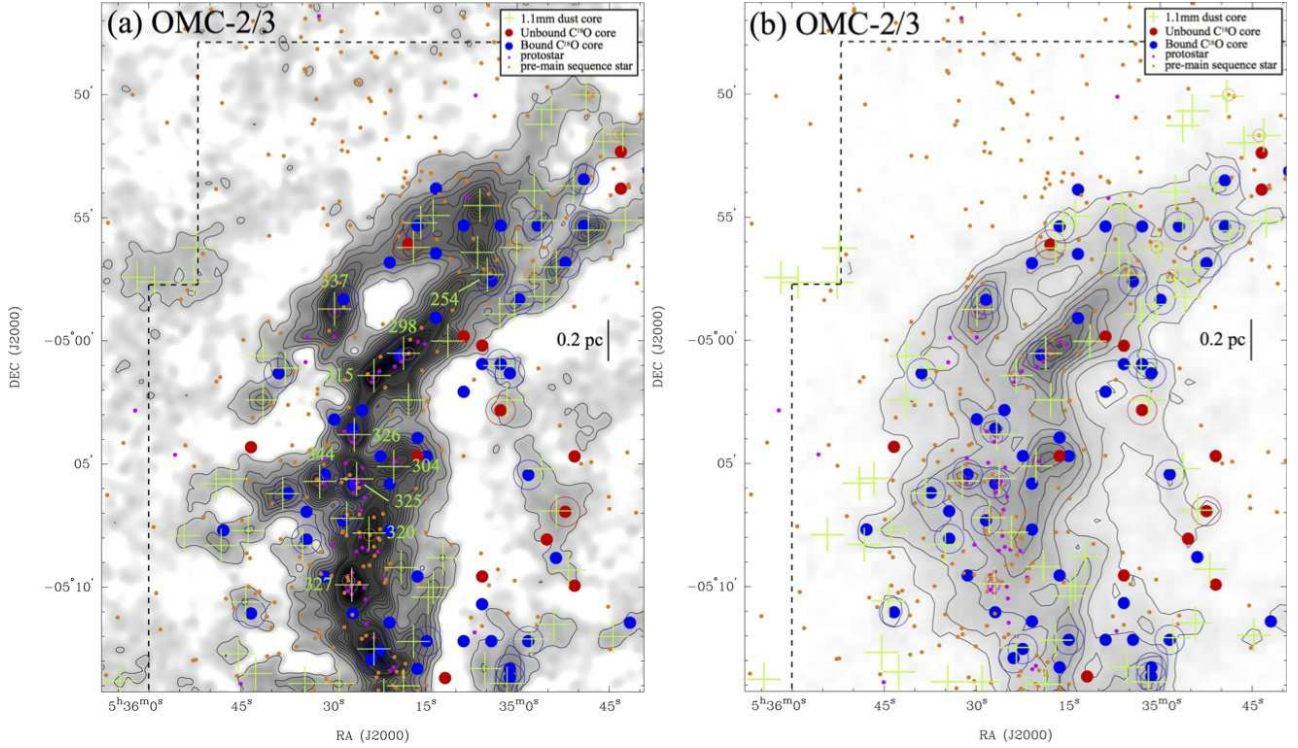}
\caption{
Our identified 1.1 mm dust and C$^{18}$O cores on (a) the AzTEC 1.1 mm and (b) the C$^{18}$O integrated intensity maps of the OMC-2/3 region. The green plus signs denote the positions of the 1.1 mm dust cores.  The red and blue filled circles show the positions of the bound ($\mathcal{R}_{\rm vir}$ $<$ 3) and unbound ($\mathcal{R}_{\rm vir}$ $\ge$ 3) C$^{18}$O cores, respectively. The red and blue open circles show the C$^{18}$O cores associated with the 1.1 mm dust cores. 
The numbers labeled in green are the ID numbers of the AzTEC cores associated with the YSOs shown in Figs. \ref{postageA}--\ref{postageC}.
The magenta and orange open circles show the positions of the Spitzer protostars and pre-main sequence stars \citep{Megeath12}, respectively, associated with the 1.1 mm dust cores. The magenta and orange filled circles show the positions of the other Spitzer sources. 
In panel (a), the contours start from the 5$\sigma$ level with an interval of 10$\sigma$ for the range of 5 - 105$\sigma$ and 50$\sigma$ for the range $>$105$\sigma$. The black dashed lines indicate the C$^{18}$O mapping area.
In panel (b), the contours start 1 K km s$^{-1}$ with an interval of 1 K km s$^{-1}$.
}
\label{omc23_id}
\end{figure*}

\clearpage
\begin{figure*}
\centering
\includegraphics[width=140mm,angle=0]{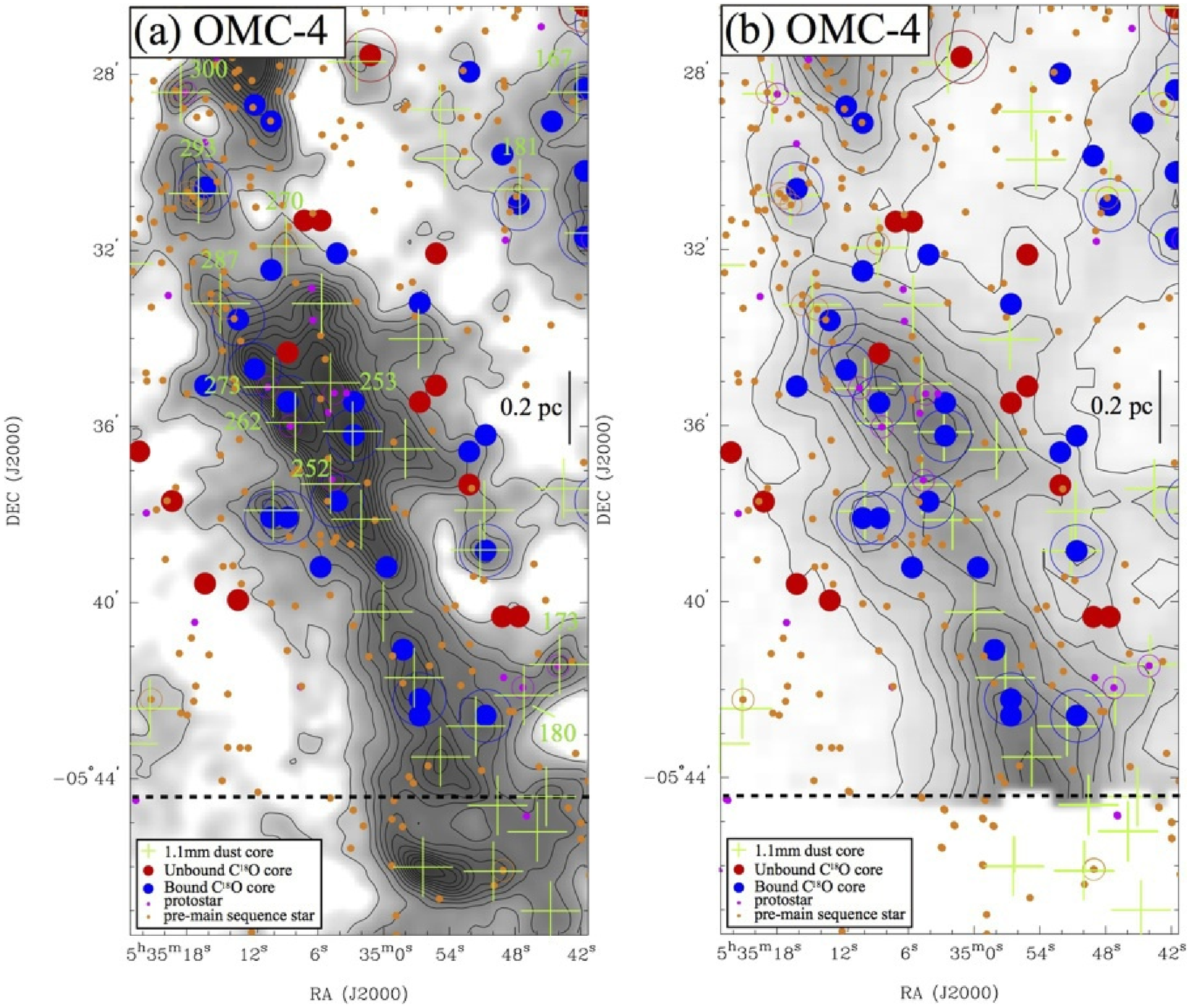}
\caption{Same as Fig. \ref{omc23_id}, but for the OMC-4 region.}
\label{omc4_id}
\end{figure*}

\begin{figure*}
\centering
\includegraphics[width=140mm,angle=0]{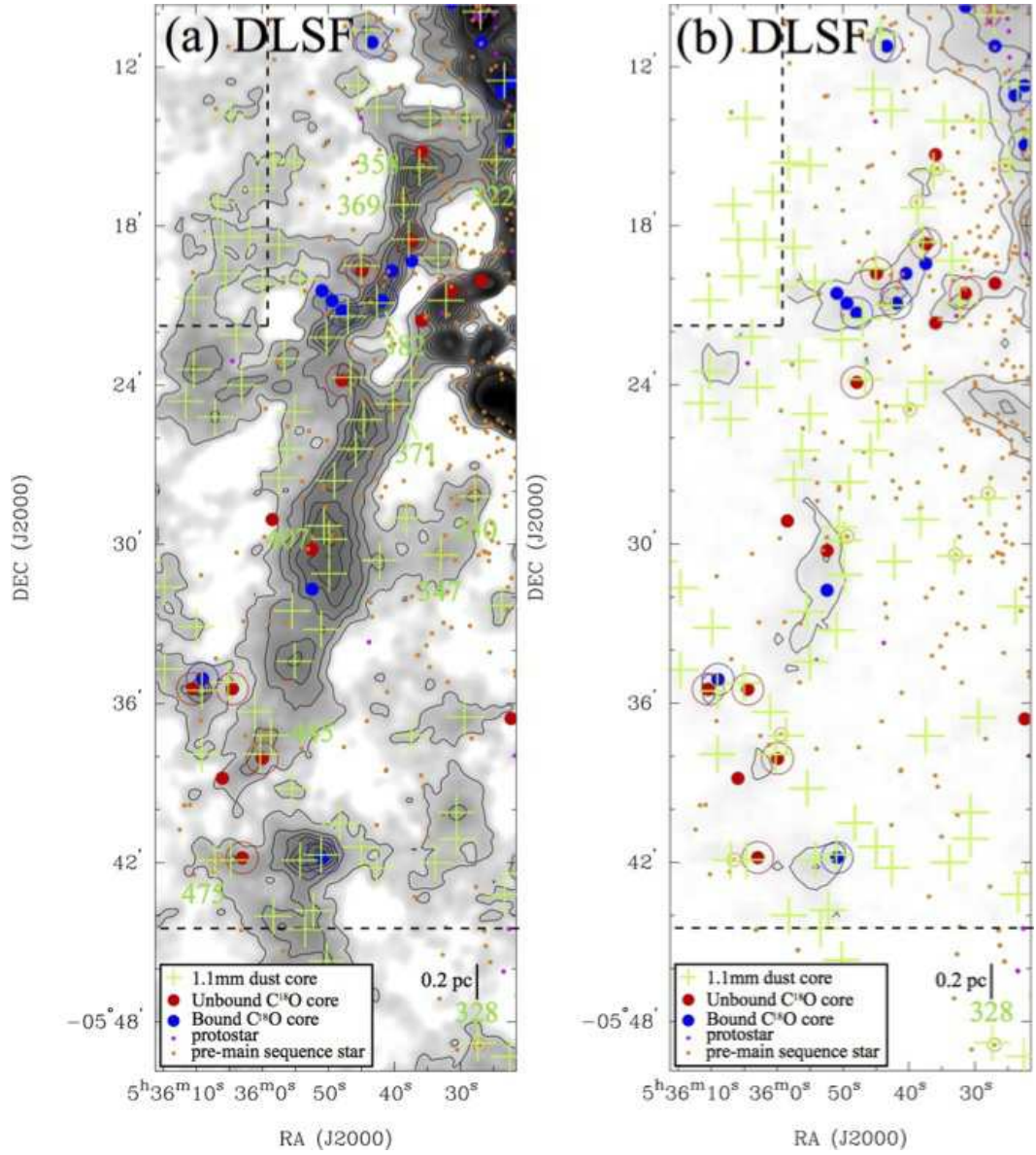}
\caption{Same as Fig. \ref{omc23_id}, but for the DLSF region.}
\label{omc4_id}
\end{figure*}

\begin{figure*}
\centering
\includegraphics[width=170mm,angle=0]{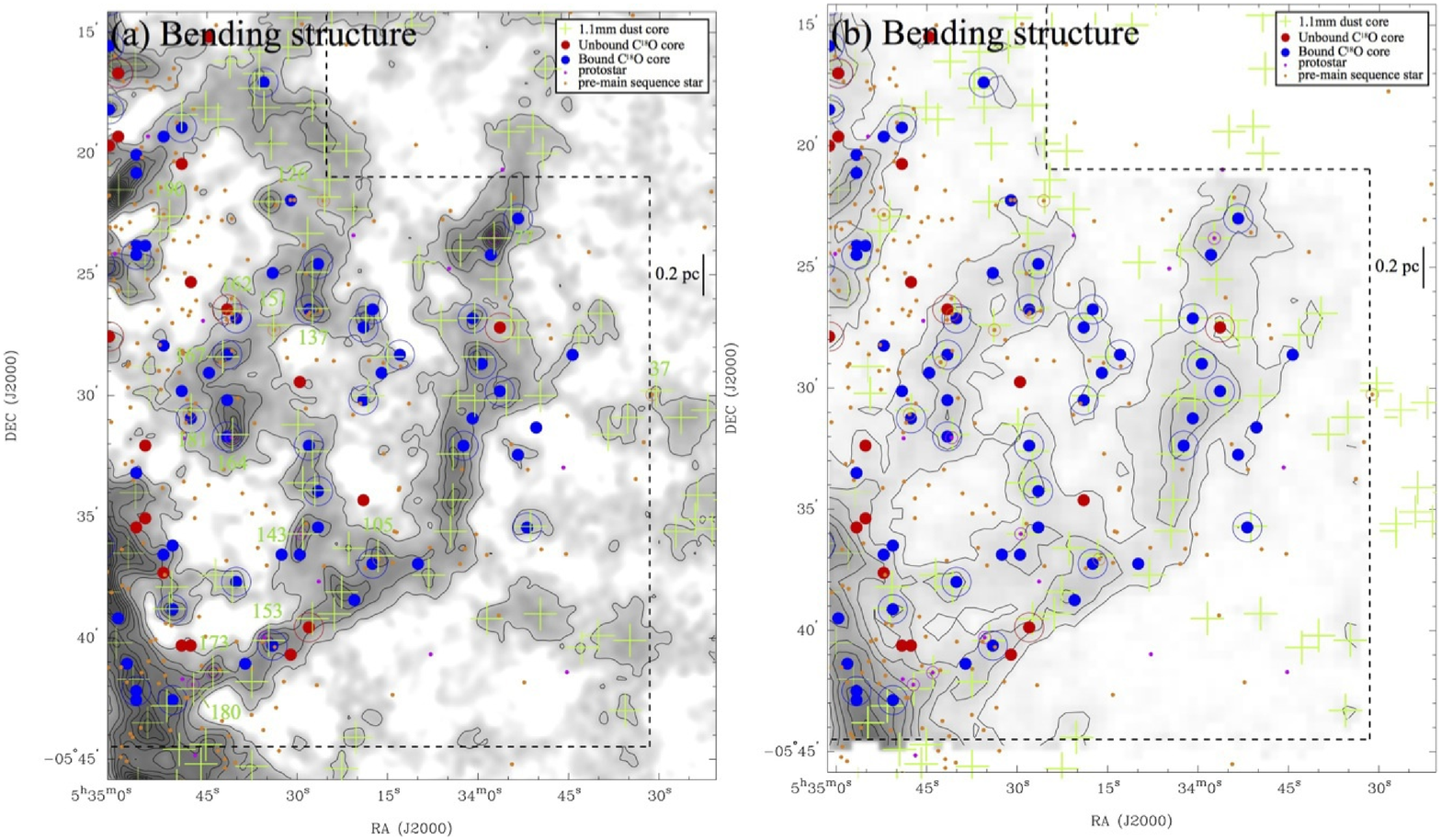}
\caption{Same as Fig. \ref{omc23_id}, but for the bending structure region.} \label{orion_w_id}
\end{figure*}

\begin{figure*}
\centering
\includegraphics[width=160mm,angle=270]{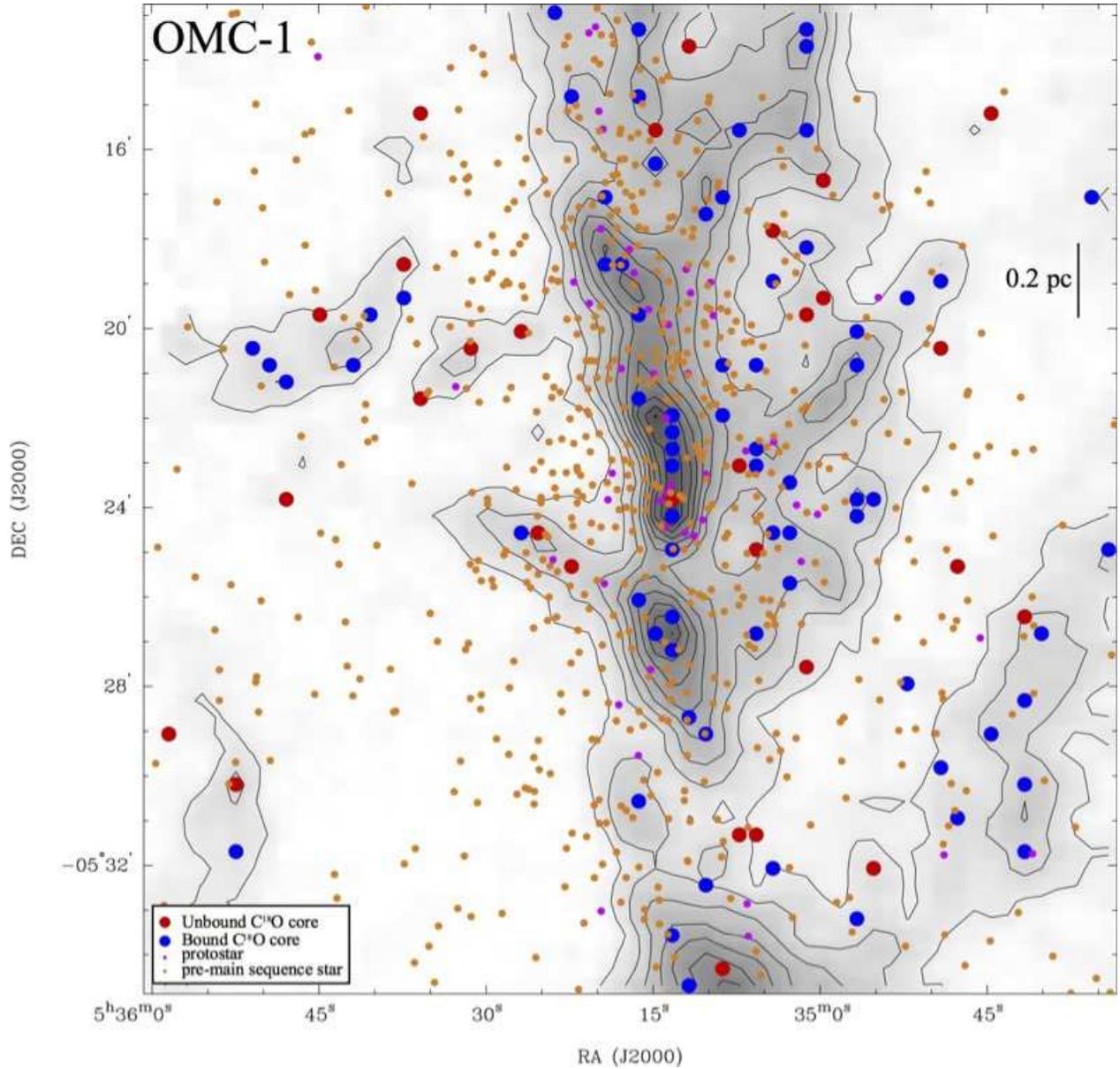}
\caption{The identified C$^{18}$O cores shown on the C$^{18}$O total integrated intensity map of the OMC-1 region. The contour levels start at 1 K km s$^{-1}$ in $T_{\rm MB}$ with intervals of 1 K km s$^{-1}$. The 1.1 mm emission around the central Orion-KL region could not be reconstructed as an accurate structure with the AzTEC data-reduction technique, because the continuum emission around Orion-KL was too bright.  The meanings of the symbols are the same as in Fig. \ref{omc23_id}.}
\label{orion_omc1_id}
\end{figure*}

\begin{figure*}
\centering
\includegraphics[width=160mm,angle=0]{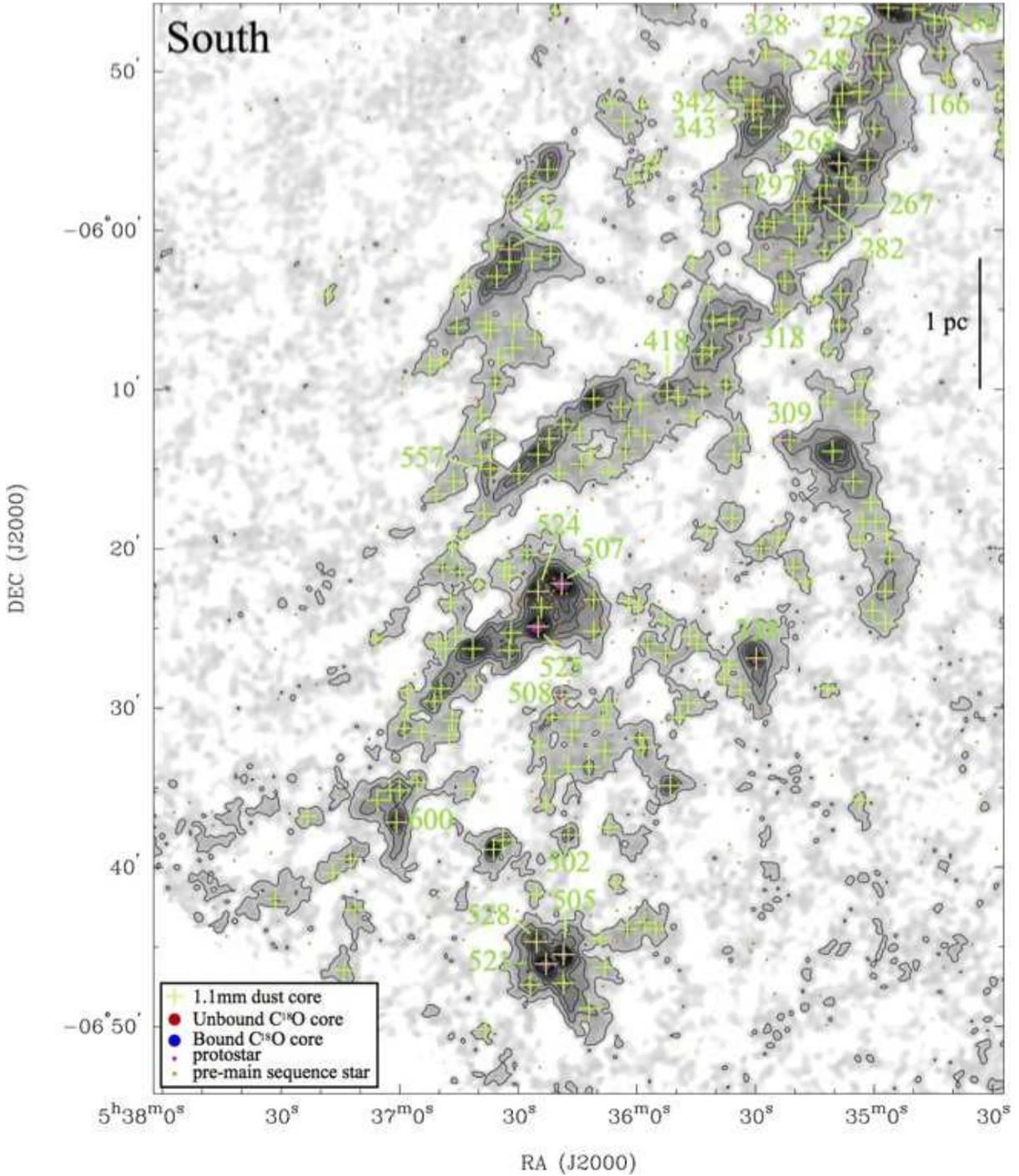}
\caption{
The identified 1.1 mm dust cores shown on the AzTEC 1.1 mm map of the South region. There is no C$^{18}$O data in this region. The symbols and the contours are the same as in Fig. \ref{omc23_id}. }
\label{orion_s_id}
\end{figure*}

\begin{figure*}
\centering
\includegraphics[width=160mm,angle=0]{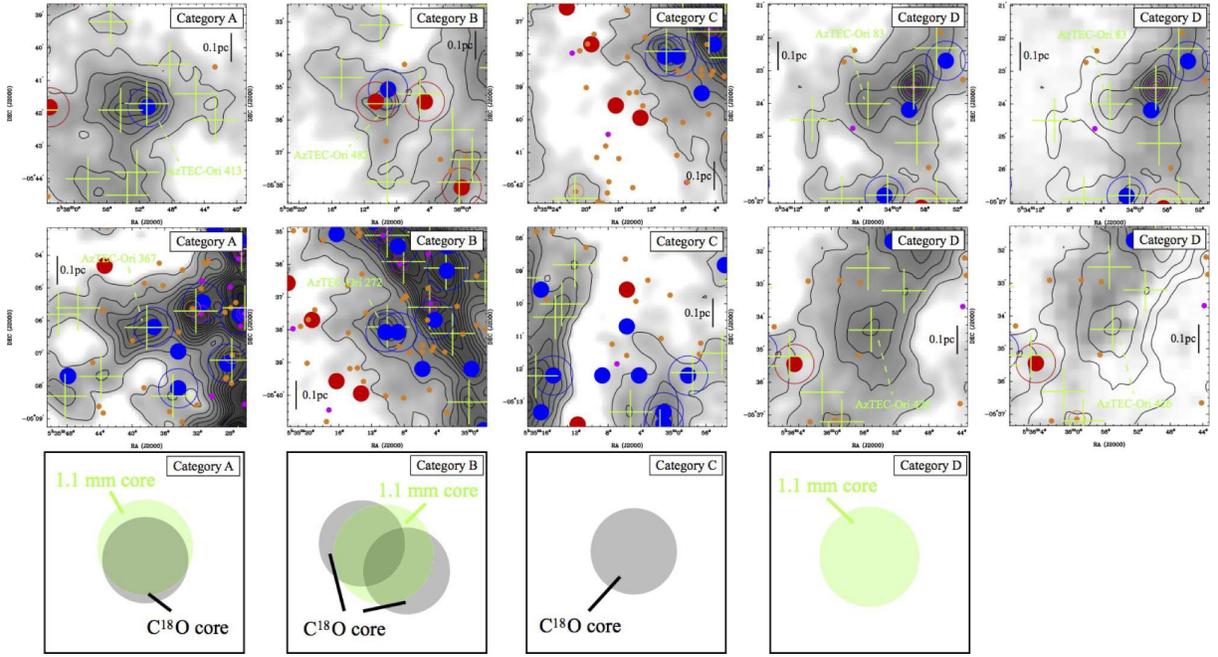}
\caption{
Typical examples in Categories A to D (from left to right). The top and middle panels show the close-up views of the 1.1 mm continuum emission images with the positions of the 1.1 mm dust and C$^{18}$O cores. The most right panels show the C$^{18}$O total intensities in gray scale with the 1.1 mm continuum emission (contours) for the two examples in Category D.  The symbols and the contours are the same as in Fig. \ref{omc23_id}. The schematic illustrations in the bottom panels indicate the spatial relations between the 1.1 mm dust and C$^{18}$O cores in the four categories. 
\label{category_sample}}
\end{figure*}
\clearpage

\begin{figure*}
\centering
\includegraphics[width=160mm,angle=0]{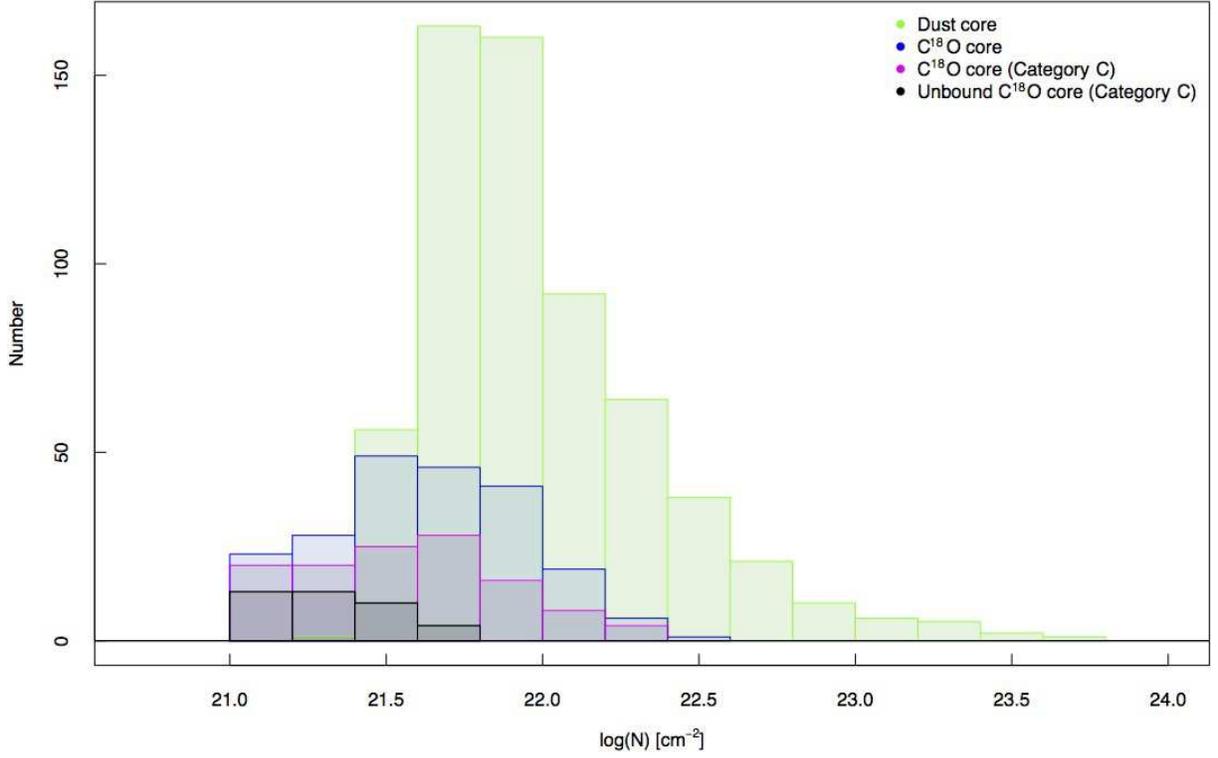}
\caption{Histograms of the column densities of the 1.1 mm dust and C$^{18}$O cores. The column densities of each cores is estimated from the equation, $N_{\rm H_2}$ = $n$ $\times$ 2 $R_{\rm core}$. The green, blue, pink, and black histograms correspond to the 1.1 mm dust core, all C$^{18}$O cores , C$^{18}$O cores in Category C, and unbound C$^{18}$O cores in Category C, respectively. The C$^{18}$O cores distributed around Orion-KL are excluded for this plot.\label{hist_column}}
\end{figure*}
\clearpage

\begin{figure*}
\centering
\includegraphics[width=120mm,angle=270]{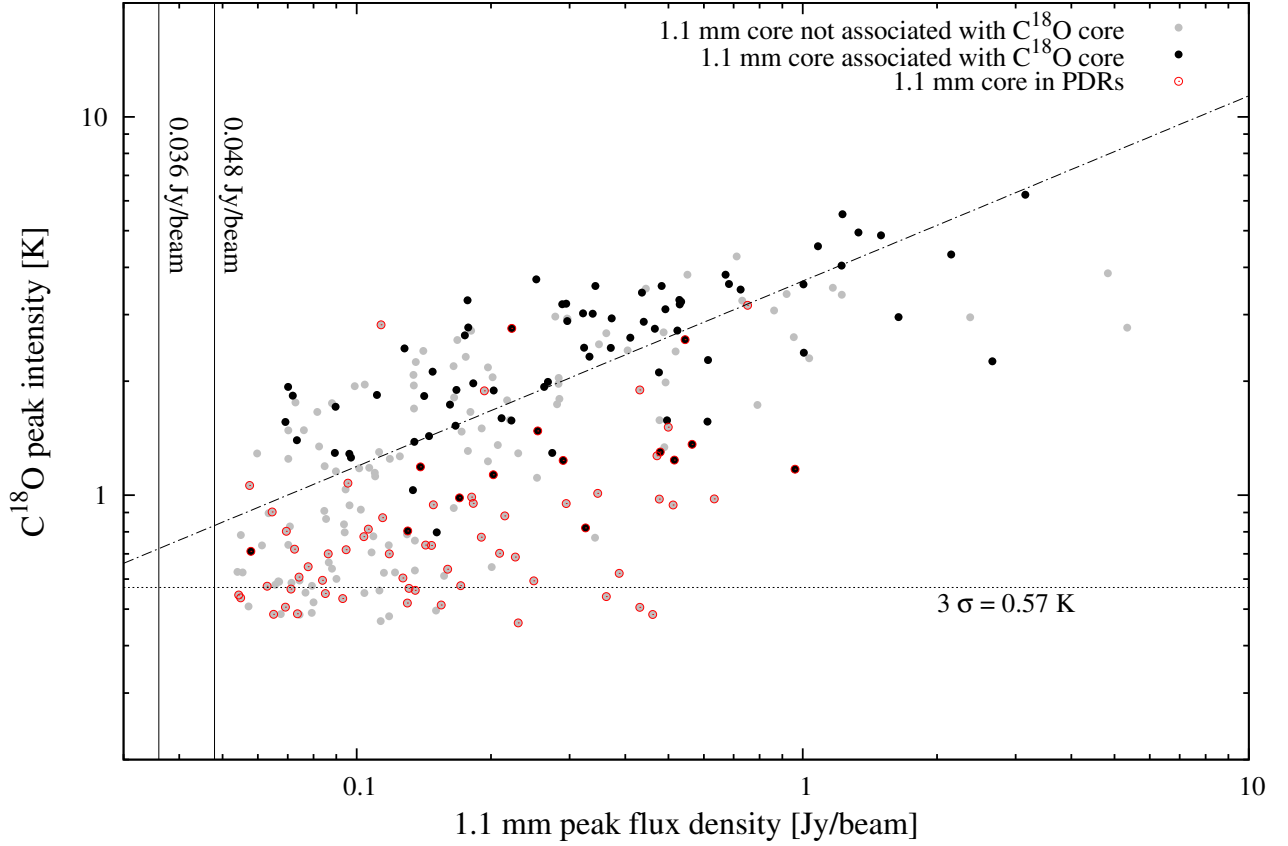}
\caption{Comparison between the 1.1 mm peak flux density and the C$^{18}$O peak intensity at the 1.1 mm dust cores. The gray filled circles are for the 1.1 mm dust cores not associated with any C$^{18}$O cores. The black filled circles are for the 1.1 mm dust cores associated with the C$^{18}$O cores. The open red circles are for the 1.1 mm dust cores in the PDRs. The horizontal line indicates the 3$\sigma$ level (1$\sigma$=0.19 K) of the C$^{18}$O data. The vertical lines indicate the 4$\sigma$ level (1$\sigma$=9 mJy beam$^{-1}$ in the central region of the 1.1 mm dust map and 12 mJy beam$^{-1}$ on the outer edge) of the 1.1 mm data.
The dashed line shows the best fit power law function for the 1.1 mm dust cores associated with the C$^{18}$O cores, log$_{10}$($T_{\rm C^{18}O}$/K) = (0.49 $\pm$ 0.14) log$_{10}$(F$_{\rm 1.1mm}$/Jy beam$^{-1}$) + (0.57 $\pm$ 0.06).
\label{flux_comp}}
\end{figure*}

\begin{figure*}
\centering
\includegraphics[width=160mm,angle=0]{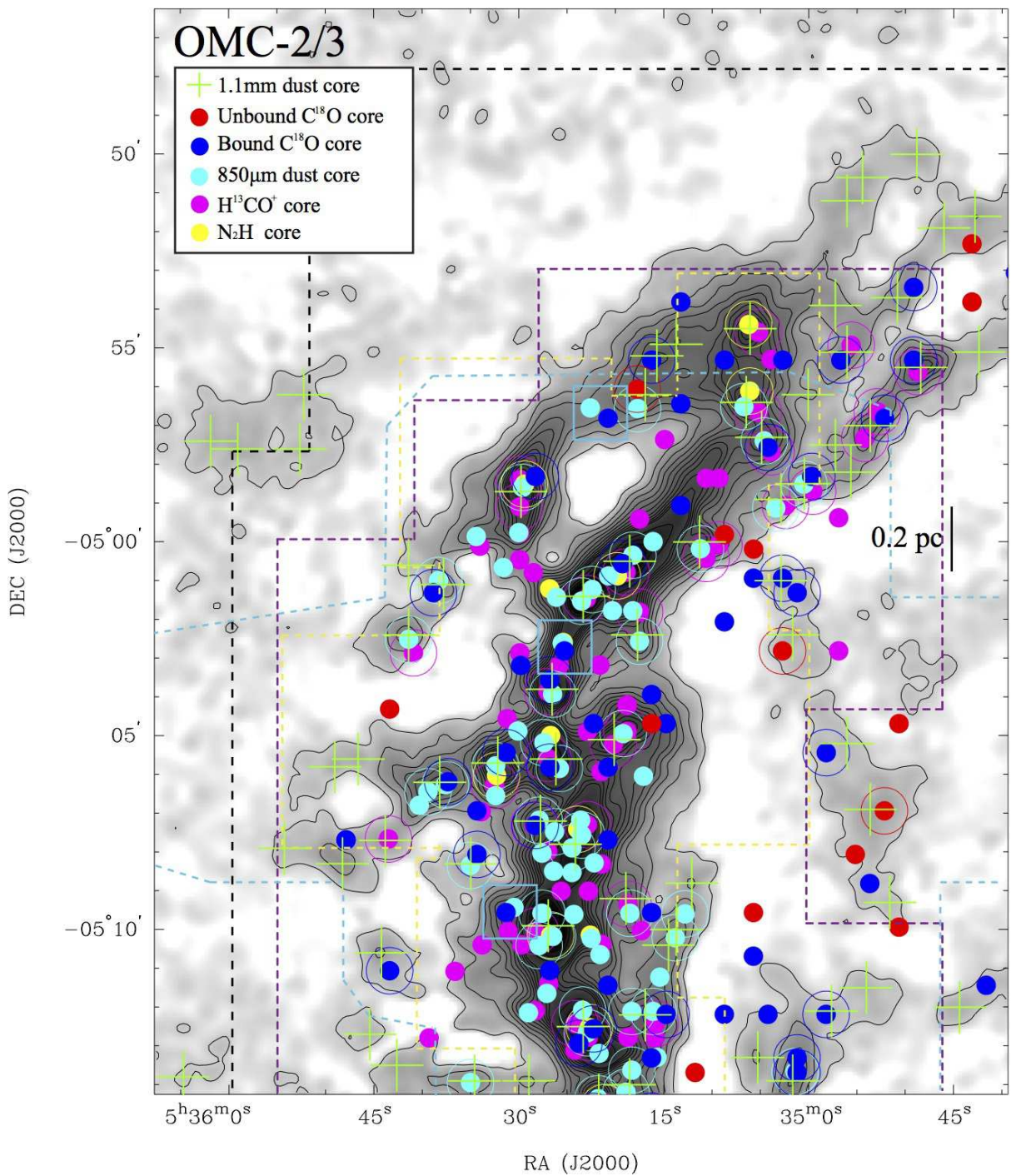}
\caption{
The identified 1.1 mm dust and C$^{18}$O cores shown together with the 850 $\mu$m, H$^{13}$CO$^+$, and N$_2$H$^+$ cores on the AzTEC 1.1 mm map of the OMC-2/3 region. The green crosses and the red and blue circles are the same as in Figure 10. The aqua-, magenta-, and yellow-filled circles denote the positions of the 850 $\mu$m, H$^{13}$CO$^+$, and N$_2$H$^+$ cores, respectively. The open circles denote the cores associated with the 1.1 mm dust cores. The black, aqua, magenta, and yellow dashed lines indicate C$^{18}$O, 850 $\mu$m, H$^{13}$CO$^+$, and N$_2$H$^+$ observed areas.  Contours for the 1.1 mm map are the same as in Fig. \ref{omc23_id}. The open aqua boxes denote the C$^{18}$O cores associated with the 850 $\mu$m dust cores, but not associated with any 1.1 mm dust cores.
}
\label{omc23_id_all}
\end{figure*}

\begin{figure*}
\centering
\includegraphics[width=100mm,angle=0]{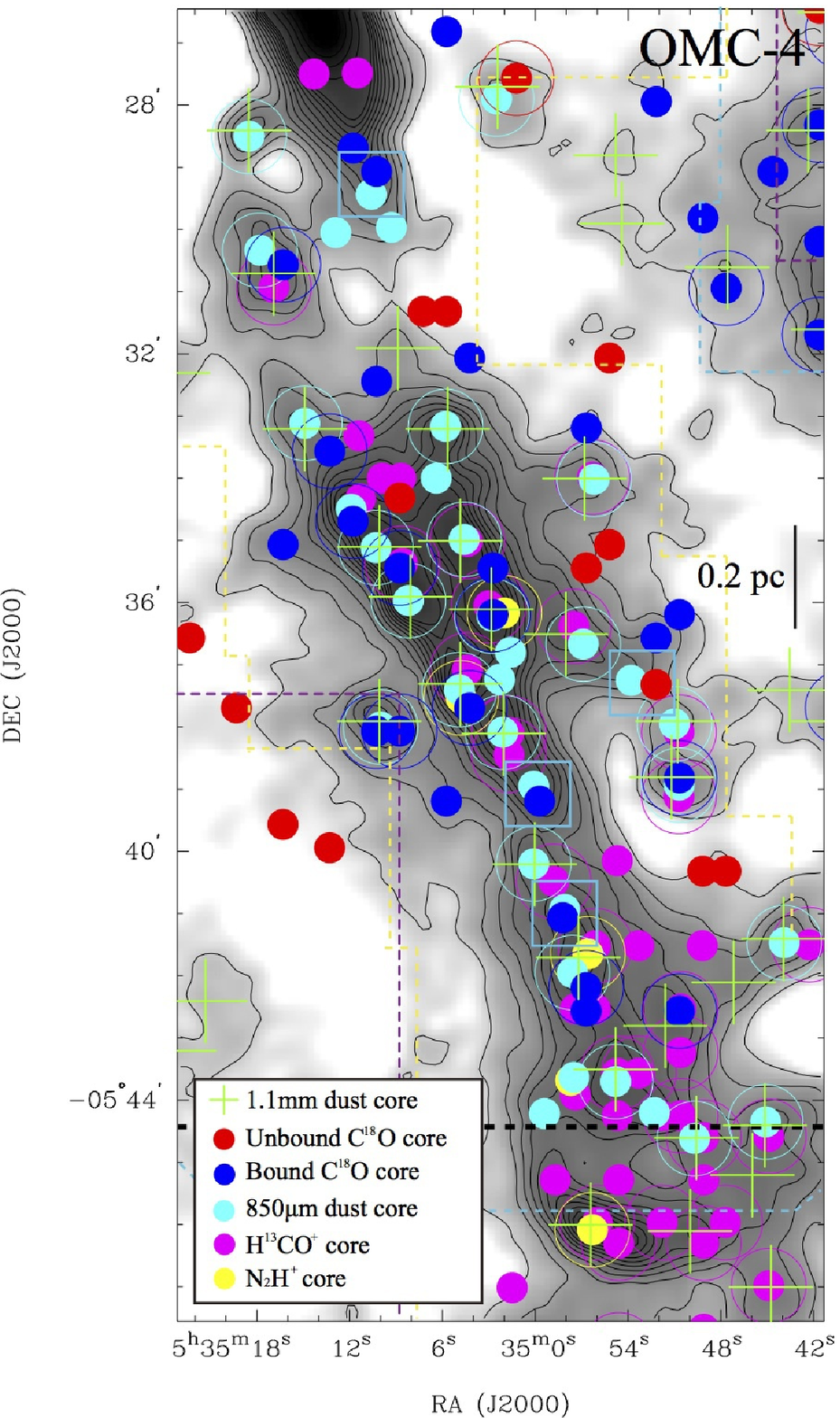}
\caption{The same as Fig. \ref{omc23_id}, but for the OMC-4 region.}
\label{omc4_id_all}
\end{figure*}

\begin{figure*}
\centering
\includegraphics[width=160mm,angle=0]{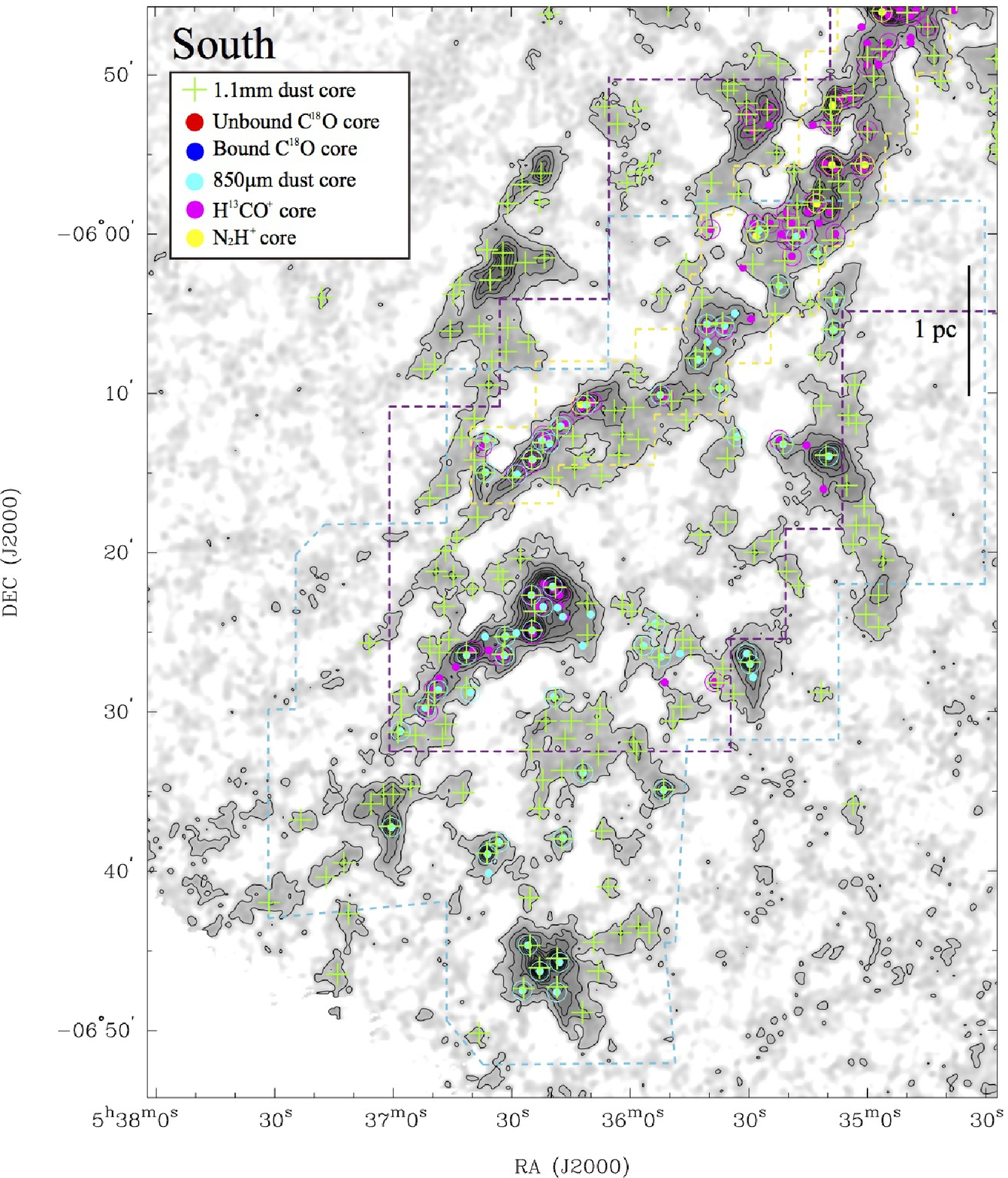}
\caption{The same as Fig. \ref{omc23_id}, but for the South region. There is no C$^{18}$O data for this region}
\label{orion_s_id_all}
\end{figure*}

\clearpage
%Tables
%Table : Physical properties of 1.1 mm dust cores
\input{Table1.tex}

%AzTEC core table
\input{Table2.tex}

%Table of physical properties of C18O cores
\input{Table3.tex}

%C18O core table
\input{Table4.tex}

%\table spizter_distribution.tex
\input{Table5.tex}

%\table KS test
\input{Table6.tex}
\input{Table7.tex}

%Category
\input{Table8.tex}

\input{Table9.tex}

\input{Table10.tex}

%%%%%%%%%%%%%%%%%%%%%%%%%%%%%%%%
\clearpage
\appendix

\section{Weak dependence of the core properties on the Clumpfind parameters \label{sect:dependence}}

\begin{figure*}
\centering
\includegraphics[width=110mm,angle=270]{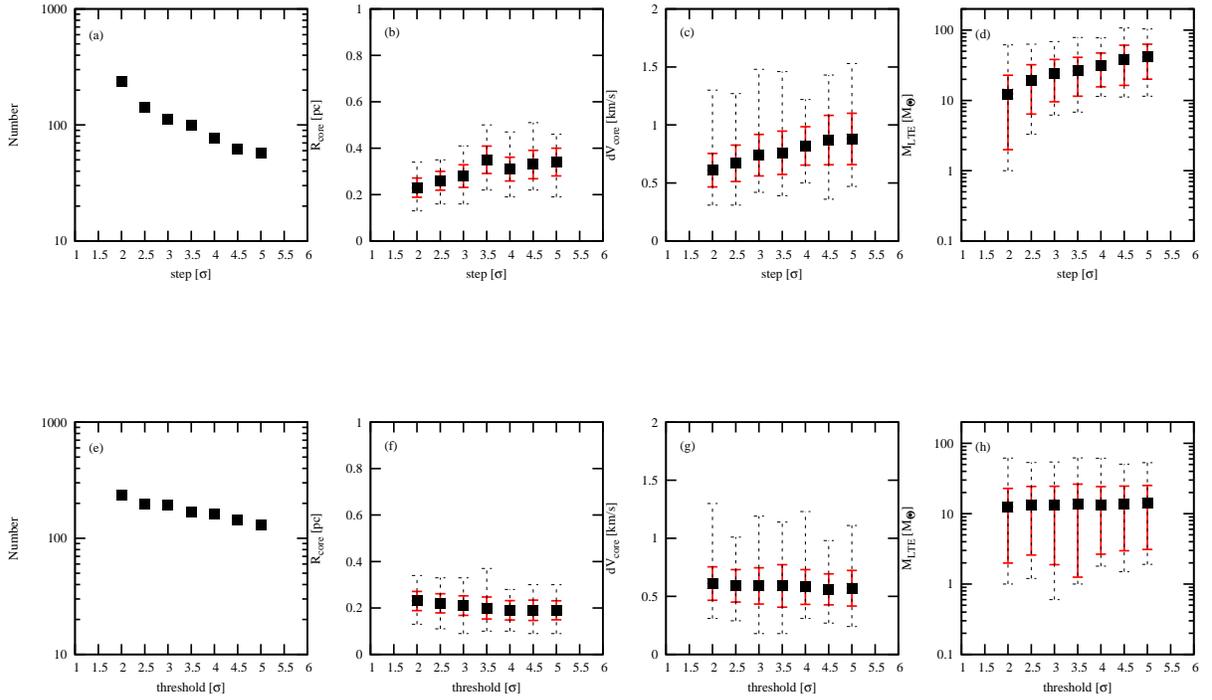}
\caption{(a)Number, (b)$R_{\rm core}$, (c) $dV_{\rm core}$, and (d) $M_{\rm LTE}$ of the identified C$^{18}$O cores as a function of the step size with a threshold level of 2$\sigma$ in the Clumpfind algorithm. (e)Number, (f)$R_{\rm core}$, (f) $dV_{\rm core}$, and (h) $M_{\rm LTE}$ as a function of the threshold level with a step size of 2$\sigma$.  In the panels (b), (c), (d), (f), (g), and (h), the back filled squares show the mean values and the lower and upper limits in black show the minimum and maximum values. The red error bars indicate the standard deviation.}
\label{dependence}
\end{figure*}

\begin{figure*}
\centering
\includegraphics[width=80mm,angle=0]{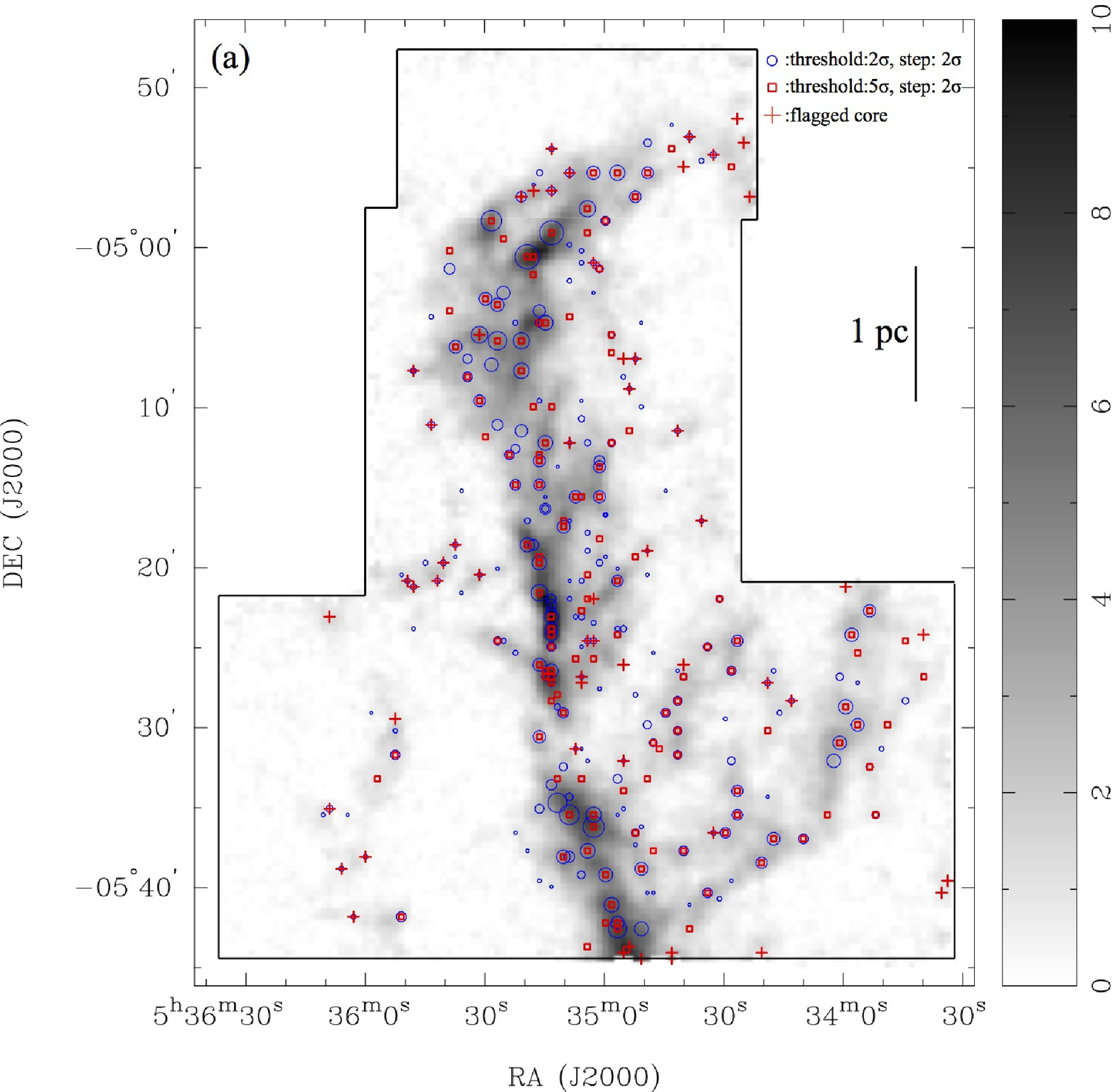}
\includegraphics[width=80mm,angle=0]{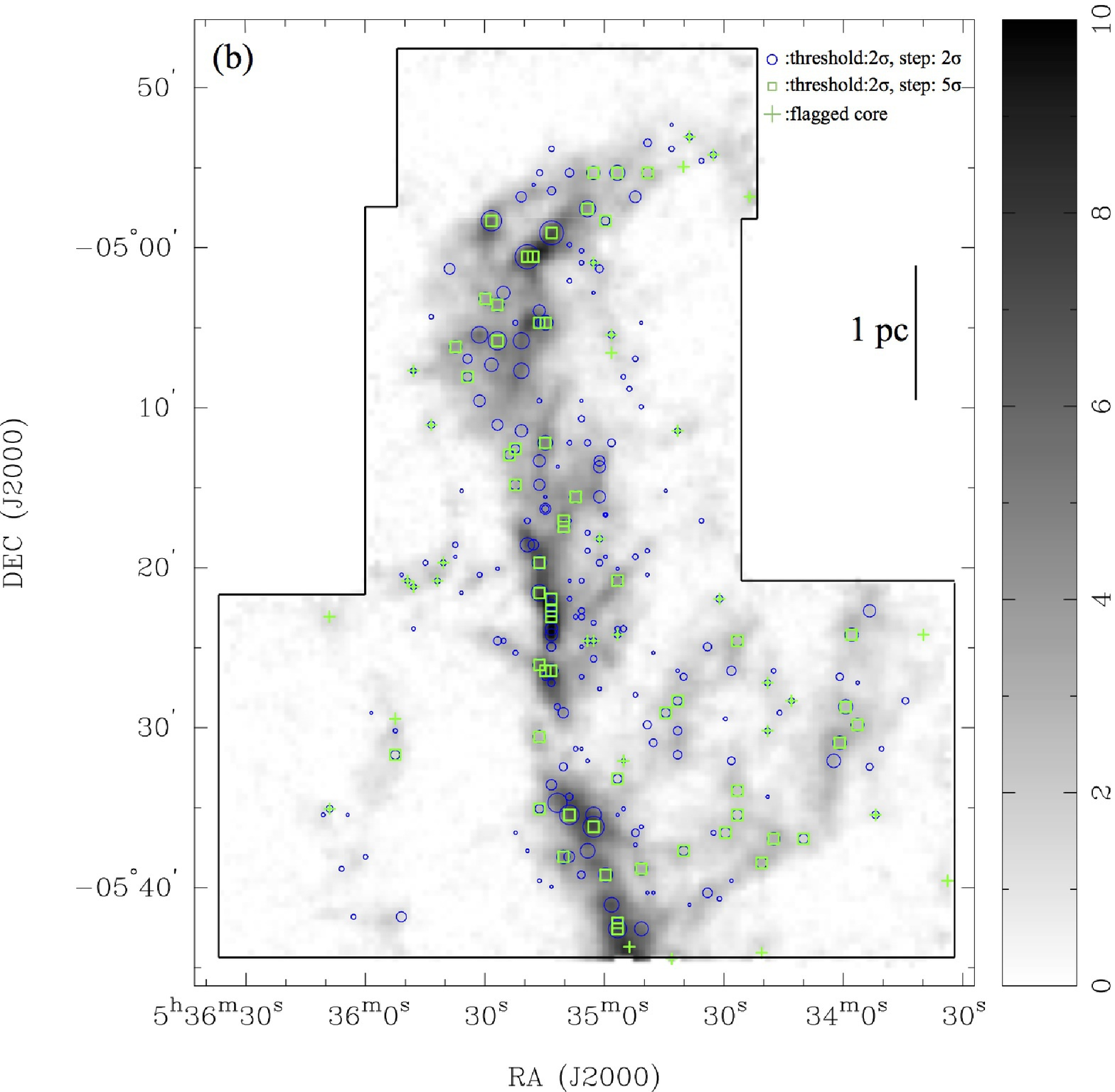}
\caption{Positions of the C$^{18}$O cores identified by the Clumpfind algorithm with various parameters on the C$^{18}$O total integrated intensity map (gray scale). The blue circles indicate the C$^{18}$O cores identified with a 2$\sigma$ threshold level and a 2$\sigma$ step size. Their sizes are set to be proportional to their peak intensities. In panel (a), the red squares indicate the C$^{18}$O cores identified with a 5$\sigma$ threshold level and a 2$\sigma$ step size. In panel (b), the green squares indicate the C$^{18}$O cores identified with a 2$\sigma$ threshold level and a 5$\sigma$ step size. In both panels, the crosses indicate the C$^{18}$O cores flagged out due to the lack of the spatial size and/or velocity width.}
\label{varing_parameter}
\end{figure*}

As described in Section \ref{1.1mm_ID}, the properties of the cores identified by the Clumpfind algorithm in some degree depend on the input parameters of the algorithm.
In order to investigate the influence of the Clumpfind parameters on the core properties, we performed the core identification with various step sizes and threshold levels in a reasonable range of 2$\sigma$ to 5$\sigma$.
Figure \ref{dependence} shows the resulting number, $R_{\rm core}$, $dV_{\rm core}$, and $M_{\rm LTE}$ of the identified C$^{18}$O cores as a function of the step size and the threshold level. The number of the identified C$^{18}$O cores significantly decreases with increasing step size, while the core number weakly depends on the threshold level. 
In contrast, the dependence of the number of cores on the threshold is week. 
The $R_{\rm core}$, $dV_{\rm core}$, and $M_{\rm LTE}$ values gradually increase with increasing step size, but do not depend on the threshold level.
In conclusion, we demonstrated the weak dependence of the core properties on both the step size and the threshold level  in a reasonable range of 2$\sigma$ to 5$\sigma$, which was also shown by \citet{Pineda09} and \citet{Ikeda09}. 
Figure \ref{varing_parameter} shows the distribution of the identified C$^{18}$O cores with various step sizes and threshold levels. Most C$^{18}$O cores having weak intensities are not identified by the Clumpfind with a larger step size and threshold level, and some C$^{18}$O cores which are close to each other are merged into one core.

\begin{figure*}
\centering
\includegraphics[width=150mm,angle=0]{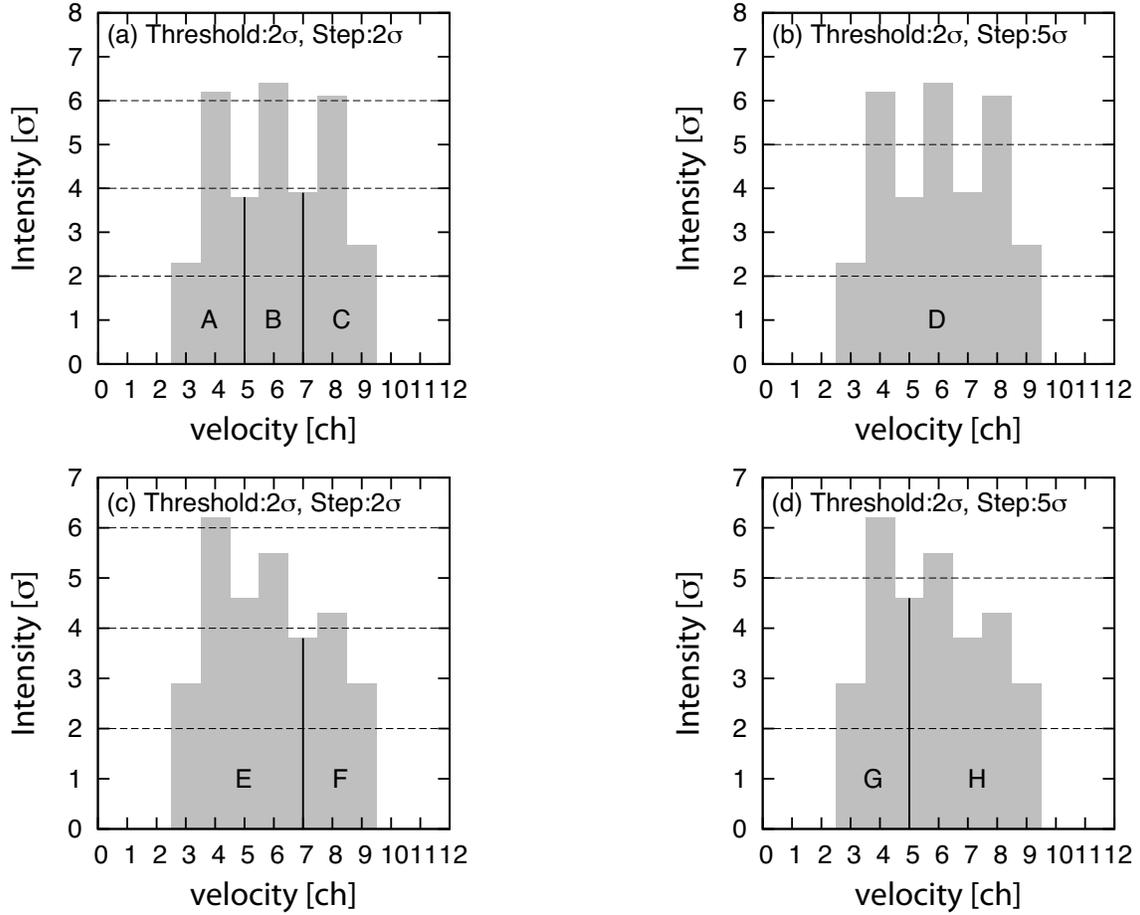}
\caption{
Plots to typically show the difference in the core identification with various parameters.
The four panels show the distribution of the identified cores along the velocity axis.
The panels (a) and (c)  are for the source extraction with a 2$\sigma$ threshold level and a 2$\sigma$ step size.
The panels (b) and (d)  are for the source extraction with a 2$\sigma$ threshold level and a 5$\sigma$ step size.
The horizontal axis shows the velocity in units of channel. The vertical axis shows the intensity in units of $\sigma$.
}
\label{schematic}
\end{figure*}

The source extraction with a 2$\sigma$ threshold level and a 5$\sigma$ step size identified some cores which are not identified with a 2$\sigma$ threshold level and a 2$\sigma$ step size. Figures \ref{schematic} (a) and (b) show the difference in the core identification with the two different step sizes and the same threshold level.
As described in Section \ref{clumpfind}, a core must contain two or more continuous velocity channels, 
and they must have at least 3 pixels whose intensities are above the 3$\sigma$ level, and 
in addition the pixels must be connected to one another in both the space and velocity domains. 
  In the case of a 2$\sigma$ threshold level and a 2$\sigma$ step size, the cores labeled as A, B, and C are flagged out due to the lack of the velocity width (pixels along the velocity axis) as shown in Fig \ref{schematic} (a). However, in the case of a 2$\sigma$ threshold level and a 5$\sigma$ step size, these cores are merged along the velocity axis and the merged cores labeled as D are identified as shown in Fig \ref{schematic} (b).
The source extraction with a 2$\sigma$ threshold level and a 5$\sigma$ step size does not identify the some cores which are identified with a 2$\sigma$ threshold level and a 2$\sigma$ step size as shown in Figs. \ref{schematic} (c) and (d) . In the case of a 2$\sigma$ threshold level and a 2$\sigma$ step size, the core labeled as E is identified. However, the core labeled as “F” is flagged out due to the lack of the velocity width and spatial size. In the case of a 2$\sigma$ threshold level and a 5$\sigma$ step size, the velocity width of each core is changed due to the different step size. The core labeled as G is flagged out due to the lack of the velocity width. The core labeled as H is also flagged out due to the lack of the spatial size.

\section{How the Clumpfind identified the dust and C$^{18}$O cores in the mapping data \label{sect:c18oID}}
In Section \ref{1.1mm_ID} and \ref{clumpfind}, we identified cores from the 1.1 mm dust continuum and C$^{18}$O emission using the Clumpfind method \citep{Williams94}. Although this algorithm has been widely used for the identification of cores and clumps, 
results of the identification are known to vary depending on the parameters used in the algorithm \citep{Pineda09}. In order to see how the algorithm works in the case of our catalog, we compare the distribution of the cores with the 1.1 mm dust continuum and C$^{18}$O maps. 

\begin{figure*}
\centering
\includegraphics[width=160mm,angle=0]{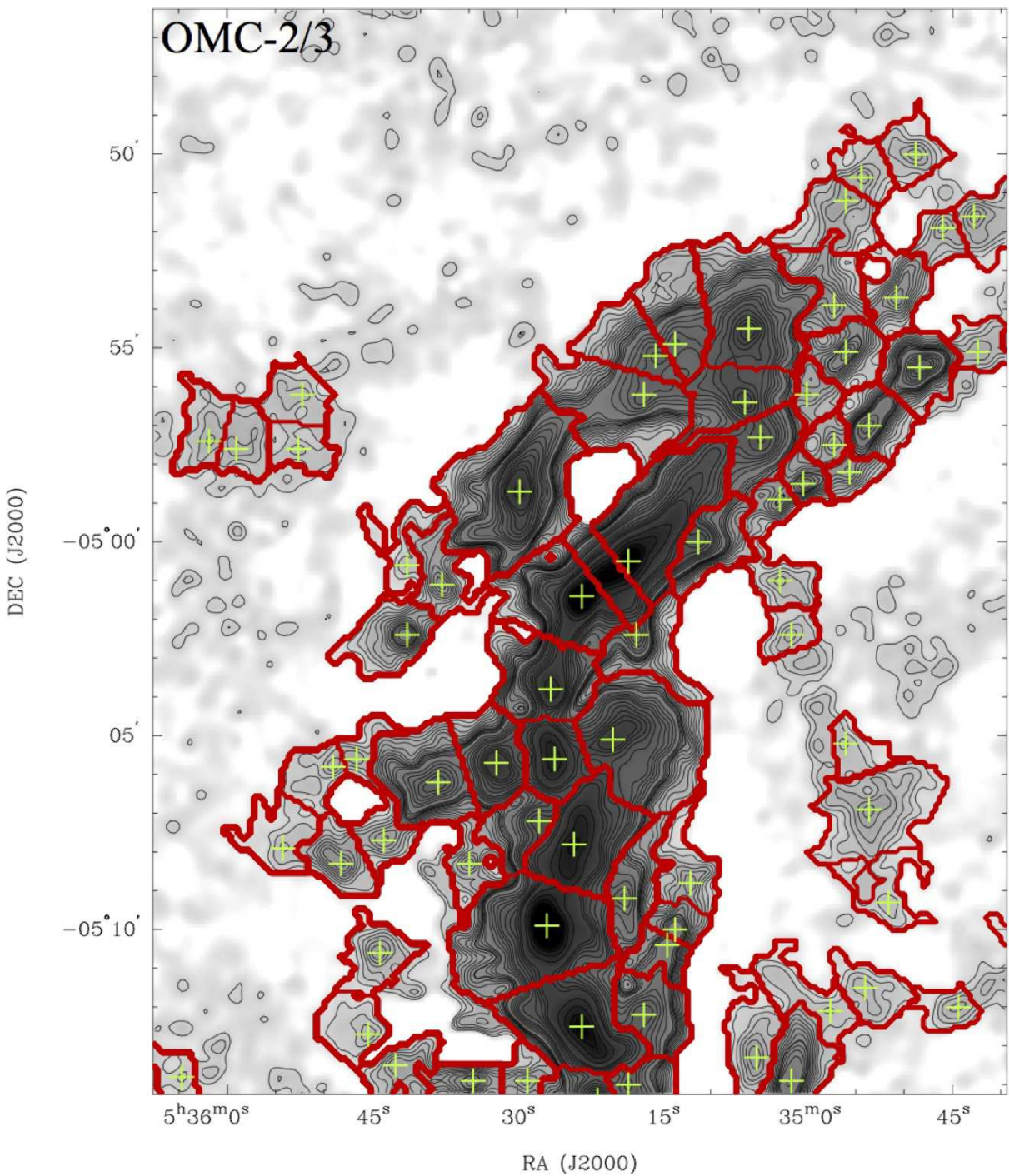}
\caption{The identified 1.1 mm dust cores on the AzTEC 1.1mm maps of the OMC-2/3 region. 
The green plus signs show the positions of the 1.1 mm dust cores, and the thick red lines denote their boundaries defined by the Clumpfind algorithm. The contours start from 4$\sigma$ noise level with an increment of 2$\sigma$ for the range 4--30$\sigma$, 5$\sigma$ for the range 30--105$\sigma$, and 50$\sigma$ for the range $>$105$\sigma$ (1$\sigma$=9 mJy beam$^{-1}$).}
\label{omc23_only_id}
\end{figure*}

\begin{figure*}
\centering
\includegraphics[width=80mm,angle=0]{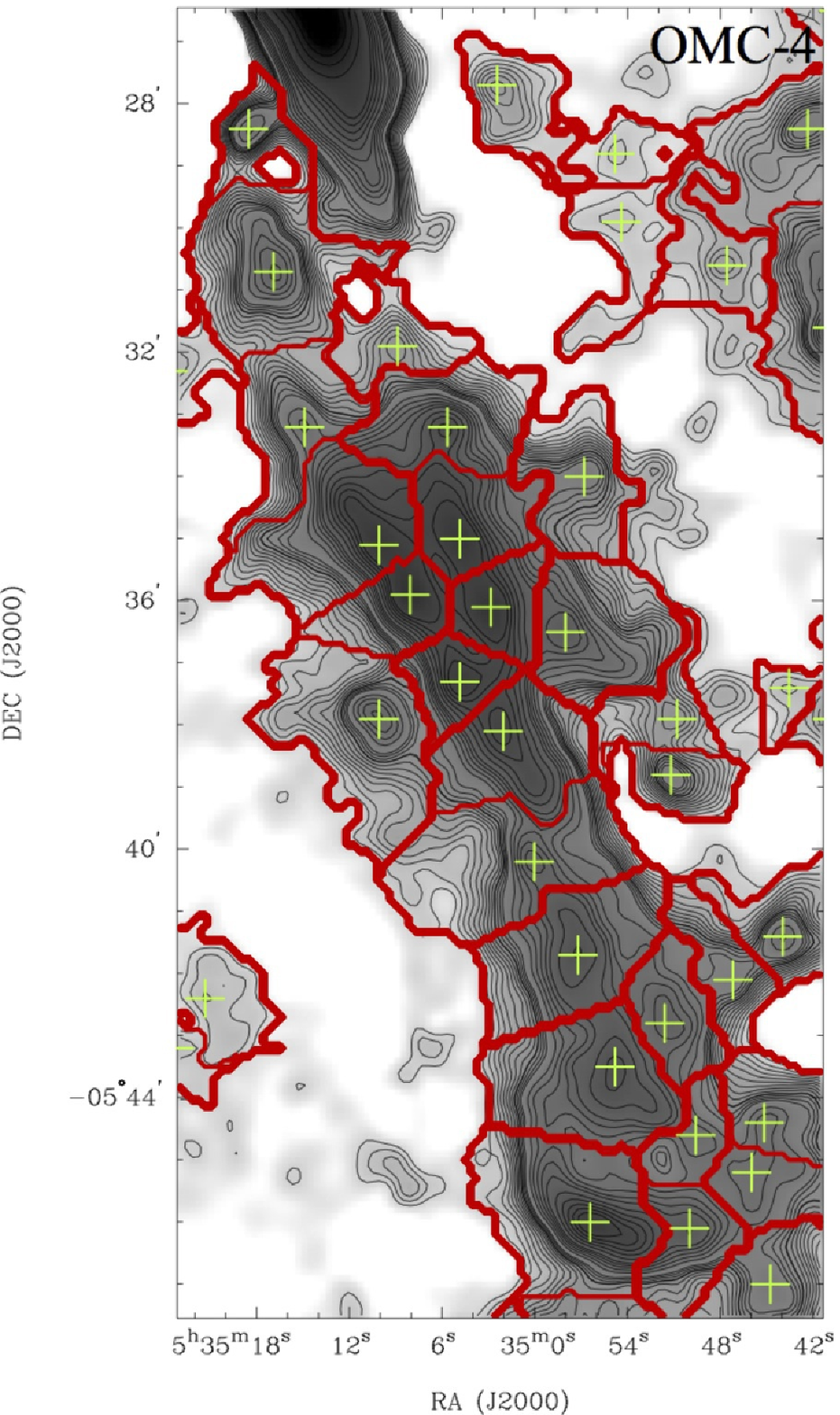}
\includegraphics[width=60mm,angle=0]{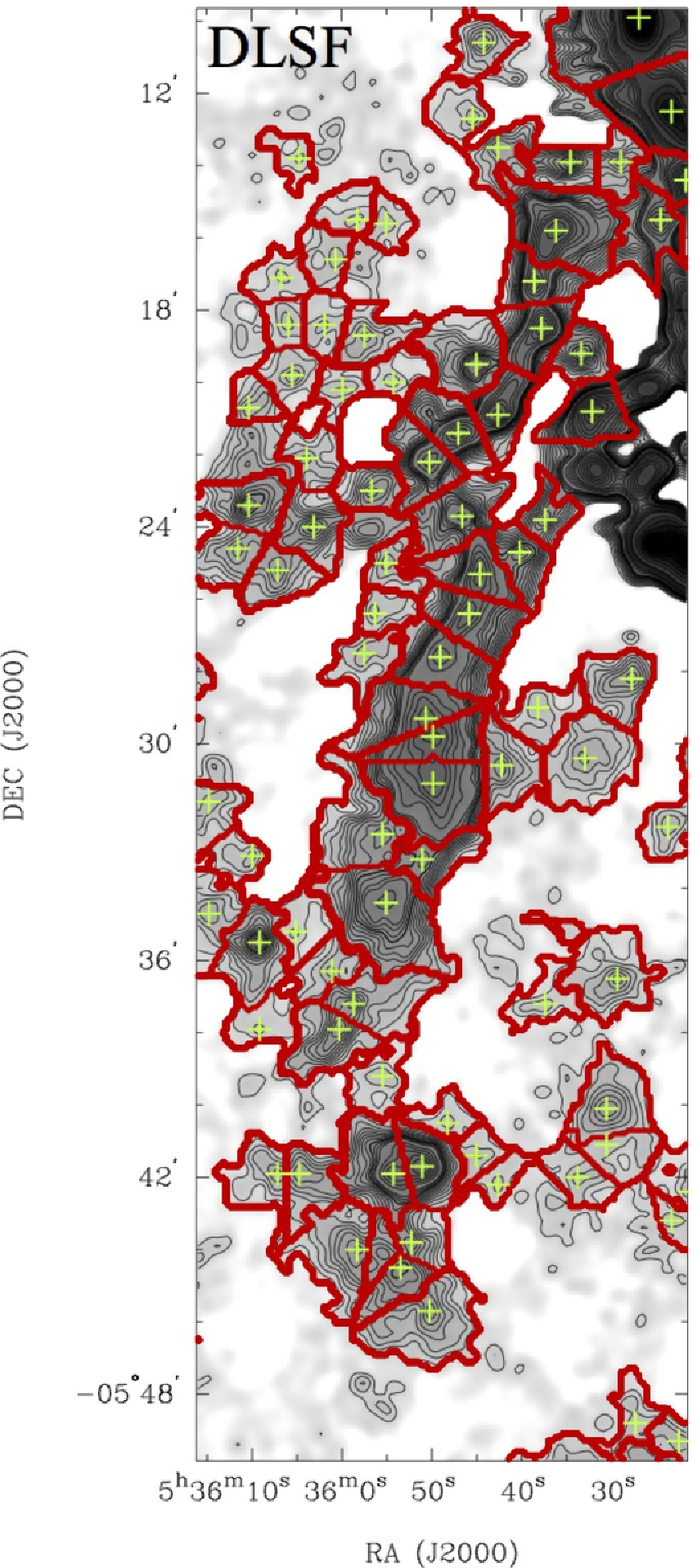}
\caption{Same as Fig. \ref{omc23_only_id}, but for the OMC-4 and DLSF regions.}
\label{omc4_dlsf_only_id}
\end{figure*}

\begin{figure*}
\centering
\includegraphics[width=160mm,angle=0]{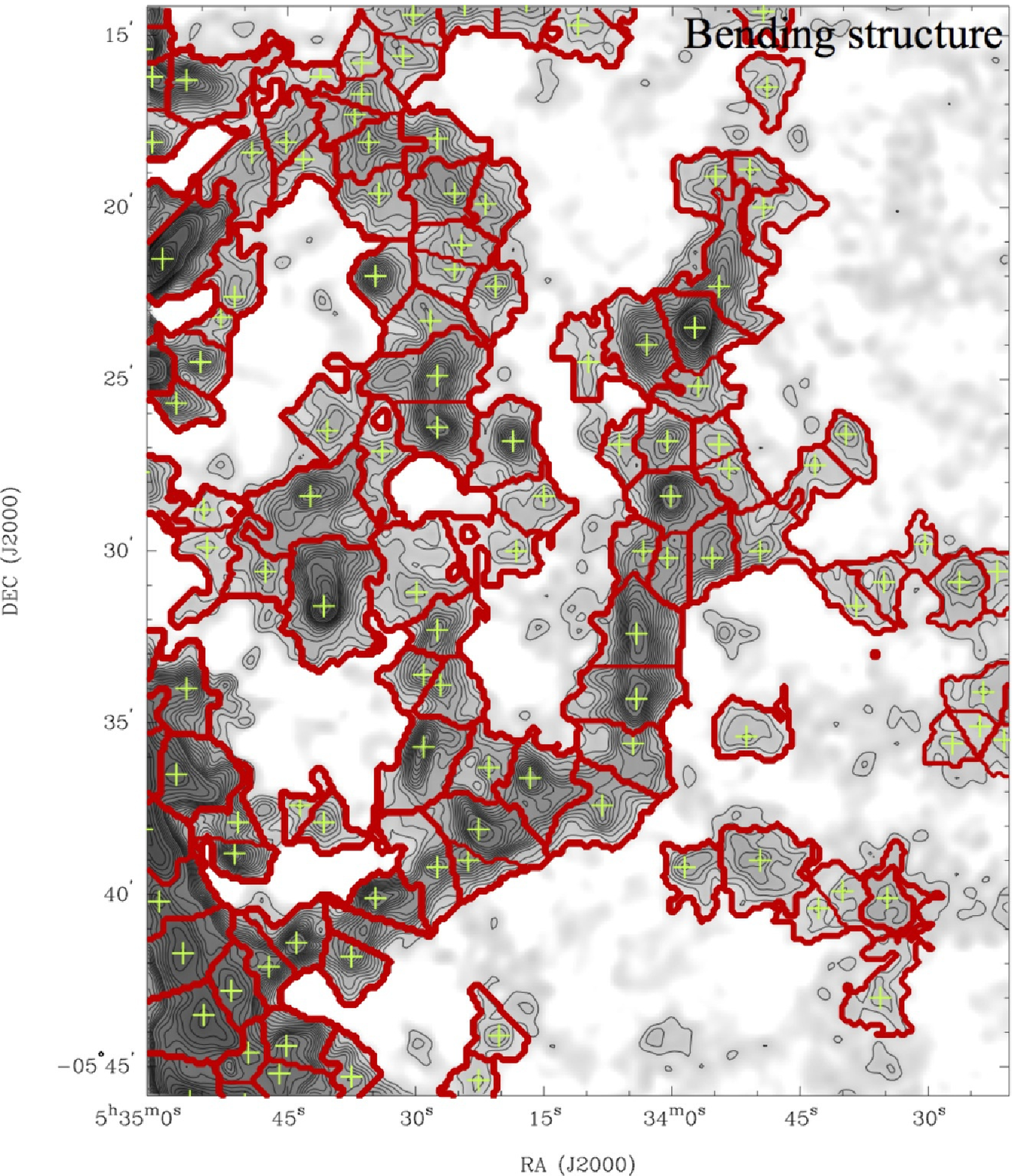}
\caption{Same as Fig. \ref{omc23_only_id}, but for the bending structure region.}
\label{bend_only_id}
\end{figure*}

\begin{figure*}
\centering
\includegraphics[width=160mm,angle=0]{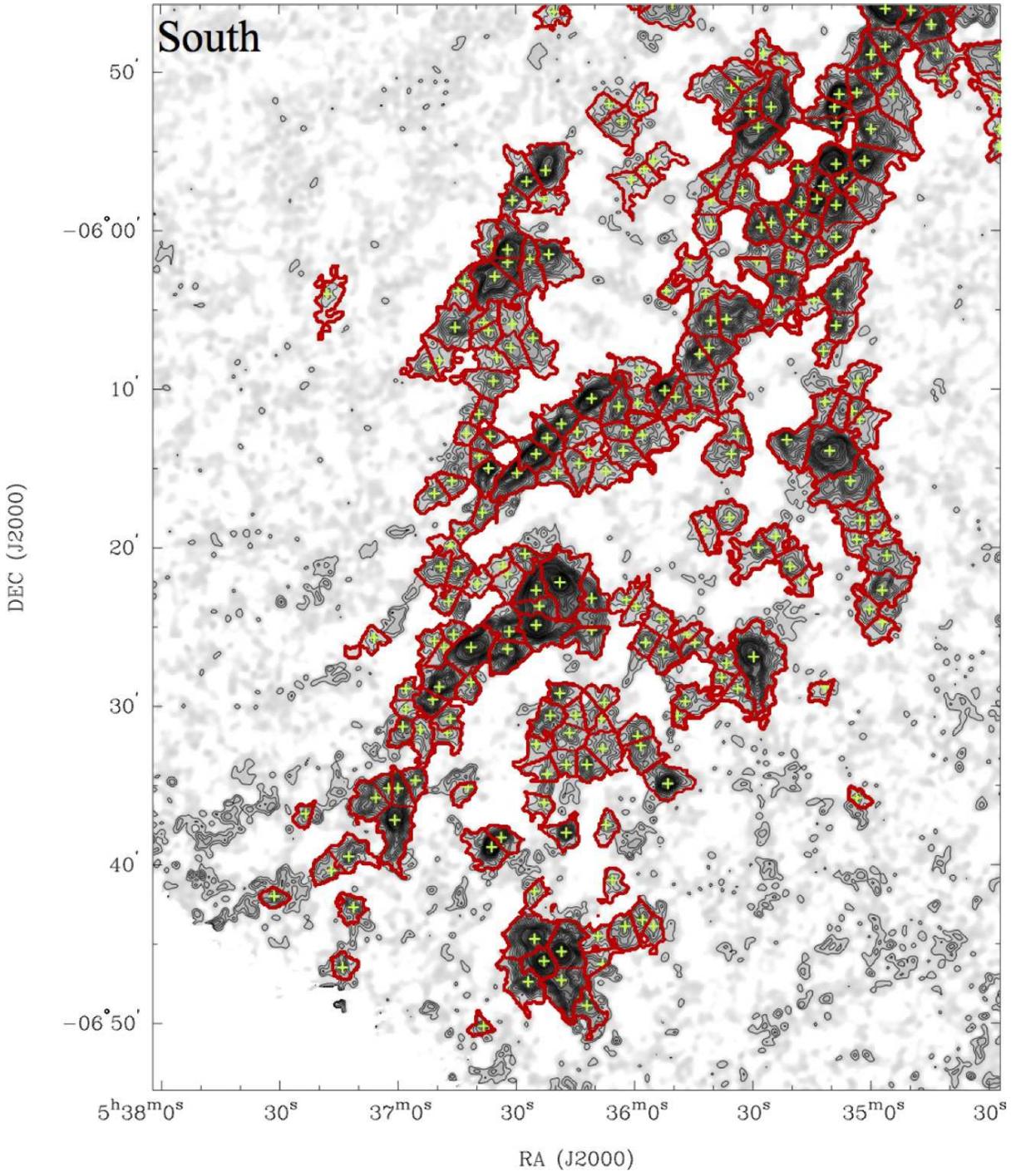}
\caption{Same as Fig. \ref{omc23_only_id}, but for the South region.}
\label{south_only_id}
\end{figure*}

Figures \ref{omc23_only_id} -- \ref{south_only_id} show the distribution of the identified 1.1 mm dust cores overlaid on the 1.1 mm dust continuum maps with the boundaries of each core. The movie of the C$^{18}$O velocity channel maps with a velocity resolution of 0.104 km s$^{-1}$ overlaid the positions of the identified C$^{18}$O cores are available at the web page of the NRO star formation project (via http://th.nao.ac.jp/MEMBER/nakamrfm/sflegacy/sflegacy.html).
The apparent distributions of the 1.1mm dust and C$^{18}$O cores have good agreements with the positions of the cores identified by the Clumpfind.

\section{Effectiveness of the FRUIT method in extracting the dust core properties from the ground-based bolometer array map \label{sect:negative}}
\begin{figure*}
\centering
\includegraphics[width=120mm,angle=270]{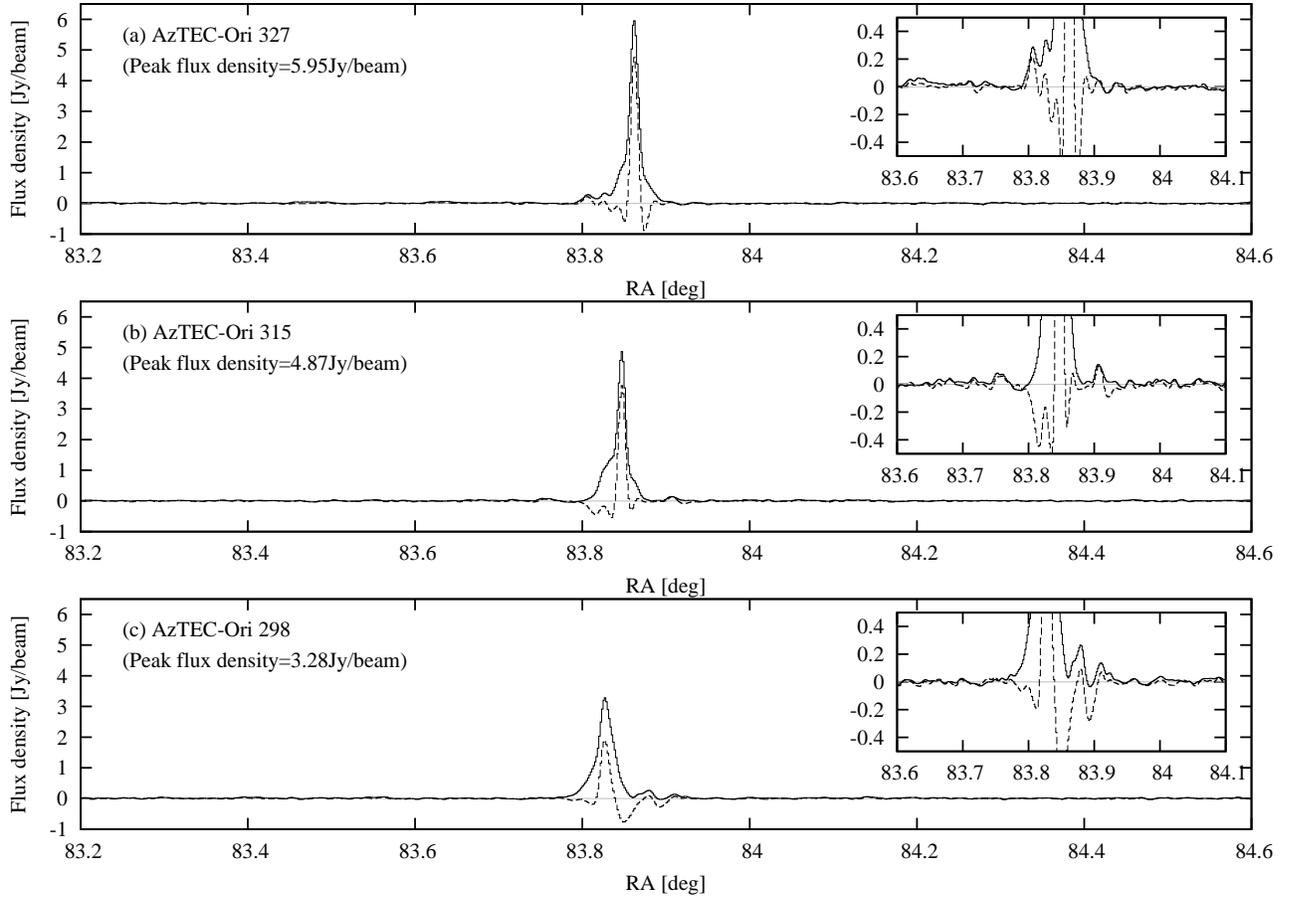}
\caption{Intensity profiles of the 1.1 mm dust continuum emission along the R.A. direction for the three brightest cores of (a) AzTEC-Ori 327 (DEC = -5$^\circ$ 9$\arcmin$ 54$\arcsec$.4), (b) 315 (DEC = -5$^\circ$ 1$\arcmin$ 24$\arcsec$.4), and (c) 298 (DEC = -5$^\circ$ 0$\arcmin$ 30$\arcsec$.4). The cut lines along the Dec. direction go through the peak positions of the cores.
The black dashed lines indicate the PCA profiles. The black solid lines indicate the FRUIT profiles. }
\label{intensity_profiles}
\end{figure*}

\citet{Shimajiri11} applied a principal component analysis (PCA) cleaning method to remove atmospheric noise in the AzTEC 1.1 mm dust continuum map. The PCA cleaning, however, removes also astronomical signals in the case that the emission has an extended structure, because the PCA method can not distinguish the extended astronomical emission over several bolometer array elements from the extended atmospheric emission. This causes the negative-flux feature around bright objects. Figure \ref{intensity_profiles} shows the intensity profiles of the 1.1mm dust continuum emission along the R.A. direction for the three brightest cores with flux densities above 3 Jy beam$^{-1}$.  The negative flux features around the bright objects are clearly recognized in the PCA profiles. 
Hence the flux density and size of the extended sources are underestimated. 

In order to recover the extended features in our Orion map after the PCA cleaning, an interactive mapping method FRUIT was applied. 
To investigate the performance of the FRUIT data reduction, Shimajiri et al. (2011) performed a simulation in which Gaussian sources with various FWHM sizes and total flux were implanted in Orion data-set as described in Section \ref{1.1mm_ID}. The larger the FWHM size of the model source, the lower the recovered fraction of the input total flux densities were artificially embedded in the Orion data. In the case that the FWHM size of the input source is below 150$\arcsec$($\sim$0.3pc), the output total flux density is underestimated by less than 20\%. Since the size range of the identified 1.1 mm cores is 0.01 -- 0.20 pc, we believe that the sizes and flux densities of the dust cores are well recovered on the 1.1 mm map.
In fact, significant negative flux features can not be recognized in the FRUIT map as seen in Figure \ref{intensity_profiles}.

\begin{figure*}
\centering
\includegraphics[width=120mm,angle=270]{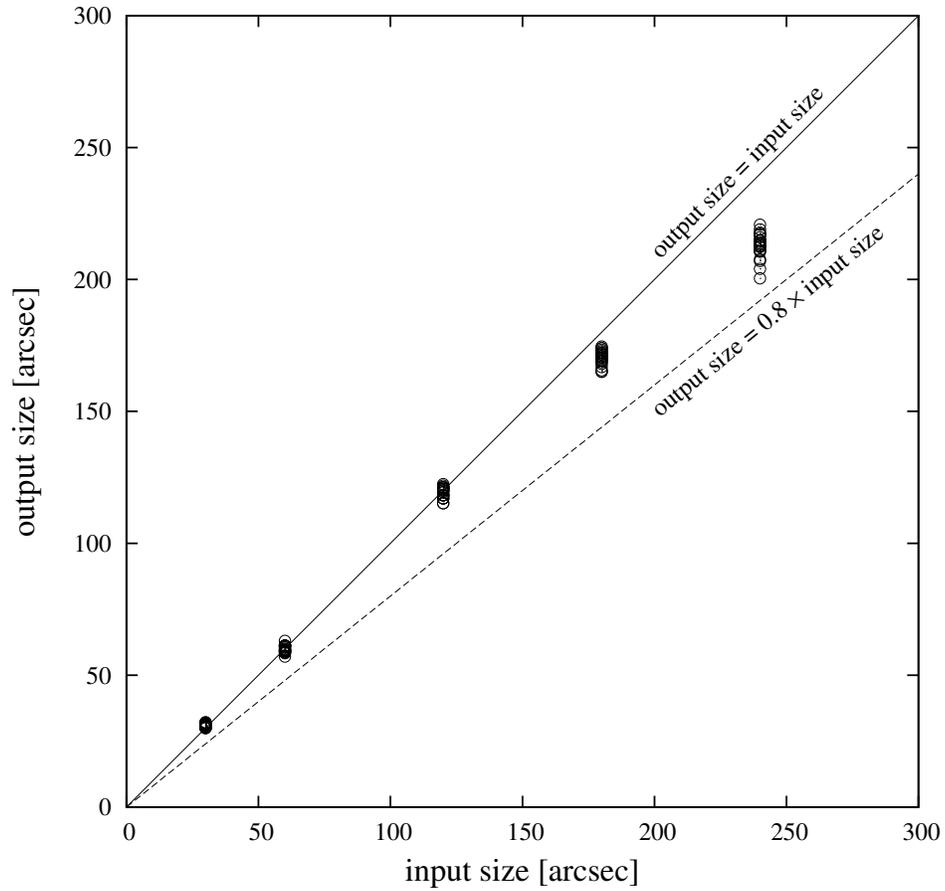}
\caption{Recovery of the source size for simulation observations. We performed a simulation in which twenty five Gaussian sources with a peak flux of 1 Jy and FWHM sizes of 30$\arcsec$, 60$\arcsec$, 120$\arcsec$, 180$\arcsec$, and 240$\arcsec$ were inserted to Orion data-set. The solid and broken lines indicate (output size = input size) and (output size = 0.8 $\times$ input size), respectively. }
\label{output_size}
\end{figure*}

We also investigated the recovery of source size using the same simulation data set in \citet{Shimajiri11}.  
Figure \ref{output_size} shows how the input source size is recovered.
In the case that the inserted source sizes are less than 120$\arcsec$ ($\sim$ 0.24 pc), the source sizes are mostly recovered.
In the case that inserted source sizes are 240$\arcsec$ ($\sim$ 0.5 pc),  the source sizes are underestimated by 20\%. In our results, the size range of the 1.1 mm dust cores is 0.01 -- 0.20 pc, and we did not detect any cores with sizes of 0.2 -- 0.5 pc. These results suggest that 
there are no large ($>$ $\sim$ 0.5 pc) cores which could not be well recovered by the FRUIT method and were misidentified as smaller ($<$ $\sim$ 0.5 pc) cores.
We conclude that the source size of the cores detected in the 1.1 mm dust continuum map is mostly recovered.

\end{document}

%% file: Table1.tex
% produced on Tue May 20 09:16:02 2014
% by flag_by_sn.pro 
\begin{deluxetable}{lccc}
%\tabletypesize{\scriptsize}
\tablecaption{Physical properties of the 1.1 mm dust cores}
\tablewidth{0pt}
\tablehead{
\colhead{Property} &
\colhead{Mean$\tablenotemark{\dag}$} & 
\colhead{Minimum} &
\colhead{Maximum}\ }
\startdata
Peak flux density [Jy beam$^{-1}$] &   0.27 $\pm$   0.45 &   0.05 &   5.95 \\
$R_{\rm core}$ [pc]  &   0.09 $\pm$   0.03 &   0.01 &   0.20 \\
$M_{\rm H_2}$  [$M_{\odot}$] &   5.52 $\pm$   9.55 &   0.60 & 116.95 \\
Density [$\times$ 10$^4$ cm$^{-3}$]  &   5.48 $\pm$  38.94 &   0.31 & 915.01 \\
\enddata
\label{Physical_properties_1mm}
\tablenotetext{\dag}{Errors are the standard deviation}
\end{deluxetable}

%% file: Table2.tex
% produced on Wed May  7 17:49:44 2014
% by flag_by_sn.pro 
% searching radius for YSO and DISK is      15.0000 arcsec.
% searching radius for C18O, H13CO+, 850 cores is      36.0000 arcsec.
\begin{deluxetable}{cllccccccl}
\tablecolumns{10}
\tabletypesize{\tiny}
\rotate
\tablecaption{Identified AzTEC 1.1 mm dust cores}
\tablewidth{0pt}
\tablehead{
\colhead{ID}
& \colhead{RA}
& \colhead{DEC}
& \colhead{Peak flux density}
& \colhead{Aspect ratio}
& \colhead{$R_{\rm core}$}
& \colhead{$M_{\rm H_2}$}
& \colhead{$n$}
& \colhead{category}
& \colhead{note$\tablenotemark{\dag}$$\tablenotemark{\ddag}$}\\
\colhead{AzTEC-Ori}
& \colhead{(J2000)}
& \colhead{(J2000)}
& \colhead{[Jy beam$^{-1}$]}
& \colhead{major/minor}
& \colhead{[pc]}
& \colhead{[$M_{\odot}$] }
& \colhead{[$\times$10$^4$cm$^{-3}$]}
& \colhead{}
& \colhead{}
}
\startdata
%      &    &   &   [Jy beam$^{-1}$]   & major axis/minor axis   &  [pc]   &   [$M_{\odot}$]   &   [$\times$10$^4$cm$^{-3}$]  &   &  \\
%\hline
 1 & 5 31   28.6 & -5 40   33.6 &   0.10 &    1.1 &   0.10 &    1.9 & 
   0.7 & \nodata & \\
2 & 5 31   48.8 & -5 23    4.2 &   0.11 &    1.2 &   0.08 &    1.5 & 
   1.2 & \nodata & \\
3 & 5 32   11.2 & -5 38   16.6 &   0.18 &    1.0 &   0.14 &    6.0 & 
   0.9 & \nodata & \\
 4 & 5 32   11.6 & -5 36   46.6 &   0.09 &    2.1 &   0.08 &    1.3 & 
   1.1 & \nodata & \\
 5 & 5 32   14.8 & -5 37   40.7 &   0.09 &    1.9 &   0.06 &    1.3 & 
   3.0 & \nodata & \\
 6 & 5 32   18.5 & -5 37   46.7 &   0.10 &    1.0 &   0.10 &    1.9 & 
   0.8 & \nodata & \\
 7 & 5 32   19.3 & -5 33   58.8 &   0.09 &    1.3 &   0.09 &    1.3 & 
   0.9 & \nodata & \\
 8 & 5 32   26.6 & -5 26   34.9 &   0.06 &    1.1 &   0.08 &    1.1 & 
   0.8 & \nodata & \\
\enddata
\label{table_cores}
\tablenotetext{\dag}{[NW2007] \#, [ISK2007] \#, and [TKU2008] \# are the SCUBA 850 $\mu$m dust core from \citet{Nutter07},  the H$^{13}$CO$^{+}$ (1--0) core from \citet{Ikeda07}, and the N$_2$H$^+$ (1--0) core from \citet{Tatematsu08}, respectively.}
\tablenotetext{\ddag}{The Spitzer YSOs and Disk sources, respectively, cataloged by \citet{Megeath12}}
\end{deluxetable}

%% file: Table3.tex
% produced on Mon Jun  2 10:33:14 2014
% by make_table.pro 
\begin{deluxetable}{lccccccccccc}
\tabletypesize{\tiny}
\tablecaption{Physical properties of the C$^{18}$O cores}
\tablewidth{0pt}
\tablehead{
\colhead{} &
\multicolumn{3}{c}{Bound} &
\colhead{} &
\multicolumn{3}{c}{Unbound} &
\colhead{} &
\multicolumn{3}{c}{Total} \\
\cline{2-4}
\cline{6-8}
\cline{10-12}\\
\colhead{Property} &
\colhead{Mean$\tablenotemark{\dag}$} & 
\colhead{Minimum} &
\colhead{Maximum} &
\colhead{} &
\colhead{Mean$\tablenotemark{\dag}$} &
\colhead{Minimum} &
\colhead{Maximum} & 
\colhead{} &
\colhead{Mean$\tablenotemark{\dag}$} & 
\colhead{Minimum} &
\colhead{Maximum}\ }
\startdata
$M_{\rm LTE}$ [$M_{\odot}$] &  15.3 $\pm$   10.7 &    1.3 &   61.8 &   & 4.4 $\pm$    2.5 &    1.0 &   13.4 &   &12.4 $\pm$   10.4 &    1.0 &   61.8 \\
$dV_{\rm core}$ [km s$^{-1}$]  &   0.59 $\pm$   0.13 &   0.31 &    1.02 &  &  0.66 $\pm$   0.18 &   0.40 &   1.31 &  &  0.61 $\pm$   0.14 &   0.31 &   1.31 \\
$R_{\rm core}$ [pc]  &   0.23 $\pm$   0.04 &   0.13 &   0.34 &  & 0.21 $\pm$   0.04 &   0.13 &   0.32 & &   0.23 $\pm$   0.04 &   0.13 &   0.34 \\
$n$ [$\times$ 10$^3$ cm$^{-3}$] &   4.8 $\pm$    2.8 &    1.1 &    17.5 &  &   2.0 $\pm$    1.1 &    0.8 &    5.6 &    & 4.1 $\pm$    2.8 &    0.8 &   17.5 \\
$\mathcal{R}_{\rm vir}$ &    1.5 $\pm$    0.7 &    0.4 &    3.0 &    & 5.0 $\pm$    2.8 &   3.1 &   20.1 &  &  2.4 $\pm$    2.2 &    0.4 &   20.1 \\
$M_{\rm J}$ [$M_{\odot}$] &    3.0 $\pm$    2.7 &    0.2 &   20.2 &  &  5.6 $\pm$    8.9 &    0.5 &   55.6 &  &  3.7 $\pm$    5.2 &    0.2 &   55.6 \\
$M_{\rm \phi}$ [$M_{\odot}$] &    0.4 $\pm$    0.1 &    0.1 &    0.8 &  &   0.3 $\pm$    0.1 &    0.1 &    0.7 &  &  0.4 $\pm$    0.1 &    0.1 &    0.8 \\
$\mathcal{R}_{\rm c}$ &    5.5 $\pm$    3.4 &    1.1 &   20.8 &  &  1.3 $\pm$    0.7 &    0.1 &    2.6 &   & 4.4 $\pm$    3.5 &    0.1 &   20.8 \\
\enddata
\label{Physical_properties_c18o}
\tablenotetext{\dag}{Errors are the standard deviation}
\end{deluxetable}

%% file: Table4.tex
% produced on Wed May 28 15:41:09 2014
% by make_table.pro 
% seraching radius for 1.1mm core is       36.0000 arcsec.
\begin{deluxetable}{cllcccccccccccccccl}
%\centering
\rotate
\tabletypesize{\tiny}
\tablecolumns{19}
\tablecaption{Identified C$^{18}$O cores}
\tablewidth{0pt}
\tablehead{
\colhead{ID}
& \colhead{RA}
& \colhead{DEC}
& \colhead{$V_{\rm LSR}$}
& \colhead{$T_{\rm peak}$}
& \colhead{Aspect}
& \colhead{$R_{\rm core}$}
& \colhead{$dV_{\rm core}$}
& \colhead{$\sigma$}
& \colhead{$M_{\rm LTE}$}
& \colhead{$M_{\rm VIR}$}
& \colhead{$\mathcal{R}_{\rm vir}$}
& \colhead{$M_{\rm J}$}
& \colhead{$M_{\rm \Phi}$}
& \colhead{$\mathcal{R}_{\rm c}$}
& \colhead{$n$}
& \colhead{area}
& \colhead{category}
& \colhead{note}\\
\colhead{C$^{18}$O-Ori}
& \colhead{(J2000)}
& \colhead{(J2000)}
& \colhead{[km s$^{-1}$]}
& \colhead{[K]}
& \colhead{}
& \colhead{[pc]}
& \colhead{[km s$^{-1}$]}
& \colhead{[km s$^{-1}$]}
& \colhead{[$M_{\odot}$]}
& \colhead{[$M_{\odot}$]}
& \colhead{}
& \colhead{[$M_{\odot}$]}
& \colhead{[$M_{\odot}$]}
& \colhead{}
& \colhead{[$\times$10$^4$cm$^{-3}$]}
& \colhead{}
& \colhead{}
& \colhead{}
}
\startdata
% &J2000      &   J2000  &     km s$^{-1}$  & K  &    & pc  & km s$^{-1}$ & km s$^{-1}$  & $M_{\odot}$  & $M_{\odot}$ & & $M_{\odot}$  &  $M_{\odot}$ &  &  10$^4$ cm$^{-3}$ &  \\
%\hline
 1 & 5 33  44.4 & -5 28  19.0 &    8.0 &    1.7 &    1.2 &   0.20
 &   0.63 &   0.27 &    8.8 &   16.7 &    1.9 &    3.0 &    0.3 &    2.7 & 
   0.5 & Bending & C &  \\
 2 & 5 33  50.4 & -5 31  19.0 &    8.4 &    1.2 &    1.3 &   0.24
 &   0.59 &   0.25 &    6.6 &   17.4 &    2.7 &    2.3 &    0.4 &    2.4 & 
   0.2 & Bending & C &  \\
 3 & 5 33  51.9 & -5 35  26.5 &    8.0 &    1.9 &    1.7 &   0.25
 &   0.55 &   0.23 &    8.7 &   15.8 &    1.8 &    1.7 &    0.4 &    4.0 & 
   0.2 & Bending & A & AzTEC-Ori      69 \\
 4 & 5 33  53.4 & -5 32  26.5 &    7.9 &    1.8 &    1.1 &   0.24
 &   0.54 &   0.23 &    8.5 &   14.8 &    1.7 &    1.6 &    0.4 &    4.2 & 
   0.3 & Bending & C &  \\
 5 & 5 33  53.4 & -5 22  41.5 &    8.1 &    3.2 &    1.3 &   0.24
 &   0.50 &   0.21 &   15.0 &   12.1 &    0.8 &    1.1 &    0.4 &    9.8 & 
   0.5 & Bending & A & AzTEC-Ori      72 \\
 6 & 5 33  56.4 & -5 29  49.0 &    8.1 &    3.3 &    1.2 &   0.20
 &   0.61 &   0.26 &   17.2 &   15.1 &    0.9 &    2.6 &    0.3 &    6.1 & 
   1.0 & Bending & A & AzTEC-Ori      75 \\
 7 & 5 33  56.4 & -5 27  11.5 &    7.8 &    0.9 &    1.1 &   0.23
 &   0.44 &   0.19 &    2.6 &    9.3 &    3.6 &    0.7 &    0.4 &    2.4 & 
   0.1 & Bending & A & AzTEC-Ori      71 \\
 8 & 5 33  57.9 & -5 24  11.5 &    8.3 &    3.5 &    1.1 &   0.26
 &   0.58 &   0.25 &   23.1 &   18.8 &    0.8 &    2.2 &    0.5 &    8.7 & 
   0.5 & Bending & C &  \\
\enddata
\label{table_c18o_cores}
\end{deluxetable}

%% file: Table5.tex
% produced on Fri May 30 12:13:30 2014
% by number_density_7th.pro
\begin{deluxetable}{lccccccccccc}
\tabletypesize{\tiny}
\tablecaption{Numbers and number densities of the 1.1 mm dust cores, protostars, and pre-main sequence stars in the six distinct regions}
\tablewidth{0pt}
\tablehead{
\colhead{} &
\multicolumn{3}{c}{Area definition} &
\colhead{} &
\multicolumn{3}{c}{Number} &
\colhead{} &
\multicolumn{3}{c}{Number density} \\
\cline{2-4}
\cline{6-8}
\cline{10-12}\\
\colhead{Region}
&\colhead{BLC}
&\colhead{TRC}
&\colhead{area}
&\colhead{}
&\colhead{1.1 mm $\tablenotemark{\dag}$}
&\colhead{protostar}
&\colhead{PMS star}
&\colhead{}
&\colhead{1.1 mm $\tablenotemark{\dag}$}
&\colhead{protostar}
&\colhead{PMS star}\\
\cline{2-4}
\cline{6-8}
\cline{10-12}\\
\colhead{} 
&\colhead{RA$_{\rm J2000}$, DEC$_{\rm J2000}$} 
&\colhead{RA$_{\rm J2000}$, DEC$_{\rm J2000}$}
&\colhead{[pc$^{2}$]} 
&\colhead{}
&\colhead{} 
&\colhead{}
&\colhead{}
&\colhead{}
&\colhead{[pc$^{-2}$]} 
&\colhead{[pc$^{-2}$]}
&\colhead{[pc$^{-2}$]}
}\startdata
OMC1 & 5 36 00, -5 32 35 & 5 34 35, -5 13 35 &    5.6 & & \nodata & 45 & 621 &  &
\nodata &    8.0 &  110.2 \\
OMC2/3 & 5 36 05, -5 14 40 & 5 34 40, -4 46 40 &    8.1 & & 55 & 51 & 286 &   & 6.8 & 
   6.3 &   35.5 \\
OMC4 & 5 35 22, -5 47 42 & 5 34 42, -5 28 42 &    2.6 & & 19 & 19 & 190 &   & 7.4 & 
   7.4 &   73.9 \\
DLSF & 5 36 15, -5 48 35 & 5 35 35, -5 24 35 &    3.2 & & 36 & 1 & 65 &  & 11.1 & 
   0.3 &   20.0 \\
Bending & 5 34 45, -5 43 30 & 5 33 30, -5 21 30 &    5.6 & &47 & 11 & 90 &  &  8.4
 &    2.0 &   16.1 \\
South & 5 38 00, -6 55 30 & 5 34 30, -5 45 30 &   49.8 && 209 & 61 & 366 &  &  4.2 & 
   1.2 &    7.4 \\
\enddata
\tablenotetext{\dag}{Numbers of 1.1 mm dust cores without YSOs.}
\label{Spitzer_dist}
\end{deluxetable}

%% file: Table6.tex
%\begin{deluxetable}{|l|ccccc|}
\begin{table}
\begin{center}
\caption{Kolmogorov-Smirnov Test for $R_{\rm core}$ and $dV_{\rm core}$ $\tablenotemark{\dag, \ddag}$ }
\begin{tabular}{|l|ccccc|}
\hline
\hline
$R_{\rm core}$ $\backslash$ $dV_{\rm core}$ & OMC1 & OMC23 & OMC4 & DLSF & Bending \\
\hline
OMC1     & \nodata       &   97.5\% &   67.5\% &   31.3\% &   97.5\% \\
OMC23    &  97.5\% &  \nodata        &   67.5\% &    3.1\% &   67.5\% \\
OMC4     &  67.5\% &   67.5\% &  \nodata       &   67.5\% &   97.5\% \\
DLSF     &  11.1\% &   11.1\% &   31.3\% &  \nodata      &   67.5\% \\
Bending  &  67.5\% &   97.5\% &   97.5\% &   11.1\% &  \nodata        \\
\hline
\end{tabular}
\label{KStest_Rcore_dV}
\tablenotetext{\dag}{Values in the upper triangular portion are the probabilities of the KS test for $dV_{\rm core}$.}
\tablenotetext{\ddag}{Values in the lower triangular portion are the probabilities of the KS test for $R_{\rm core}$.}
\end{center}
\end{table}
%\end{deluxetable}

%% file: Table7.tex
%\begin{deluxetable}{|l|c|c|c|c|c|}
\begin{table}
\begin{center}
\caption{Kolmogorov-Smirnov Test for $R_{\rm vir}$ and $M_{\rm LTE}$ $\tablenotemark{\dag, \ddag}$}
\begin{tabular}{|l|ccccc|}
\hline
\hline
$\mathcal{R}_{\rm vir}$ $\backslash$ $M_{\rm LTE}$ & OMC1 & OMC23 & OMC4 & DLSF & Bending \\
\hline
OMC1     & \nodata       &   97.5\% &   11.1\% &    0.7\% &   67.5\% \\
OMC23    &  97.5\% &  \nodata        &   11.1\% &    0.7\% &   67.5\% \\
OMC4     &  67.5\% &   31.3\% &  \nodata       &   67.5\% &   67.5\% \\
DLSF     &  31.3\% &   11.1\% &   31.3\% &  \nodata      &   11.1\% \\
Bending  &  97.5\% &   67.5\% &   97.5\% &   67.5\% &  \nodata        \\
\hline
\end{tabular}
\label{KStest_Ratio_MLTE}
\tablenotetext{\dag}{Values in the upper triangular portion are the probabilities of the KS test for $M_{\rm LTE}$.}
\tablenotetext{\ddag}{Values in the lower triangular portion are the probabilities of the KS test for $R_{\rm vir}$.}
\end{center}
\end{table}
%\end{deluxetable}

%% file: Table8.tex
\begin{deluxetable}{lcccc}
%\tabletypesize{\scriptsize}
\tablecaption{Numbers and fractions of the 1.1 mm dust and C$^{18}$O cores in each category}
\tablewidth{0pt}
\tablehead{
\colhead{} 
&\colhead{1.1 mm core } 
&\colhead{}
&\colhead{C$^{18}$O} 
&\colhead{} \\
\cline{3-5}
\\
\colhead{} 
&\colhead{} 
&\colhead{bound}
&\colhead{unbound}
&\colhead{total} 
}\startdata
Category A &       69/257 (26.8 \%) &      56/156 (35.9 \%) &
      13/57 (22.8 \%) &      69/213 (32.4 \%) \\
Category B &       10/257 (3.9 \%) &      19/156 (12.2 \%) &
       4/57 (7.0 \%) &      23/213 (10.8 \%) \\
Category C &  \nodata &       81/156 (51.9 \%) &      40/57 (70.2
 \%) &     121/213 (56.8 \%) \\
Category D &      178/257 (69.3 \%) &\nodata   &\nodata   &\nodata \\
\enddata
\label{category_table}
\end{deluxetable}

%% file: Table9.tex
% produced on Fri May 30 12:13:30 2014

\begin{deluxetable}{lcc}
\tablecaption{Area definition of PDRs}
\tablewidth{0pt}
\tablehead{
\colhead{} &
\multicolumn{2}{c}{Area definition} \\
\cline{2-3}\\
\colhead{Region}
&\colhead{BLC}
&\colhead{TRC}\\
\cline{2-3}\\
\colhead{} 
&\colhead{RA$_{\rm J2000}$, DEC$_{\rm J2000}$} 
&\colhead{RA$_{\rm J2000}$, DEC$_{\rm J2000}$}
}\startdata
DLSF          & 5 36 15, -5 48 35 & 5 35 35, -5 24 35 \\
M43          &  5 36 00, -5 24 35 & 5 35 25, -5 13 06   \\
Region A    & 5 34 48, -5 20 54 & 5 34 04, -5 13 06  \\
Region B    & 5 34 08, -5 26 06 & 5 33 49, -5 20 36 \\
Region C    & 5 34 01, -5 41 30 & 5 35 27, -5 37 30  \\
Region D    & 5 35 35, -5 42 18 & 5 35 24, -5 38 00  \\
\enddata
\label{area_definition}
\end{deluxetable}

%% file: Table10.tex
\begin{deluxetable}{ccccccccc}
\tabletypesize{\scriptsize}
\tablecaption{Numbers and fractions of the 850 $\mu$m, C$^{18}$O, H$^{13}$CO$^+$, and N$_2$H$^+$ cores associated with the 1.1 mm dust cores}
\tablewidth{0pt}
\tablehead{
\colhead{850 $\mu$m $\tablenotemark{\dag}$ } 
&\colhead{} 
&\colhead{} 
&\colhead{C$^{18}$O (1--0)}
&\colhead{} 
&\colhead{} 
&\colhead{H$^{13}$CO$^+$ (1--0)$\tablenotemark{\dag}$} 
&\colhead{} 
&\colhead{N$_2$H$^{+}$ (1--0)$\tablenotemark{\dag}$} 
\\
\cline{1-1}
\cline{3-5}
\cline{7-7}
\cline{9-9}
\\
\colhead{} 
&\colhead{} 
&\colhead{bound} 
&\colhead{unbound}
&\colhead{total}
&\colhead{} 
&\colhead{} 
&\colhead{} 
&\colhead{}  }
\startdata
     133/215 (61.9 \%) & & 
      75/156 (48.1 \%) &
      17/57 (29.8 \%) &
      92/213 (43.2 \%) & &
     119/210 (56.7 \%) & &
      21/27 (77.8 \%) \\
\enddata
\label{association_rate}
\tablenotetext{\dag}{
Catalogs of the 850 $\mu$m, H$^{13}$CO$^+$, and N$_2$H$^{+}$ cores are from \citet{Nutter07}, \citet{Ikeda07}, and \citet{Tatematsu08}, respectively.}
\end{deluxetable}